\documentclass[11pt,a4paper]{article}
\pdfoutput=1
\usepackage{epsfig,amsfonts,amssymb}
\usepackage{graphicx}
\usepackage{mathrsfs}
\usepackage{comment}
\usepackage{cite}
\usepackage{graphicx}
\usepackage{float}
\usepackage[normalem]{ulem}
\usepackage[utf8]{inputenc}
\usepackage[T1]{fontenc}
\usepackage{lmodern, textcomp}
\usepackage[english]{babel}
\usepackage{csquotes,cancel}
\usepackage[left=1in,right=1in,top=.8in,bottom=.9in]{geometry}

\usepackage{eurosym, xspace}
\usepackage{graphicx, subfig, float}
\usepackage{verbatim}

\usepackage{amsmath, amsfonts, amssymb,upgreek}
\usepackage{cancel}
\usepackage{mathtools}
\usepackage{slashed}

\usepackage[usenames, dvipsnames]{xcolor}
\usepackage[amsmath, hyperref, framed]{}%{ntheorem}
\usepackage{hhline}
\usepackage{bm}
\usepackage{array, caption}
\usepackage{array,multirow}

\newcolumntype{C}[1]{>{\centering\arraybackslash}m{#1}}

\usepackage[pdftex, breaklinks=true, linktocpage,
	colorlinks=true, urlcolor=blue, linkcolor=blue, citecolor=red
]{hyperref}
\makeatletter
\newsavebox{\@brx}
\newcommand{\llangle}[1][]{\savebox{\@brx}{\(\m@th{#1\langle}\)}%
  \mathopen{\copy\@brx\kern-0.5\wd\@brx\usebox{\@brx}}}
\newcommand{\rrangle}[1][]{\savebox{\@brx}{\(\m@th{#1\rangle}\)}%
  \mathclose{\copy\@brx\kern-0.5\wd\@brx\usebox{\@brx}}}
\makeatother

\def\ZZZ{{\hbox{ Z\kern-1.6mm Z}}}
\def\RRR{{\hbox{ R\kern-2.4mm R}}}
\def\CCC{{\hbox{ C\kern-2.0mm C}}}
\def\zzz{{\hbox{z\kern-1mm z}}}

\newcommand{\qeq}{{\hbox{=\kern-2.3mm ? \kern.5mm }}}
\renewcommand{\qeq}{=}

\newcommand{\be}{\begin{eqnarray}}
\newcommand{\ee}{\end{eqnarray}}

\newcommand{\vp}{\varphi}

\newcommand{\ben}{\begin{eqnarray}\displaystyle}
\newcommand{\een}{\end{eqnarray}}

\newcommand{\p}{\partial}
\renewcommand{\d}{\partial}

\def\one{{\hbox{ 1\kern-.8mm l}}}
\def\zero{{\hbox{ 0\kern-1.5mm 0}}}

\newcommand{\bea}[1]{\begin{eqnarray}\label{#1} }
\newcommand{\eea}{\end{eqnarray}}

\usepackage{bm}
\newcommand\non{\nonumber}

\newcommand\f{\frac}

\def\figone{

\def\JPicScale{0.8}
\ifx\JPicScale\undefined\def\JPicScale{1}\fi
\unitlength \JPicScale mm
\begin{picture}(135,80)(0,0)
\linethickness{0.3mm}
\multiput(40,80)(0.12,-0.18){167}{\line(0,-1){0.18}}
\linethickness{0.3mm}
\multiput(30,70)(0.18,-0.12){167}{\line(1,0){0.18}}
\linethickness{0.3mm}
\put(30,50){\line(1,0){30}}
\linethickness{0.3mm}
\multiput(30,30)(0.18,0.12){167}{\line(1,0){0.18}}
\linethickness{0.3mm}
\multiput(40,20)(0.12,0.18){167}{\line(0,1){0.18}}
\linethickness{0.3mm}
\put(60,50){\line(1,0){40}}
\linethickness{0.3mm}
\multiput(100,50)(0.12,0.18){167}{\line(0,1){0.18}}
\linethickness{0.3mm}
\multiput(100,50)(0.18,0.12){167}{\line(1,0){0.18}}
\linethickness{0.3mm}
\put(100,50){\line(1,0){30}}
\linethickness{0.3mm}
\multiput(100,50)(0.18,-0.12){167}{\line(1,0){0.18}}
\linethickness{0.3mm}
\multiput(100,50)(0.12,-0.18){167}{\line(0,-1){0.18}}

\put(30,80){\makebox(0,0)[cc]{$\zeta Q_B A_1^c$}}

\put(25,70){\makebox(0,0)[cc]{$A_2^c$}}

\put(25,50){\makebox(0,0)[cc]{$A_n^c$}}

\put(20,30){\makebox(0,0)[cc]{$A_1^o$}}

\put(35,20){\makebox(0,0)[cc]{$A_p^o$}}

\put(35,60){\makebox(0,0)[cc]{$\vdots$}}

\put(40,30){\makebox(0,0)[cc]{$\vdots$}}

\put(125,80){\makebox(0,0)[cc]{$B_1^c$}}

\put(135,70){\makebox(0,0)[cc]{$B_2^c$}}

\put(135,50){\makebox(0,0)[cc]{$B_m^c$}}

\put(135,30){\makebox(0,0)[cc]{$B_1^o$}}

\put(120,15){\makebox(0,0)[cc]{$B_q^o$}}

\put(115,35){\makebox(0,0)[cc]{$\vdots$}}

\put(120,57){\makebox(0,0)[cc]{$\vdots$}}

\put(66,47){\makebox(0,0)[cc]{$\psi_r^c\vp_r$}}

\put(93,47){\makebox(0,0)[cc]{$\psi_s^c\vp_s$}}

\end{picture}

}

\def\figtwo{

\def\JPicScale{0.8}
\ifx\JPicScale\undefined\def\JPicScale{1}\fi
\unitlength \JPicScale mm
\begin{picture}(135,80)(0,0)
\linethickness{0.3mm}
\multiput(40,80)(0.12,-0.18){167}{\line(0,-1){0.18}}
\linethickness{0.3mm}
\multiput(30,70)(0.18,-0.12){167}{\line(1,0){0.18}}
\linethickness{0.3mm}
\put(30,50){\line(1,0){30}}
\linethickness{0.3mm}
\multiput(30,30)(0.18,0.12){167}{\line(1,0){0.18}}
\linethickness{0.3mm}
\multiput(40,20)(0.12,0.18){167}{\line(0,1){0.18}}
\linethickness{0.3mm}
\put(60,50){\line(1,0){40}}

\put(30,80){\makebox(0,0)[cc]{$\zeta Q_B A_1^c$}}

\put(25,70){\makebox(0,0)[cc]{$A_2^c$}}

\put(25,50){\makebox(0,0)[cc]{$B_m^c$}}

\put(20,30){\makebox(0,0)[cc]{$A_1^o$}}

\put(35,20){\makebox(0,0)[cc]{$B_q^o$}}

\put(35,60){\makebox(0,0)[cc]{$\vdots$}}

\put(40,30){\makebox(0,0)[cc]{$\vdots$}}

\put(66,47){\makebox(0,0)[cc]{$\psi_r^c\vp_r$}}

\put(93,45){\makebox(0,0)[cc]{$\tilde\psi_s^c\tilde\vp^s$}}

\put(100,50){\makebox(0,0)[cc]{$\times$}}

\end{picture}

}

\definecolor{armygreen}{rgb}{0.29, 0.33, 0.13}

\begin{document}

\baselineskip 24pt

\begin{center}
{\Large \bf  %Momentum Space Holographic Toolkits for Massive spin-2 Fields\\
%From AdS to Flat Space: Momentum Space Holography for Massive Spin-2\\
From AdS to Flat Space: Massive Spin-2 Fields\\}
%Toward Flat-Space Physics from AdS: Momentum-Space Holography of Massive Spin-2\\
%Momentum-Space AdS/CFT for Massive Spin-2 Fields and Flat Limit}

\end{center}

\vskip .6cm
\medskip

\vspace*{4.0ex}

\baselineskip=18pt

\begin{center}

{\large 
\rm   Vijay Nenmeli${}^a$, Arvind Shekar${}^b$ and Mritunjay Verma${}^c$ }

\end{center}

\vspace*{4.0ex}

\centerline{ \it \small ${}^a$ School of Mathematics and Maxwell Institute for Mathematical Sciences,}
\centerline{ \it \small University of Edinburgh, Peter Guthrie Tait road, Edinburgh EH9 3FD, UK}

\centerline{ \it \small ${}^b$ Mathematical Sciences and STAG Research Centre, University of Southampton,}
\centerline{\it \small  Highfield, Southampton SO17 1BJ, UK}

\centerline{\it \small ${}^c$ Indian Institute of Technology Indore, Khandwa Road, Simrol, Indore 453552, India}
\vspace*{1.0ex}
\centerline{\small E-mail:  V.V.Nenmeli@sms.ed.ac.uk, A.Shekar@soton.ac.uk, 
mritunjay@iiti.ac.in }

\vspace*{5.0ex}

\centerline{\bf Abstract} \bigskip

We analyze a bulk effective field theory in AdS containing a U(1)-charged massive spin-2 field coupled to a gauge field, by performing the required holographic renormalization, and computing the one and two-point functions. We then compute the renormalized bulk three-point function involving two massive spin-2 fields and one gauge field. Matching with the CFT 3-point correlator of two non-conserved spin-2 operators and a conserved current, we obtain explicit mappings between the bulk minimal and gyromagnetic couplings and the boundary OPE data. Finally, we take the flat-space limit of the momentum space CFT correlator and verify that the resulting amplitude matches the expected flat-space structure.

\vfill

\vfill \eject

\baselineskip18pt

\tableofcontents

\section{Introduction } 
Massive higher spin fields are expected to play an important role in the classification of consistent gravitational theories (see \cite{Rahman:2015pzl, Sleight:2017krf} and references therein for review and literature on the higher spin fields in general backgrounds, including AdS). In particular, it can be argued that an infinite tower of massive higher spin fields are needed for the gravitational theories to be consistent with causality at high energies \cite{Camanho:2014apa}. However, the requirement of causality only forces the inclusion of a higher spin field above a certain energy scale. In an effective field theory (EFT), an interacting theory of a single massive higher spin field can be consistent below some energy scale when the particle masses are large compared to the exchanged momenta. Such consistent EFTs of massive higher spin fields can be useful for some purposes, e.g., for describing the higher spin resonances in colliders.

\vspace*{.07in}In spite of the utility of massive higher spin fields, it is very tricky to construct a consistent EFT of single massive higher spin fields. An EFT obtained by truncating the degrees of freedom above some energy scale will, in general, still be inconsistent in a given background. A classic example of this is provided by the electromagnetic couplings of the massive higher spin fields in the flat spacetimes. It is possible to write down the kinetic terms for the massive higher spin fields in flat spacetimes \cite{singh1, singh2}. However, including interactions is subtle. E.g., the minimal electromagnetic coupling of massive higher spin fields (with spins greater than 1) introduces new unphysical degrees of freedom and leads to superluminal propagation of signals. This is the so called Velo-Zwanziger instability \cite{velo1,velo2,velo3,velo4,velo5,velo6,velo7,velo8,velo9,velo10,velo11,Porrati:2010hm}. The current understanding is that these instabilities appear when we use the EFT beyond its domain of validity \cite{Porrati:2010hm}. The analysis involving spin-2 fields indicates that the propagation of the correct number of degrees of freedom requires constraints on the gyromagnetic ratio. However, the gyromagnetic ratio required for this purpose, which is $\f{1}{2}$, is in contradiction with the value $2$ required for the unitary behaviour of Compton scattering of these fields in UV \cite{Ferrara:1992yc
} (see \cite{2102.13180} for a recent discussion of the gyromagnetic ratio of massive higher spin fields). As argued in \cite{Porrati:2010hm}, this apparent contradiction can be resolved if we note that the extra degrees of freedom arise only if we push the theory to higher orders in the perturbation theory. Further, the restoration of causality requires adding specific nonlinear terms \cite{Porrati:2010hm}.  

A consistent interacting theory involving a single higher massive spin field and gravity is also very difficult to construct. It was argued in \cite{Camanho:2014apa} that 3-point higher derivative graviton couplings not present in Einstein's theory would lead to non causal propagation of gravitons in non trivial backgrounds, and to restore causality, one needs to include an infinite tower of massive higher spin fields. On the boundary CFT side, this is reflected, e.g., in the analysis of 4 point function involving two stress tensor and 2 scalar fields. In the Regge limit, this correlator can get contributions from the exchange of stress energy tensor. This contribution is controlled by the 3-point function involving the stress energy tensor and contains causality violating terms \cite{Afkhami-Jeddi:2016ntf}. One can show that including non-conserved spin-2 operators changes the Regge behaviour, but it is not enough to cancel the causality violating terms. One needs to include an infinite tower of non-conserved higher spin operators in the spectrum \cite{Afkhami-Jeddi:2016ntf}. 

\vspace*{.07in}The root cause of the above issue is the application of the EFT beyond its regime of validity. E.g., in the CEMZ analysis, the acausal behaviour arises if the higher derivative terms become important in the EFT regime and dominate Einstein's term \cite{Camanho:2014apa}. This shows that we can talk about the EFT of single massive higher spin fields coupled to gravity at least in some regimes. This is the spirit in which we shall work in this paper. However, even if the causality problem is taken care of, putting massive higher spin fields on a generic curved background also has other problems, such as non unitary behaviour. Even at the free field level, in a curved background, the derivatives of the equations of motion may involve second and higher derivatives of the massive fields and may no longer remain constraint equations. This changes the expected number of degrees of freedom of the massive higher spin field. It is possible to prevent the constraint equations from becoming dynamical by putting suitable restrictions on the curved background. In particular, in the $d+1$ dimensional Einstein spaces satisfying the condition $R_{MN}=\f{R}{d+1}G_{MN}$ (where $R_{MN}$ and $G_{MN}$ denote the Ricci and metric tensors respectively), one can prevent the constraint equations from becoming dynamical \cite{Buchbinder:1999be, Buchbinder:2000fy}.    

\vspace*{.07in}Due to the above feature of AdS, it is a useful background for studying EFTs of massive higher spin fields and their interactions. The cubic couplings of higher spin fields have been studied in \cite{Sleight:2016dba, Sleight:2016xqq}. However, a systematic analysis of their correlators using the AdS/CFT dictionary is missing. In particular, the holographic renormalization of massive higher spin fields, which is needed to deal with the boundary UV divergences, is missing. Such results would be useful for those analysis which require correlators of massive higher spin fields, such as the study of causality properties. For developing the tools for massive higher spins in AdS, the spin-2 case is a useful starting point (see, e.g., \cite{Afkhami-Jeddi:2016ntf}). Motivated by this, in this paper, we shall consider some aspects of massive spin-2 fields in the AdS background (see, e.g., \cite{Hinterbichler:2011tt, deRham:2014zqa} for reviews on massive spin-2 fields). Since it satisfies the conditions of Einstein spaces, one can write down the kinetic term for the massive spin-2 fields in this background. It turns out that there is a one-parameter family of allowed kinetic terms in Einstein spaces parametrized by a number $\xi$ \cite{Buchbinder:1999be}. We shall work with the simplest choice of this parameter, namely $\xi=1$. For this value, one obtains the expected AdS relation $m^2L^2= \Delta(\Delta-d)$, for the massive spin-2 fields. Our main goal will be to develop the momentum space tools for studying massive spin-2 fields in AdS. In particular, we shall study the free field solution of the massive spin-2 field and obtain its bulk-to-boundary (Btb) propagator in the momentum space. This will then be used to perform the holographic renormalization of the massive spin-2 field and compute its one and two point functions.

\vspace*{.07in}We shall also consider the interaction of complex massive spin-2 fields with the gauge field in AdS. We shall use the AdS/CFT prescription and holographic renormalization to study the 3-point correlator in this theory. To obtain the CFT 3-point function, we only need to turn on the boundary sources infinitesimally which are switched off after computing the correlators. Hence, the backreaction due to the matter fields can be ignored. Further, since the field strength of the gauge field will be taken to be very small, the causal propagation of the fields will not be affected, and we won't have to worry about the Velo-Zwanziger causality problem. 

\vspace*{.07in}Our focus will be on the 3-point correlator of two massive spin-2 fields and a gauge field. This will be a function of the bulk couplings. For the massive spin-2 field interacting with a gauge field, involving up to 3 derivatives, there are 5 bulk couplings: minimal coupling $g$, gyromagnetic coupling $\alpha$, and 3 higher derivative couplings $\beta_1,\beta_2$, and $\beta_3$. The higher derivative couplings are expected to play an important role in the Regge limit \cite{Camanho:2014apa, Afkhami-Jeddi:2016ntf}. However, in this work, we shall mostly focus on the minimal and gyromagnetic couplings. The 3-point function computed in the bulk would be a function of these couplings. We shall map these to the CFT coefficients which appear in the 3-point CFT correlator of two non-conserved spin-2 and a conserved spin-1 current in the boundary theory. We shall be using the momentum space CFT techniques for the non-conserved CFT operators in our analysis, following \cite{Marotta:2022jrp, Marotta:2024sce}. 

\vspace*{.07in}We shall also consider the flat limit of the above AdS 3-point function. The polarizations of the massive spin-2 field in the flat limit will be dictated by the Btb propagator. In the flat limit, we shall show that the AdS 3-point function involving two massive complex spin-2 fields and a gauge field matches the expected 3-point function in the flat space-time. We shall follow the approach developed in \cite{Marotta:2024sce} for taking the flat limit of CFT correlators in the momentum space. %The resulting 3-point function agrees with the expected result in the flat spacetimes. 

\vspace*{.07in}The rest of the paper is organised as follows. In section \ref{sec:rev}, we review some results which are needed in this work. In particular, we summarise the momentum space CFT 3-point function of a conserved current and two spin-2 non-conserved operators following \cite{Marotta:2022jrp}. We also review the analysis of massive spin-2 fields in a curved background having constant curvature, following \cite{Buchbinder:1999be, Buchbinder:2000fy}. Even though this paper is about the massive spin-2 fields, in section \ref{sec3}, we shall consider massive scalar $\phi^3$ theory in AdS and show how the flat limit of AdS, using the momentum space CFT techniques, works for the massive fields. In particular, we shall show how the tree level AdS 3-point function reduces to the 3-point vertex factor in the flat spacetime. In section \ref{sec4}, we shall analyze the free massive spin-2 field in AdS and derive its Btb propagator in the Fefferman Graham coordinates. In section \ref{sec5}, we shall consider the holographic renormalization of the massive spin-2 field and compute one and two point functions. In section \ref{s3a}, we shall compute the bulk 3-point function involving two massive spin-2 fields and a gauge field. We shall also match this with the boundary result, obtaining relations between the bulk gyromagnetic coupling and the boundary OPE coefficients. In section \ref{sec7}, we shall consider the flat limit of the AdS Btb propagator and the 3-point function using the momentum space CFT techniques and show that we obtain the expected 3-point function in the flat space. We shall end with some discussion in section \ref{s4}. The appendices contain some useful results and details of some calculations done in this work. In appendix \ref{geo768}, we summarise our notations and conventions and note some useful identities. In appendix \ref{sec5:s2}, we shall note the detailed expression of the form factors which appear in the CFT 3-point function of two non-conserved spin-2 operators and a conserved current in terms of the triple K integrals. In appendix \ref{appenF}, we shall give the expressions of the Btb propagator of the massive spin-2 field in the Poincare coordinates. In appendices \ref{G} and \ref{secf}, we shall give some details of the bulk 3-point function and the flat limit computations respectively. Finally, in appendix \ref{expflat3point}, we shall review the expected 3-point function of two massive spin-2 fields and a gauge field in the flat space.

\section{Review}
\label{sec:rev}
\subsection{CFT correlators involving spin-2 operators}
\label{sec2review3er}

In this section, we summarise the momentum space CFT 3-point function involving non-conserved spin-2 operators and a conserved current following \cite{Marotta:2022jrp}. Extracting the momentum conserving delta function, we can write 
\be
\mathcal{A}_3^{\mu_1\mu_2\mu_3\mu_4\mu_5}&=& (2\pi)^d \delta^d({\bf p}_1+{\bf p}_2+{\bf p}_3) \;\llangle[\Bigl]\mathcal{O}_{1}^{*\mu_1\mu_2}({ \bf p}_1) {\cal J}^{\mu_3}({\bf p}_2) \mathcal{O}_{3}^{\mu_4\mu_5}({\bf p}_3)   \rrangle[\Bigl].  
\ee
The double bracket notation just means that we have extracted the overall momentum conserving delta function. The operators $\mathcal{O}_i$ have conformal dimensions $\Delta_i$ and spin-2. Below, we shall only consider the case where $\Delta_1=\Delta_3\equiv \Delta$ since in the bulk they would be dual to the massive spin-2 particles of same mass. The conserved current $\mathcal{J}$ has conformal dimension $d-1$. For stating the results, it is convenient to make use of the auxiliary polarization tensors and avoid the explicit Lorentz indices as in \cite{Marotta:2022jrp}. Hence, we express the reduced correlator as 
\be
\mathcal{A}_{3}(\textbf{p}_1,\textbf{p}_2)&\equiv&\llangle[\Bigl] \epsilon_1\cdot\mathcal{O}^*_{1}({\bf p}_1)\epsilon_2\cdot J({\bf p}_2) \epsilon_3\cdot\mathcal{O}_{3}({\bf p}_3)   \rrangle[\Bigl]\,,
\ee
where $\epsilon\cdot\mathcal{O}= \epsilon^{\mu\nu}\mathcal{O}_{\mu\nu}$ and $\epsilon\cdot{J}= \epsilon^{\mu}{J}_{\mu}$. Since we shall be working with the symmetric traceless operators $\mathcal{O}_i$, we shall write $\epsilon^{\mu\nu}=\epsilon^\mu \epsilon^\nu$ and impose the condition $\epsilon^\mu \epsilon_\mu=0$. For more details on use of auxiliary polarization tensors in this context, see e.g., \cite{Marotta:2022jrp, costa, Bergman}. Next, we split the reduced correlator in terms of the transverse and longitudinal parts \cite{Marotta:2022jrp, Bzowski:2013sza} and write
\be
\mathcal{A}_{3}(p_1,p_2)&=& \llangle[\Bigl] \epsilon_1\cdot\mathcal{O}^*({\bf{p}_1}) \epsilon_2\cdot j({\bf p}_2) \epsilon_3\cdot\mathcal{O}({\bf p}_3)\rrangle[\Bigl]\;\;+\;\;\f{\epsilon_2\cdot p_2}{p_2^2}\llangle[\Bigl]  \epsilon_1\cdot\mathcal{O}^*({\bf p}_1)p_{2\nu} J^\nu({\bf p}_2) \epsilon_3\cdot\mathcal{O}({\bf p}_3)\rrangle[\Bigl]\non\\
&\equiv&\mathcal{A}^\perp +\mathcal{A}^{||}\,. \label{3.21}
\ee
The $\mathcal{A}^\perp$ denotes the transverse and $\mathcal{A}^{||}$ denotes the longitudinal part. The $j^\mu$ denotes the transverse part of the current defined as $j^\mu = J^\nu\,\pi_{2\nu}^{~\;\mu}$, where the projector $\pi^{\mu\nu}$ is defined as  
\be
\pi_2^{\mu\nu} =\delta^{\mu\nu}-\f{p_2^\mu p_2^\nu}{p_2^2}\qquad;\qquad p_2^\mu\; \pi_{2\,\mu\nu} =0\quad\implies\quad p_\mu j^\mu=0\,.
\ee
The transverse part $\mathcal{A}^\perp$ can be expressed as 
\be
\mathcal{A}^\perp({\bf p}_1,{\bf p}_2) &=&(\epsilon_2\cdot \pi_2\cdot p_1)A+ (\epsilon_2\cdot \pi_2\cdot \epsilon_1)B_1 +(\epsilon_2\cdot \pi_2\cdot \epsilon_3)B_2\,,
\label{ansatzs}
\ee
with the functions $A, B_1$ and $B_2$ parametrized as
\be
A&=&z^2 A_0^{(0,0)}+\zeta _2 \xi _2 z A_1^{(0,0)}+ \frac{1}{2} \zeta _2^2 \xi _2^2 A_2^{(0,0)}+\zeta _1 \xi _2 z A_1^{(0,1)}+\zeta _2 \xi _1 z A_1^{(1,0)}+\zeta _1 \xi _1 z A_1^{(1,1)}\non\\
&&+\frac{1}{2} \zeta _1 \zeta _2 \xi _2^2 A_2^{(0,1)}+\frac{1}{4} \zeta _1^2 \xi _2^2 A_2^{(0,2)}+\frac{1}{2} \zeta _2^2 \xi _1 \xi _2 A_2^{(1,0)}+\frac{1}{2} \zeta _1 \zeta _2 \xi _1 \xi _2 A_2^{(1,1)}+\frac{1}{4} \zeta _1^2 \xi _1 \xi _2 A_2^{(1,2)}\non\\
&&+\frac{1}{4} \zeta _2^2 \xi _1^2 A_2^{(2,0)}+\frac{1}{4} \zeta _1 \zeta _2 \xi _1^2 A_2^{(2,1)}+\frac{1}{8} \zeta _1^2 \xi _1^2 A_2^{(2,2)},\non\\[.3cm]
B_1&=&\xi _1 z {B}_{1;0}^{(1,0)}+\xi _2 z {B}_{1;0}^{(0,0)}+\zeta _2 \xi _2 \xi _1 {B}_{1;1}^{(1,0)}+\zeta _1 \xi _2 \xi _1 {B}_{1;1}^{(1,1)}+\zeta _2 \xi _2^2 {B}_{1;1}^{(0,0)}+\zeta _1 \xi _2^2 {B}_{1;1}^{(0,1)}\non\\
&&+\frac{1}{2} \zeta _2 \xi _1^2 {B}_{1;1}^{(2,0)}+\frac{1}{2} \zeta _1 \xi _1^2 {B}_{1;1}^{(2,1)},\non\\[.3cm]
B_2&=&\zeta _1 z {B}_{2;0}^{(0,1)}+\zeta _2 z {B}_{2;0}^{(0,0)}
+\zeta _2 \zeta _1 \xi _2 {B}_{2;1}^{(0,1)}+\zeta _2 \zeta _1 \xi _1 {B}_{2;1}^{(1,1)}+\zeta _2^2 \xi _2 {B}_{2;1}^{(0,0)}+\zeta _2^2 \xi _1 {B}_{2;1}^{(1,0)}\non\\
&&+\frac{1}{2} \zeta _1^2 \xi _2 {B}_{2;1}^{(0,2)}+\frac{1}{2} \zeta _1^2 \xi _1 {B}_{2;1}^{(1,2)}.\label{8103e}
\ee
The tensor structures $z, \xi_i$ and $\zeta_i$ are defined as 
\be
z=\epsilon_1\cdot \epsilon_3\quad,\qquad \xi_1=\epsilon_3\cdot p_2\quad,\quad \xi_2=\epsilon_3\cdot (p_1+p_2)\quad,\quad \zeta_1=\epsilon_1\cdot p_2\quad,\quad \zeta_2=\epsilon_1\cdot p_1.
\ee
The 30 form factors $A_n^{(p,q)},\, B_{1;n}^{(p,q)}$ and $B_{2;n}^{(p,q)}$ appearing in \eqref{8103e} depend on the magnitudes of  the three momenta\footnote{In equation \eqref{ansatzs}, the momentum conserving delta function has been used to express $p_3^\mu=-p_1^\mu-p_2^\mu$.} $p_j=|{\bf p}_j| = \sqrt{{\bf p}_j^2}\  (j=1, 2, 3)$ and can be obtained by solving the Ward identities. They are given in terms of the triple K integrals. E.g., the form factor $A_0^{(0,0)}$ is given by
\be
A_0^{(0,0)}&=&a_0^{(0,0)} J_{1\{0,0,0\}}-a_1^{(1,1)} J_{2\{0,1,0\}}+\f{1}{4}a_2^{(2,2)} J_{3\{0,2,0\}},\label{28tyr}
\ee
where the triple K integrals $J_{N\{k_1,k_2,k_3\}}$ are defined by (see \cite{Marotta:2022jrp, Bzowski:2013sza, 1510.08442, Bzowski:2015yxv} for more details and useful properties of these integrals)
\begin{eqnarray}
J_{N\{k_1,k_2,k_3\}}(p_1,p_2,p_3)&\equiv&\int_0^\infty dx\,x^{\frac{d}{2}+N-1} \prod_{i=1}^3 p_i^{\Delta_i-\frac{d}{2}+k_i}\,K_{ \Delta_i-\frac{d}{2}+k_i}(x p_i)\, .	\label{B.51}
\end{eqnarray}
The coefficients $a_0^{(0,0)}, a_1^{(1,1)}$ and $a_2^{(2,2)}$ appearing in \eqref{28tyr} are independent of momenta but depend upon the conformal dimension $\Delta$ and the CFT spacetime dimension $d$. 

\vspace*{.07in}The other form factors in \eqref{8103e} are summarized in appendix \ref{sec5:s2}. The 3-point function of the conserved current and two spin-2 operators having the same conformal dimension $\Delta$ is given in terms of 5 independent parameters. This means that not all the coefficients $a_p^{(i,j)}$ are independent. In \eqref{8103e}, a total of 30 coefficients appear. The Ward identities fix 25 of them in terms of the remaining 5 independent coefficients. The relations between these 30 coefficients is also given in appendix \ref{sec5:s2}. 

\vspace*{.07in}The two point function of the spin-2 operators in momentum space is given by \cite{Marotta:2022jrp, Arkani-Hamed:2015bza}
\be
\llangle[\Bigl]  \epsilon_1\cdot\mathcal{O}^*({\bf p})\epsilon_2\cdot\mathcal{O}(-{\bf p})   \rrangle[\Bigl]&=&  \epsilon_1^{\mu\sigma}\epsilon_2^{\nu\rho} \left[a_0\delta_{\mu\nu}\delta_{\rho\sigma} +a_1\f{\delta_{\mu\nu}p_\rho p_\sigma}{ p^{2}}+a_2\f{p_{\mu}p_{\nu}p_\rho p_\sigma}{ 2p^{4}}\right]p^{2\Delta-d}\label{210trfd}
\ee
The coefficients $a_1$ and $a_2$ are given in terms of $a_0$ as
\be
a_1 = -\f{2(2\Delta-d)}{\Delta}a_0\qquad;\quad a_2 = \f{2(2\Delta-d)(2\Delta-d-2)}{\Delta(\Delta-1)}a_0
\ee
Since we only use the conformal properties to derive the above correlators, the two point function in \eqref{210trfd} reduces to the 2-point function of conserved stress energy tensor if we specialise the conformal dimension to $\Delta=d$. To see this, we note that we can express the 2-point function in \eqref{210trfd} in the form
\be
\llangle[\Bigl]  \epsilon_1\cdot\mathcal{O}^*({\bf p})\epsilon_2\cdot\mathcal{O}(-{\bf p})   \rrangle[\Bigl]&=& \epsilon_1^{\mu\sigma}\epsilon_2^{\nu\rho} a_0\left[\Pi_{\mu\sigma,\nu\rho}(p)+\f{(d-\Delta)p_\mu p_\nu}{\Delta(\Delta-1)(d-1)p^2}\Bigl( E\,\delta_{\sigma\rho} + F\f{p_\sigma p_\rho}{p^2}  \Bigl) \right]\label{2.20iyt}
\ee
where, $E=2(d-1)(\Delta-1), F= (2\Delta-3d\Delta +d^2+d-2)$ and 
\be
\Pi_{\mu\sigma,\nu\rho}(p) =\f{1}{2}\Bigl[\pi_{\mu\nu}(p)\pi_{\sigma\rho}(p)+\pi_{\mu\rho}(p)\pi_{\nu\sigma}(p)\Bigl]-\f{1}{d-1}\pi_{\mu\sigma}(p)\pi_{\nu\rho}(p).
\ee
The second term in \eqref{2.20iyt} vanishes for the conserved spin-2 currents satisfying $\Delta=d$ and the expression reduces to the 2-point function of stress energy tensor in momentum space (see, e.g., \cite{Bzowski:2013sza}). 

\vspace*{.07in}The Ward identities also relate the CFT coefficients appearing in 2 and 3-point functions. For the above described 2 and 3 point functions, the relation is given by
 \begin{eqnarray}
 a_0= \,2^{\frac{d}{2} -6}\,\frac{(d-2\Delta)}{g\,(d-2)} \,\Gamma\left(\frac{2\Delta-d}{2}\right)\Gamma\left( \frac{d-2\Delta}{2} \right)\,\Gamma\left(\frac{d}{2}\right)\,\left[4 a_0^{(0,0)}
+(d-2) (-4a_1^{(1,1)} +d a_2^{(2,2)})\right]\label{214rtyd}
\end{eqnarray} 
This identity is also a consequence of the relation between the longitudinal part of the 3-point function in \eqref{3.21} and the 2-point function of the spin-2 operators. The generating functional of the boundary CFT correlators involving the spin-2 operators and a conserved spin-1 current is given by 
\be
Z[{A}^{}_{(0)}, \phi_{(0)},\phi^{*}_{(0)}]= \int \mathcal{D}\Phi\;\exp\biggl[-S_{CFT}\; -\int d^dx \Bigl(\mathcal{J}^\mu {A}^{}_{(0)\mu} + \mathcal{O}^{*\mu\nu}\phi^{}_{(0) \mu\nu}+ \mathcal{O}^{\mu\nu}\phi^{*}_{(0) \mu\nu}\Bigl)\biggl]
\ee
The ${A}^{}_{(0) \mu}, \phi^{}_{(0) \mu\nu}$ and $\phi^{*}_{(0) \mu\nu}$ are the sources for the CFT operators $\mathcal{J}^\mu, \mathcal{O}^{*\mu\nu} $ and $\mathcal{O}^{\mu\nu}$, respectively and determine the boundary conditions of the corresponding bulk fields in the AdS CFT correspondence. The generating function is invariant under the $U(1)$ transformation 
\be
\delta A_{(0) \mu}(\textbf{x}) = \p_\mu\lambda(\textbf{x})\;;\quad \delta \phi^{}_{(0)\mu\nu}(\textbf{x}) = i g\lambda(\textbf{x}){\phi^{}_{(0) \mu\nu}}(\textbf{x}) \;;\quad \delta \phi^{*}_{(0)\mu\nu}(\textbf{x}) = -i g\lambda(\textbf{x}){\phi^{*}_{(0) \mu\nu}}(\textbf{x})
\ee
This invariance gives the conservation Ward identity
\begin{equation}
    \partial^\mu \langle \mathcal{J}^\mu(\textbf{x}) \rangle_s 
    = i g \left(
    \phi^{}_{(0) \mu\nu}(\textbf{x}) \langle \mathcal{O}^{*\mu\nu}(\textbf{x}) \rangle_s
    -\phi^{*}_{(0) \mu\nu}(\textbf{x}) \langle \mathcal{O}^{\mu\nu}(\textbf{x}) \rangle_s
    \right)\, ,
\end{equation}
where the expectation value in the above equation are valid in the presence of the sources. If we differentiate w.r.t. $\phi^{}_{(0) {\mu_1\mu_2}}(\textbf{x}_1), \phi^{}_{(0) \mu_4\mu_5}(\textbf{x}_3)$ and Fourier transforming to momentum space gives,
\be
&&\hspace*{-1.8cm}\llangle[\Bigl] \mathcal{O}^{*\mu_1\mu_2}({\bf p}_1)\, {\bf p}_{2\mu_3}{\cal J}^{\mu_3}({\bf p}_2) \,\mathcal{O}^{\mu_4\mu_5}({\bf p}_3)    \rrangle[\Bigl]\non\\[.2cm]
&=&  g
\llangle[\Bigl]  \mathcal{O}^{*\mu_1\mu_2}(-{\bf p}_3) \,\mathcal{O}^{\mu_4\mu_5}({\bf p}_3) \rrangle[\Bigl]\;-\;g\llangle[\Bigl]\mathcal{O}^{*\mu_1\mu_2}({\bf p}_1) \,\mathcal{O}^{\mu_4\mu_5}(-{\bf p}_1) \rrangle[\Bigl] \,  \label{decnb4e}
\ee
The relation \eqref{214rtyd} is consistent with the above equation.

\subsection{Massive spin-2 fields in curved backgrounds}
\label{sec2.2}
In this subsection, we review the Lagrangian formulation of the free massive spin-2 field in $d+1$ dimensional curved backgrounds following \cite{Buchbinder:1999be, Buchbinder:2000fy}. As mentioned in the introduction, at the free field level, one can describe the massive higher spin fields in a Lagrangian framework in the flat spacetimes. However, if we put the fields on a curved background, the constraint equations may involve second derivatives of the massive field and may no longer remain constraint equations. Fortunately, in the  Einstein spaces satisfying the condition $R_{MN}=\f{R}{d+1}G_{MN}$, the constraint equations are prevented from becoming dynamical. In \cite{Buchbinder:1999be, Buchbinder:2000fy}, a detailed analysis is done for the real massive spin-2 field. Generalizing this to complex massive spin-2 fields is straightforward. The free field action of these fields in the Einstein spaces depends on a single real parameter, $\xi$, and takes the form
\begin{align}
S=&\int d^{d+1}x\;\sqrt{G}\biggl[\frac{1}{2}\nabla_M\phi^*_{NP}\nabla^M{\phi}^{NP} 
-\nabla_M{\phi^*}_{NP}\nabla^P\phi^{NM} -\frac{1}{2} \nabla_M{\phi^*}\nabla^M\phi+\frac{1}{2}\nabla_M{\phi^*}^{MN}\nabla_N {\phi}\nonumber\\
&+\frac{1}{2}\nabla_M{\phi}^{MN}\nabla_N {\phi^*}-\frac{\xi}{d+1}R\,\phi^*_{MN}{\phi}^{MN}-\frac{1-2\xi}{2(d+1)}R\,\phi^*\,\phi+\frac{m^2}{2} (\phi^*_{MN}{\phi}^{MN} -{\phi^*}\phi)\biggl]
\end{align}
 The $\phi^*_{MN}$ is the complex conjugate of $\phi_{MN}$ and will be treated as an independent field. The $\phi = G^{MN}\phi_{MN}$ denotes the trace of the spin-2 field. By varying with respect to $\phi^*_{MN}$, we obtain the equation of motion for $\phi_{MN}$. The variation of the action is given by 
\begin{align}
    \delta S=-\,\frac{1}{2}\int_{\mathcal{M}} d^{d+1} x \,\sqrt{G}\,E_{MN} \,\delta\phi^{*MN}+\int_{\d {\cal M}}\sqrt{\gamma}\, n_M \delta B^M
\end{align}
where the $B^M$ appearing in the boundary term is given by
\begin{align}
    B^M=\frac{1}{2}\bigg({\phi^*}_{NP}\nabla^M\phi^{NP}-2{\phi^*}_{NP}\nabla^P \phi^{NM}-\phi^* \nabla^M \phi+{\phi^*}^{MN} \nabla_N \phi+\phi^* \nabla_N \phi^{NM}\bigg)
\end{align}
and the equation of motion is $E_{MN}=0$ where
\begin{align}
    E_{MN} =& \nabla^2\phi_{MN}-2 \nabla^P \nabla_{(N}\phi_{M)P}-G_{MN}\nabla^2\phi+\nabla_M\nabla_N\phi+G_{MN}\nabla^P\nabla^Q\phi_{PQ}\non\\
    &+\frac{2\xi}{d+1}R\,\phi_{MN}+\frac{1-2\xi}{(d+1)}R\,\phi\,G_{MN}-{m^2}\left( \phi_{MN}- \phi\,G_{MN}\right)
\end{align}
The trace of the above equation gives
\begin{align}
   \nabla_M \nabla_N \phi^{MN}-\Box \phi+\frac{d(1-2\xi)+1}{(d+1)(d-1)}R\,\phi+d\,m^2\phi=0.
\end{align}
On the other hand, the single and double divergences (i.e. acting with $\nabla^{M}$) give the following subsidiary conditions
\be
\nabla^M E_{MN}&=&m^2(\nabla_N\phi-\nabla_M\phi^M_{\;\;N})  +R_{NM}(\nabla^M\phi-2\nabla_P\phi^{MP})+\frac{2\xi}{d+1}\,(\phi_{MN}\nabla^M R+R\nabla^M \phi_{MN})\non\\
&&+\frac{1-2\xi}{d+1}\left(\phi\nabla_N R+R\nabla_N\phi\right)-\phi^{MP}(\nabla_P R_{NM}+\nabla^Q R_{N MPQ})
\ee
\be
&&\hspace*{-.3in}\nabla^M\nabla^N E_{MN}\non\\
&=&m^2(\Box\phi-\nabla_M\nabla_N\phi^{MN})  +R^{MN}(\nabla_M\nabla_N\phi-2\nabla_M\nabla^P\phi_{NP})+\nabla_MR^{MN}(\nabla_N\phi -2\nabla^P\phi_{NP})\non\\
&&+\frac{1}{d+1}\Big\{
2\xi\Big[(\nabla^M\nabla^N R)\phi_{MN}
+2(\nabla^M R)\nabla^N\phi_{MN}
+R\,\nabla^M\nabla^N\phi_{MN}\Big]\non\\
&&+(1-2\xi)\Big[(\nabla^2 R)\phi+2(\nabla^N R)(\nabla_N\phi)+R\,\nabla^2\phi\Big]
\Big\}+\phi^{MN}(\nabla^P\nabla^Q R_{ MQNP}-\nabla^P\nabla_N R_{MP})\non\\
&&-\nabla^M\phi^{NP}(\nabla_P R_{ MN}-\nabla^Q R_{NMPQ})=0
\ee
For the Einstein spaces satisfying the condition $R_{MN}=\f{R}{d+1}G_{MN}$, the terms involving the double derivatives of $\phi_{MN}$ in the above constraint equations vanish and they simplify. They take their simplest form for $\xi=1$ and are given by
\begin{align}
    \Box\phi_{MN}+2R^P{}_M{}^Q{}_N\,\phi_{PQ}-m^2\phi_{MN}=0,\qquad\phi^M{}_{M}=0,\,\,\,\,\nabla^M\phi_{MN}=0\,.
\end{align}
In the AdS background, these equations become
\be
\left(\Box -m^2+\f{2}{L^2}\right)\phi_{MN}=0\qquad;\qquad \phi^M{}_{M} =0=\nabla^M\phi_{MN}\,.
\ee

\section{Flat limit of massive scalars using momentum space CFT}
\label{sec3}
In this section, we shall illustrate how the flat limit of AdS correlators involving massive scalar fields is taken in the momentum space. The approach follows the momentum space techniques developed in \cite{Marotta:2024sce}. Consider a scalar field of mass $m$ in Euclidean AdS$_{d+1}$ (of radius $L$) with a cubic interaction described by the action
\[
S=\int d^{d+1}x\sqrt{G}\left(\frac{1}{2}\d_M\phi\,\d^M \phi+\frac{1}{2}m^2\phi^2+\frac{1}{3!}\lambda\, \phi^3\right),
\]
where the index $M$ runs from $0$ to $d$. We work in the Poincar\'e patch with coordinates $\{z,x^\mu\}$ where $\mu$ runs from 1 through $d$. The corresponding AdS line element is listed in~\eqref{stanpoin54}.  

\vspace*{.07in}By Fourier transforming with respect to the coordinates $x^\mu$, we can extract the regular solution of the free field equation of motion ($(\Box-m^2)\phi=0$) as
\begin{align}
    \phi(z,\textbf{k})=\mathcal{K}(z,k)\,\hat\phi^{(0)}(\bf k),\label{3.14pi}
\end{align}
where $\hat\phi^{(0)}(\bf k)$ is related to the boundary value of the field as $\hat\phi^{(0)}(\textbf{k})=\lim_{z\rightarrow 0}z^\Delta \phi(z,\textbf{k})$, and $\mathcal{K}$ is the scalar bulk-to-boundary (Btb) propagator given in the Poincar\'e coordinate by
\begin{equation}
\mathcal{K}(z,k)= \frac{2^{\f{d}{2}-\Delta+1}}{L^{d-\Delta-2}\Gamma[\Delta-\f{d}{2}]}\,k^{\Delta-\f{d}{2}}z^{d/2}\,K_{\Delta-\f{d}{2}}(k\,z).
\end{equation}
Here, the mass of the field $\phi$ is related to the conformal dimension $\Delta$ of the dual boundary operator as $\Delta(\Delta-d)=m^2L^2$. The $k$ denotes the magnitude of the boundary momenta defined by $k^2= \delta^{\mu\nu}k_\mu k_\nu$ (where $\delta^{\mu\nu}$ denotes the induced (inverse) Euclidean metric on the conformal boundary). The $K_{\beta}(z)$ is the modified Bessel function of the second kind. 

\vspace*{.07in}The tree level CFT 3-point function in the momentum space computed using the AdS/CFT correspondence is given by \cite{1510.08442}
\be
\langle O(\textbf{k}_1)O(\textbf{k}_2)O(\textbf{k}_3)\rangle 
 = (2\pi)^d\delta^d(\textbf{k}_1+\textbf{k}_2+\textbf{k}_3)\lambda L^{d+1}\left(\frac{2^{\f{d}{2}-\Delta+1}}{\Gamma[\Delta-\f{d}{2}]}\right)^3 \,\,I_{\frac{d}{2}-1\{\Delta-\f{d}{2},\Delta-\f{d}{2},\Delta-\f{d}{2}\}},\label{3.15}
\ee
where $I_{\alpha\{\beta_1,\beta_2,\beta_3\}}$ is the triple K integral defined by
\be
I_{\alpha\{\beta_1,\beta_2,\beta_3\}}(k_1,k_2,k_3)\equiv \int_0^\infty dz \,z^\alpha \prod_{i=1}^3 k^{\beta_i} K_{\beta_i}(k_i\,z)\,,
\ee
whenever this integral converges. To take the flat limit, it is useful to consider the variable $\uptau$ defined by $z=L\,e^{\uptau /L}$. Further, if we want to keep the mass $m$ of the particle fixed while taking the flat limit, we need to send both   $\Delta$ and $L$ to infinity keeping their ratio fixed to $m$. In this limit, the modified Bessel function behaves as \cite{Marotta:2024sce}
\begin{align}
  K_{\Delta-\frac{d}{2}+s}(kz)= \left(\frac{\pi}{2EL}\right)^{\tfrac{1}{2}} 
\left(\frac{k}{m+E}\right)^{-mL - s} e^{-EL-E\uptau }\left(1+O\left(\frac{1}{L}\right)\right),\label{ekle}
\end{align}
with $E=\sqrt{k^2+m^2}$. The uplifted Euclidean $d+1$ dimensional momenta is defined to be $q^a=(\pm iE,k^\mu)$ with $q^2= \delta^{ab}q_aq_b=-m^2$. The choice of signs will corresponds to in and out going states when we subsequently Wick rotate. Since we work only at tree level, we shall take all particles to be incoming and hence shall exclusively choose the $+$ sign.

\vspace*{.07in}Using the identity \eqref{ekle}, we find that in the large $L$ limit, the bulk-to-boundary propagator takes the form,
\[\mathcal{K}(\uptau ,k) =\frac{e^{-E(\uptau +L)}}{Z^{1/2}}\left(1+O\left(\frac{1}{L^2}\right)\right)\quad;\qquad\frac{1}{Z^{1/2}}\equiv\left(\frac{2^{-mL}\,e^{mL}\,(L(m+E))^{m L}}{(EL)^{1/2}\,(m L)^{mL-\frac{1}{2}}}\right).\]
The factor $\f{1}{Z^{1/2}}$ diverges as we send $L$ to infinity . Naively, due to this, the field $\phi$ in \eqref{3.14pi} also diverges in this limit. To resolve this issue, we note that the source $\hat\phi^{(0)}(\textbf{k})$ is arbitrary. Therefore, we can define a renormalised source as
\begin{align}
    \hat\phi^{(0)}_R(\textbf{k})=\frac{\hat\phi^{(0)}(\textbf{k})}{Z^{1/2}},
\end{align}
and scale $\hat\phi^{(0)}(\textbf{k}) \to 0$ so as to keep $\hat\phi^{(0)}_R(\textbf{k})$ finite. This can be done since $\hat\phi^{(0)}(\textbf{k})$ is a free parameter. The conservation of spatial components of momenta is manifest from \eqref{3.15} due to the presence of the $d$ dimensional delta function $\delta^d(\textbf{k}_1+\textbf{k}_2+\textbf{k}_3)$. The energy conserving delta function emerges by writing $E_i = -iq_i^0$ and integrating over $\uptau$ variable as
\be\int_{-\infty}^\infty\,d\uptau \, e^{-\uptau(E_1+E_2+E_3)}=\int_{-\infty}^\infty\,d\uptau \, e^{i\uptau(q^0_1+q^0_2+q^0_3)}=(2\pi)\delta(q^0_1+q^0_2+q^0_3).
\ee
To make contact with Lorentzian scattering amplitudes, we Wick rotate. The $d+1$ dimensional flat space Lorentzian momenta are hence defined as $q^a = (E,k^\mu)$. With this, in the flat limit, the AdS 3-point function reduces to
\[\sqrt{Z_{1}Z_{2}Z_{3}}\Bigl\langle O(\textbf{k}_1)O(\textbf{k}_2)O(\textbf{k}_3)\Bigl\rangle\Bigg\rvert_{L\to\infty}=\,\,(2\pi)^{d+1}\delta^d(\textbf{k}_1+\textbf{k}_2+\textbf{k}_3)\,\delta(E_1+E_2+E_3)i\lambda,\]
which is the expected tree level 3-point vertex in the $\phi^3$ theory in the $d+1$ dimensional flat space.

\section{Massive spin-2 fields in AdS}
\label{sec4}
The free field action of a complex massive spin-2 field, which would give an expected number of degrees of freedom in the Einstein background, is described in section \ref{sec2.2}. Specializing to AdS$_{d+1}$ and working with the choice $\xi=1$ for the free parameter, the complex massive spin-2 field in the AdS background can be described by the action 
\begin{eqnarray}
S&=&\int d^{d+1}x\;\sqrt{G}\biggl[\frac{1}{2}\nabla_M\phi^*_{NP}\nabla^M{\phi}^{NP} 
-\nabla_M{\phi^*}_{NP}\nabla^P\phi^{NM} -\frac{1}{2} \nabla_M{\phi^*}\nabla^M\phi+\frac{1}{2}\nabla_M{\phi^*}^{MN}\nabla_N {\phi}\nonumber\\
&& +\frac{1}{2}\nabla_M{\phi}^{MN}\nabla_N {\phi^*}+\frac{d}{L^2} (\phi^*_{MN}{\phi}^{MN} -\f{1}{2}{\phi^*}\phi)+\frac{m^2}{2} (\phi^*_{MN}{\phi}^{MN} -{\phi^*}\phi)\biggl],
\end{eqnarray}
where the indices $M,N$ etc. run from $0$ to $d$.

\vspace*{.07in}For doing the bulk calculations and matching with the boundary CFT, we shall Fourier transform the boundary directions as
\begin{eqnarray}
f(u,\,\textbf{k})=\int d^d x %\frac{d^dx}{(2\pi)^d}
\;e^{-i k\cdot x}\; f(u,\,\textbf{x}),
\end{eqnarray}  
where $f$ can be any bulk quantity and may have arbitrary spacetime index structure. The $u$ denotes the bulk coordinate. We shall use both position and momentum space expressions in our derivations. To go from one to other, we need to use the replacement rule $\p_\mu\rightarrow ik_\mu$.

\subsection{Free field solution}

Below, we shall be performing the holographic renormalization of the massive spin-2 field. For this, we need to analyse its free field equations of motion. The free field equation of motion for $\phi^*_{MN}$ is given by
\be
\left(\Box-m^2+\f{2}{L^2}\right)\phi_{MN}=0\quad;\qquad \nabla_M\phi^M_{\;\;\;N}=0 = G^{MN}\phi_{MN}.\label{323er}
\ee
The constraint conditions $\nabla_M\phi^{M}_{\;\;\;N}=0$ in the AdS background can be written more explicitly in the Fefferman Graham coordinates $\{\rho,x^\mu\}$ (see \eqref{FG} for details on metric variables in these coordinates) as 
\begin{subequations}
\begin{align}
&(N=\rho)\,:\hspace{1cm}\f{4\rho^2}{L^2} \p_\rho\phi_{\rho \rho}+\f{\rho}{L}\delta^{\mu\nu}\p_\mu\phi_{\nu \rho} +\f{2(4-d)}{L^2}\rho \phi_{\rho \rho} + \f{1}{2L}\delta^{\sigma\nu}\phi_{\sigma\nu}=0,\label{opi}\\[.2cm]
&(N=\mu)\,:\hspace{1cm}\f{4\rho^2}{L^2} \p_\rho\phi_{\rho \mu}+\f{\rho}{L}\delta^{\nu\sigma}\p_\nu\phi_{\sigma \mu} +\f{2(2-d)}{L^2}\rho \phi_{\rho \mu} =0.\label{opi2}
\end{align}
\end{subequations}
On the other hand, the tracelessness condition $G^{MN}\phi_{MN}=0$ gives 
\be
 \f{4\rho}{L}\phi_{\rho\rho} +\delta^{\mu\nu}\phi_{\mu\nu}=0\label{opi1}.
\ee
The auxiliary conditions \eqref{opi} and \eqref{opi1} can be combined to obtain 
\be
iL\delta^{\mu\nu}k_\mu\phi_{\nu\rho}-2(d-3)\phi_{\rho\rho}+4\rho\partial_\rho\phi_{\rho\rho}=0.\label{6.71}
\ee
This form of the constraint will be useful later.

\vspace*{.07in}Using the above conditions along with \eqref{Box}, the equation of motion for different components in the Fefferman Graham coordinates are given by

\begin{subequations}
\begin{align}
({\rho\rho})\;\;:\qquad 4\rho^2 \p_\rho^2\phi_{\rho\rho}+2(6-d)\rho\p_\rho\phi_{\rho\rho} -(\rho Lk^2+m^2L^2+2d-4)\phi_{\rho\rho}&=&0\,,\label{eqrhorho}\\[.3cm]
({\mu\rho})\;\;:\qquad 4\rho^2 \p_\rho^2\phi_{\mu\rho}+2(6-d)\rho\p_\rho\phi_{\mu\rho} -(\rho Lk^2+m^2L^2+2d-4)\phi_{\mu\rho}&=&4i\rho k_\mu \phi_{\rho\rho}\,,\label{eqmurho}\\[.3cm]
({\mu\nu})\;\;:\qquad4\rho^2 \p_\rho^2\phi_{\mu\nu}+2(6-d)\rho\p_\rho\phi_{\mu\nu} -(\rho Lk^2+m^2L^2+2d-4)\phi_{\mu\nu}&&\non\\
&&\hspace*{-2.2in}=4i\rho (k_\mu \phi_{\nu\rho}+k_\nu \phi_{\mu\rho})-\f{8\rho}{L}\delta_{\mu\nu}\phi_{\rho\rho}\,.
\label{eqmunu}
\end{align}
\end{subequations}
The general solution of \eqref{eqrhorho} is given by 
\be
\phi_{\rho\rho}&=& c_1(\textbf{k}) \rho^{\f{d-4}{4}}K_\beta(k\sqrt{\rho L})+ c_2(\textbf{k}) \rho^{\f{d-4}{4}}I_\beta(k\sqrt{\rho L})\,,
\ee
where $I_\beta$ and $K_\beta$ are the modified Bessel functions of the first and second kind, respectively, and $k$ denotes the magnitude of the momenta $k^\mu$. In the limit $\rho\rightarrow\infty$ (which corresponds to deep AdS interior), the Bessel function $I_\beta(k\sqrt{\rho L})$ diverges. Hence, we discard this solution, obtaining
\be
\phi_{\rho\rho}(\textbf{k},\rho)&=& c_1(\textbf{k}) \rho^{\f{d-4}{4}}K_\beta(k\sqrt{\rho L})\,.
\ee
The order $\beta$ is related to the mass of the field and the conformal dimension of the dual operator as
\be
\beta = \sqrt{m^2L^2+\f{d^2}{4}}\equiv \Delta-\f{d}{2}\quad\implies\qquad m^2L^2=\Delta(\Delta-d)\,.
\ee
Next, the solution of \eqref{eqmurho} which is regular as $\rho\rightarrow\infty$ is given by 
\be
\phi_{\mu\rho}(\textbf{k},\rho)&=& b_\mu(\textbf{k}) \rho^{\f{d-4}{4}}K_\beta(k\sqrt{\rho L}) -\f{2i c_1(\textbf{k})}{k\sqrt{L}} k_\mu \rho^{\f{d-2}{4}} K_{\beta+1}(k\sqrt{\rho L})\,,
\ee
where $b_\mu$ is at this point an arbitrary vector. Imposing the auxiliary conditions \eqref{opi} and \eqref{opi1}, we find the relation between $b_\mu$ and $c_1$
\be
c_1(\textbf{k}) = -\f{i k^\mu b_\mu L}{2\beta -d+2}\,.
\ee
Finally, the solution of \eqref{eqmunu} is given by
\be
\phi_{\mu\nu}(\textbf{k},\rho)&=& a_{\mu\nu}(\textbf{k}) \rho^{\f{d-4}{4}}K_\beta(k\sqrt{\rho L}) -\f{2i(k_\mu b_\nu+k_\nu b_\mu)}{k \sqrt{L}}\rho^{\f{d-2}{4}}K_{\beta+1}(k\sqrt{\rho L})\non\\
&&+\f{4c_1(\textbf{k})}{kL^{\f{3}{2}}}\delta_{\mu\nu} \rho^{\f{d-2}{4}} K_{\beta+1}(k\sqrt{\rho L})-\f{4c_1(\textbf{k})}{k^2L}k_{\mu}k_{\nu} \rho^{\f{d}{4}} K_{\beta+2}(k\sqrt{\rho L})\,,
\ee
where $a_{\mu\nu}(\textbf{k})$ is a symmetric tensor. Substituting this solution in \eqref{opi2}, we find 
\be\label{bmu}
b_\mu(\textbf{k}) = \f{iL\;k^\nu a_{\mu\nu}}{d-2\beta}\quad\implies\qquad c_1(\textbf{k}) = -\f{ k^\mu k^\nu a_{\mu\nu} L^2}{(2\beta -d+2)(2\beta-d)}\,.
\ee
Similarly, substituting in \eqref{opi1}, we find $\delta^{\mu\nu}a_{\mu\nu}=0$ (where, we raise the indices of the momenta using the boundary metric $\delta^{\mu\nu}$). Thus, we have determined the free field solution of massive spin-2 fields in AdS in terms of a single traceless symmetric tensor $a_{\mu\nu}$(\textbf{k}).

\subsection{Bulk-to-boundary Propagator}
Using the free field solution of the massive spin two field $\phi_{MN}$ (which is determined by the symmetric traceless tensor $a_{\mu\nu}$), we can obtain its bulk-to-boundary (Btb) propagator. For this, we need to relate $a_{\mu\nu}(\textbf{k})$ with the boundary value of the field $\phi_{\mu\nu}$. Using standard expressions for the modified Bessel function $K_\beta$, we find that near the boundary $\rho\rightarrow 0$, the solutions behave as 
\begin{subequations}
\begin{align}
\lim_{\rho\rightarrow 0}\phi_{\rho\rho}
&=\frac{2^{\beta -1}  \Gamma (\beta ) k^{-\beta }  L^{2-\frac{\beta }{2}} }{(d-2 \beta ) (2 \beta -d+2)} k^\mu k^\nu a_{\mu\nu}\rho^{-\frac{\beta }{2}+\frac{d}{4}-1}\;\;\equiv\;\; \phi_{(0)\rho\rho}(\textbf{k})\;\rho^{-\frac{\beta }{2}+\frac{d}{4}-1}\,,
\\
\lim_{\rho\rightarrow 0}\phi_{\rho\mu}
&=\f{i 2^{\beta -1} \Gamma (\beta )  L^{1-\frac{\beta }{2}}}{(d-2\beta)k^\beta}\left(\delta_{\mu}^{\;\;\sigma}k^\nu - \f{4\beta k^\sigma k^\nu k_\mu}{k^2 (2\beta+2-d)}\right) a_{\sigma\nu}  \rho^{-\frac{\beta }{2}+\frac{d}{4}-1}\;\;\equiv\;\; \phi_{(0)\rho\mu}\;\rho^{-\frac{\beta }{2}+\frac{d}{4}-1}\,,
\\
\lim_{\rho\rightarrow 0}\phi_{\mu\nu}
&=2^{\beta -1} \Gamma (\beta ) k^{-\beta } L^{-\frac{\beta }{2}}\biggl[\delta_{\mu}^{\;\;\sigma}\delta_{\nu}^{\;\;\uptau }+\f{4\beta}{k^2 (d-2\beta)}(\delta_{\nu}^{\;\;\uptau }k_\mu k_\sigma+\delta_{\mu}^{\;\;\uptau }k_\nu k_\sigma)\non\\
&- \f{8\beta k^\sigma k^\uptau }{k^2(d-2\beta) (d-2-2\beta)}\left(\delta_{\mu\nu}-\f{k_\mu k_\nu}{k^2}(2+2\beta)  \right)\biggl]   a_{\sigma\uptau } \rho^{-\frac{\beta }{2}+\frac{d}{4}-1}\non\\
&\equiv \phi_{(0)\mu\nu}(\textbf{k})\;\rho^{-\frac{\beta }{2}+\frac{d}{4}-1}\,.\label{phimunu}
\end{align}
\end{subequations}
For relating the integration constant $a_{\mu\nu}(\textbf{k})$ with the boundary value $\phi_{(0)\mu\nu}(\textbf{k})$ of the field, we must invert the above expressions. To do this, we first decompose $a_{\mu\nu }(\textbf{k})$ as 
\be\label{ast}
a_{\mu\nu}&=& b_{\mu\nu }+t_\mu k_\nu +t_\nu  k_\mu+bk_{\mu}k_{\nu }+c\delta_{\mu\nu }.\label{astq}
\ee
Substituting this in \eqref{phimunu}, we find that the terms proportional to $\delta_{\mu\nu}$ and the terms linear and quadratic in the momenta $ k_\mu$ can be gotten rid of by choosing 
\be
t_\mu &=& -\frac{4 \beta b_{\nu\mu}k^\nu}{k^2(d+2\beta)}\;\;;\quad b=\frac{16 (\beta -1) \beta  k_\mu k_\nu b^{\mu\nu} }{k^4 (2 \beta +d-2) (2 \beta +d)}\;\;;\quad c=\frac{8 \beta  k_\mu k_\nu b^{\mu\nu}}{k^2 (2 \beta +d-2) (2 \beta +d)}. 
\ee
Then, the near boundary behaviour of $\phi_{\mu\nu}$ takes the form
\be\label{bhatphi0rel}
\lim_{\rho\rightarrow 0}\phi_{\mu\nu}(\rho,\textbf{k})= 2^{\beta -1} \Gamma (\beta ) k^{-\beta } L^{-\frac{\beta }{2}}b_{\mu\nu}(\textbf{k}) \;\rho^{-\frac{\beta }{2}+\frac{d}{4}-1}\;\equiv\;\hat\phi^{(0)}_{\mu\nu}(\textbf{k}) \left(\frac{\rho}{L}\right)^{-\frac{\beta }{2}+\frac{d}{4}-1}\,.
\ee
The boundary behaviours of $\phi_{\rho\rho}$ and $\phi_{\rho\mu}$ also simplify to 
\begin{subequations}
\begin{align}
\lim_{\rho\rightarrow 0}\phi_{\rho\rho}(\rho,\textbf{k})&= -\frac{ 2^{\beta -1} b_{\mu\nu}(\textbf{k})k^\nu k^\nu\Gamma (\beta ) k^{-\beta } L^{2-\frac{\beta }{2}} }{(2 \beta +d)(2\beta+d-2)} \rho^{-\frac{\beta }{2}+\frac{d}{4}-1}\,,
\\
\lim_{\rho\rightarrow 0}\phi_{\rho\mu}(\rho,\textbf{k})&= \frac{i 2^{\beta -1} b_{\mu\nu}(\textbf{k})k^\nu \Gamma (\beta ) k^{-\beta } L^{1-\frac{\beta }{2}} }{2 \beta +d} \rho^{-\frac{\beta }{2}+\frac{d}{4}-1}\,.
\end{align}
\end{subequations}
Contracting \eqref{astq} with $\delta^{\mu\nu }$ and demanding it to be zero, we see that $b_{\mu\nu}(\textbf{k})$ must also be traceless, i.e., $\delta^{\mu\nu}b_{\mu\nu}(\textbf{k})=0$. Using \eqref{ast}-\eqref{bhatphi0rel}, we can express $a_{\mu\nu}$ in terms of the boundary value of $\phi_{\mu\nu}$ namely, $\hat\phi^{(0)}_{\mu\nu}$ and the solutions can be expressed as 
\be
\phi_{\mu\nu}\;\;=\;\; K_{\mu\nu}^{\;\;\;\;\uptau \sigma}\; \hat\phi^{(0)}_{\uptau \sigma}\;\;\;;\qquad
\phi_{\rho\mu}\;\;=\;\; K_{\rho\mu}^{\;\;\;\;\uptau \sigma}\; \hat\phi^{(0)}_{\uptau \sigma}\;\;\;;\qquad
\phi_{\rho\rho}\;\;=\;\; K_{\rho\rho}^{\;\;\;\;\uptau \sigma}\; \hat\phi^{(0)}_{\uptau \sigma}\,.
\ee
where $K_{MN}^{\;\;\;\;\;\;\;\uptau \sigma}$ denote the components of the Btb propagator. We have
\begin{subequations}
\begin{align}
K_{\mu\nu}^{\;\;\;\;\uptau \sigma}(\rho,\textbf{k})
&=  \f{L^{-\gamma}}{\omega}\rho^{\f{d-4}{4}} \biggl[  \delta_\mu^{\uptau }\delta_\nu^\sigma K_\beta (k\sqrt{\rho L})- \f{4\sqrt{\rho L} \delta_{\mu\nu} k^\uptau  k^\sigma }{k (d+2\beta)(d+2\beta-2)}K_{\beta-1}(k\sqrt{\rho L})\non\\
&+ \f{4\sqrt{\rho L} k^\uptau \delta_{(\mu}^{\;\;\sigma}  k_{\nu)} }{k (d+2\beta)}K_{\beta-1}(k\sqrt{\rho L}) + \f{4 k_\mu k_\nu k^\uptau  k^\sigma }{k^4 (d+2\beta)(d+2\beta-2)}\non\\
&\Bigl\{ (k^2\rho L+4\beta(\beta-1))K_{\beta}(k\sqrt{\rho L})-2k(\beta-1)\sqrt{\rho L}K_{\beta+1}(k\sqrt{\rho L})\Bigl\} \biggl]\,,\\[.3cm]
K_{\rho\mu}^{\;\;\;\;\uptau \sigma}(\rho,\textbf{k})
&= \f{iL^{1-\gamma} k^\uptau  }{\omega(d+2\beta)}\left( \delta_\mu^{\;\sigma} K_\beta(k\sqrt{\rho L}) + \f{2\sqrt{\rho L} k^\sigma k_\mu}{k(2\beta+d-2)}K_{\beta-1}(k\sqrt{\rho L}) \right) \rho^{\f{d-4}{4}}\,, \\[.3cm]
K_{\rho\rho}^{\;\;\;\;\uptau \sigma}(\rho,\textbf{k})
&= - \f{L^{2-\gamma} k^\uptau  k^\sigma }{\omega(d+2\beta)(d+2\beta-2)} \rho^{\f{d-4}{4}}  K_\beta(k\sqrt{\rho L})\,,
\end{align}
\end{subequations}
where we have defined $\omega \equiv 2^{\beta -1} \Gamma (\beta ) k^{-\beta } $ and $\gamma=-\beta +\frac{d}{4}-1$.

For the computations later, it will be useful to consider the free field solution and the Btb propagator in different bulk coordinates. The expressions of Btb propagator in the Poincar\'e coordinates \eqref{stanpoin54} is given in appendix \ref{appenF}. In general, if we use $B_{MN}$ as a proxy for both the bulk fields $\phi_{MN}$ and the bulk-to-boundary propagator $K_{MN}{}^{\mu\nu}$, then under a change of coordinate $u(\rho)$ with boundary coordinates $x^\mu$ unchanged, we have
\be
   B_{uu}(u, \textbf{k})   &=&\left(\frac{\partial\rho}{\partial u}\right)^2  B_{\rho\rho}\bigl(\rho(u), \textbf{k}\bigl),\non
   \\[.3cm]
   B_{u\mu}(u, \textbf{k})&=&\left(\frac{\partial\rho}{\partial u}\right)B_{\rho\mu}\bigl(\rho(u), \textbf{k}\bigl),\non
   \\[.3cm]
   B_{\mu\nu}(u, \textbf{k})&=&B_{\mu\nu}\bigl(\rho(u), \textbf{k}\bigl).\non
\ee

\section{Holographic renormalization of massive spin-2 field}
\label{sec5}
In this section, we shall study the holographic renormalization of the massive spin-2 field. We shall follow the approach described in \cite{Marotta:2024sce} for massive spin-1 field (for a review on holographic renormalization, see \cite{Skenderis:2002wp}). For spin-2 and higher, an additional complication arises due to the more number of components along the bulk direction. The equations of motion relate the bulk components with the boundary components. For the holographic renormalization, we need to perturbatively solve the equations of motion near the boundary $\rho\rightarrow 0$. As is standard from the procedure of holographic renormalization, coefficients of certain powers of $\rho$ act as vacuum expectation value (VEV) and sources. These can be identified by solving the equations of motion for the $x^\mu$ independent configurations of $\phi_{\rho\rho}, \phi_{\rho\mu}, \phi_{\mu \nu}$. In this case, the equations of motion are given by \eqref{eqrhorho} - \eqref{eqmunu} with $k$s set to zero. More specifically, we have
\be
4\rho^2 \p_\rho^2\phi_{MN}+2(6-d)\rho\p_\rho\phi_{MN} -(m^2L^2+2d-4)\phi_{MN}&=&-\f{8\rho}{L}\delta_{M\mu }\delta_{N\nu}\phi_{\rho\rho}\,,
\ee
with the solutions given by
\be
\phi_{MN}&=& c_{MN} \rho ^{\frac{d}{2}-\frac{\Delta }{2}-1}+d_{MN} \rho.
   ^{\frac{\Delta }{2}-1}\,.
\ee
The coefficients $c_{MN}$ and $d_{MN}$ are integration constants. The coefficient of $\rho ^{\frac{d}{2}-\frac{\Delta }{2}-1}$ plays the role of source and the coefficient of $\rho ^{\frac{\Delta }{2}-1}$ plays the role of VEV. This also shows that the sources will appear at $O(\rho^{\f{d}{2}-\f{\Delta}{2}-1})$ whereas VEVs appear at $O(\rho^{\f{\Delta}{2}-1})$ in the asymptotic solution. With this in mind, we will now start analyzing the full equations (i.e. with the $x^\mu$ dependence reinserted) asymptotically.

\subsection{Asymptotic solutions}

 \vspace*{.07in}For the holographic renormalization, we need to solve for $\phi_{MN}$ using the equations of motion \eqref{eqrhorho}-\eqref{eqmunu}, order by order in $\rho$ around $\rho=0$. Since the leading power is $\frac{d}{2}-\frac{\Delta }{2}-1$, we shall factor out $\rho ^{\frac{d}{2}-\frac{\Delta }{2}-1}$ from $\phi_{MN}$ and write
\be
\phi_{MN}(\rho,\textbf{x})=\left(\frac{\rho}{L}\right) ^{\frac{d}{2}-\frac{\Delta }{2}-1} \hat\phi_{MN}(\rho,\textbf{x})\,.\label{ansatz}
\ee
Substituting this in the equations of motion, at the leading order, we recover the relation between mass $m$ and the conformal dimension $\Delta$, namely $m^2L^2=\Delta(\Delta-d)$. For subleading orders, the $\hat\phi_{MN}$ satisfy the equations
\begin{subequations}
\begin{align}
& 4 \rho^2  \partial_{\rho}^2\hat\phi _{\rho \rho }(\rho )+2 \rho (d-2 \Delta +2) \partial_{\rho}\hat\phi_{\rho \rho }(\rho )- \left(\Delta  (d-\Delta
   )+k^2 L \rho +L^2 m^2\right)\hat\phi_{\rho \rho }(\rho )=0\,,
\\
& 4 \rho^2  \partial_{\rho}^2\hat\phi _{\mu \rho }(\rho )+2 \rho (d-2 \Delta +2) \partial_{\rho}\hat\phi_{\mu \rho }(\rho )- \left(\Delta  (d-\Delta
   )+k^2 L \rho +L^2 m^2\right)\hat\phi_{\mu \rho }(\rho )-4 i \rho k_\mu \hat\phi_{\rho\rho} =0\,,
\\
& 4 \rho^2  \partial_{\rho}^2\hat\phi _{\mu \nu }(\rho )+2 \rho (d-2 \Delta +2) \partial_{\rho}\hat\phi_{\mu \nu }(\rho )- \left(\Delta  (d-\Delta
   )+k^2 L \rho +L^2 m^2\right)\hat\phi_{\mu \nu }(\rho )\non
   \\
   & -4 i \rho ( k_\mu \hat \phi_{\nu\rho} + k_\nu\hat\phi_{\mu\rho}) +\frac{8\rho \delta_{\mu\nu} \hat\phi_{\rho\rho}}{L}=0\,.
   \end{align}
\end{subequations}
The asymptotic solutions of these equations are given by 
\be
\hat\phi_{MN}(\rho,\textbf{x})=\sum_{j=0}^{\lfloor\Delta-\f{d}{2}\rfloor}\left(\frac{\rho}{L}\right)^j\hat\phi^{(2j)}_{MN}(\textbf{x})+\left(\frac{\rho}{L}\right)^{\Delta-\f{d}{2}}\left(\hat\phi^{(2\Delta-d)}_{MN}(\textbf{x})+\delta_{\Delta,\f{d}{2}+n}\chi^{(2\Delta-d)}_{MN}(\textbf{x})\log\left(\frac{\rho}{L}\right)\right)\,.
\ee
The symbol $\lfloor x\rfloor$ denotes the integer part of $x$. The logarithmic terms only appear if the quantity $\Delta-\f{d}{2}$ is an integer $n$. The coefficients are given by
\begin{subequations}
\begin{eqnarray}
\hat\phi^{(2j)}_{\mu\nu} &=&C_j \left[(k L)^{2j}\hat\phi^{(0)}_{\mu\nu}+4ijk^{2(j-1)}L^{2j-1} (k_\mu\hat\phi^{(0)}_{\nu\rho}+k_\nu\hat\phi^{(0)}_{\mu\rho})\right.\non\\
&&\left.-8j(kL)^{2(j-1)}\left( \delta_{\mu\nu}+2(j-1)\f{k_\mu k_\nu}{k^2}\right)\hat\phi^{(0)}_{\rho\rho}\right]\,,\\
\chi^{(2\Delta-d)}_{\mu\nu} 
&=&N\left[(kL)^{2n}\hat\phi^{(0)}_{\mu\nu}+4ink^{2(n-1)}L^{2n-1} (k_\mu\hat\phi^{(0)}_{\nu\rho}+k_\nu\hat\phi^{(0)}_{\mu\rho}) \right.\\
&&\left.-8n(kL)^{2(n-1)}\left(\delta_{\mu\nu}+2(n-1)\f{k_\mu k_\nu}{k^2}\right)\hat\phi_{\rho\rho}^{(0)}\right]\,,\\
\hat\phi^{(2j)}_{\mu\rho} &=&C_j \left((k L)^{2j}\hat\phi^{(0)}_{\mu\rho}(x)+4ij(kL)^{2(j-1)}Lk_\mu\hat\phi^{(0)}_{\rho\rho}(x)\right)\,,\\
\chi^{(2\Delta-d)}_{\mu\rho} 
&=& N \left((Lk)^{2n}\hat\phi^{(0)}_{\mu\rho}+4in(Lk)^{2(n-1)}Lk_\mu\hat\phi^{(0)}_{\rho\rho}\right)\,,\\
\hat\phi^{(2j)}_{\rho\rho} &=&C_j (k L)^{2j}\hat\phi^{(0)}_{\rho\rho}(x)\,,\\
\chi^{(2\Delta-d)}_{\rho\rho} &=& N (Lk)^{2n}\hat\phi^{(0)}_{\rho\rho}\,,
\end{eqnarray}
\end{subequations}
where, $n=\Delta-\f{d}{2}$ and we have defined
\be
C_j=\frac{1}{2^j j! (2j+d-2\Delta)\dots(4+d-2\Delta)(2+d-2\Delta)}\;\;;\quad N \equiv \frac{(-1)^{n-1}}{2^{2n}\Gamma(n)\Gamma(n+1)}\,.
\ee
The constraint equation \eqref{6.71} can be used to solve for the coefficients $\{\hat\phi^{(0)}_{\rho\rho},\hat\phi^{(2\Delta-d)}_{\rho\rho}\}$ in terms of $\{\hat\phi^{(0)}_{\mu\rho},\hat\phi^{(2\Delta-d)}_{\mu\rho}\}$. Plugging \eqref{ansatz} into \eqref{6.71} and solving the equation order by order in $\rho$, we get
\begin{subequations}
\begin{align}
\hat\phi^{(0)}_{\rho\rho}&=\frac{iLk^\mu}{2(\Delta-1)}\hat\phi^{(0)}_{\mu\rho}\,,\\
\chi^{(2\Delta-d)}_{\rho\rho}&=\frac{-iLk^\mu}{2(\Delta-d+1)}\chi^{(2\Delta-d)}_{\mu\rho}\,,\\
\hat\phi^{(2\Delta-d)}_{\rho\rho}
&=-N\frac{ik^\mu L(kL)^{2n}}{(\Delta-d+1)(\Delta-1)}\hat\phi^{(0)}_{\mu\rho}-\frac{ik^\mu L}{2(\Delta-d+1)}\hat\phi^{(2\Delta-d)}_{\mu\rho}\,.
\end{align}
\end{subequations}
%\A{I replaced -N here.\\}
The Auxiliary condition \eqref{opi2} can be used to solve for the coefficients $\{\hat\phi^{(0)}_{\mu\rho},\hat\phi^{(2\Delta-d)}_{\mu\rho}\}$ in terms of $\{\hat\phi^{(0)}_{\mu\nu},\hat\phi^{(2\Delta-d)}_{\mu\nu}\}$. Plugging \eqref{ansatz} into \eqref{opi2} and solving the equations order by order in $\rho$, we get
\begin{subequations}
    \begin{align}
\hat\phi^{(0)}_{\mu\rho}&=\frac{iLk^\nu}{2\Delta}\hat\phi^{(0)}_{\mu\nu}\,,\\
\chi^{(2\Delta-d)}_{\mu\rho}&=\frac{-iLk^\nu}{2(\Delta-d)}\chi^{(2\Delta-d)}_{\mu\nu}\,,\\
\hat\phi^{(2\Delta-d)}_{\mu\rho}
&=\frac{ik^\nu L}{2(d-\Delta)}\hat\phi^{(2\Delta-d)}_{\mu\nu}+N\frac{k^{2(n-1)}L^{2n}ik^\nu L}{\Delta(\Delta-d)(\Delta-1)}\Bigl(2n k^\sigma k_\mu\hat\phi^{(0)}_{\sigma\nu}-k^2(\Delta-1)\hat\phi^{(0)}_{\mu\nu}\Bigl)\,.
\end{align}
\end{subequations}
The above equations can be used to express the $(\rho\rho)$ components in terms of $(\mu\nu)$ components and we obtain
\begin{subequations}
\begin{align}
\hat\phi^{(0)}_{\rho\rho}
&=
-\frac{L^2}{4\,\Delta(\Delta-1)}\,k^\mu k^\nu\,\hat\phi^{(0)}_{\mu\nu}\,,\\
\chi^{(2\Delta-d)}_{\rho\rho}
&=
-\frac{L^2}{4(\Delta-d+1)(\Delta-d)}\,k^\mu k^\nu\,\chi^{(2\Delta-d)}_{\mu\nu}
\,,\\
\hat\phi^{(2\Delta-d)}_{\rho\rho}
&=
-\frac{L^2k^\mu k^\nu}{4(\Delta-d+1)(\Delta-d)}\left[\hat\phi^{(2\Delta-d)}_{\mu\nu}-
N\,
\frac{2(2\Delta-2d+1)}{\Delta(\Delta-1)}\,
\,
(kL)^{2\Delta-d} \hat\phi^{(0)}_{\mu\nu}\right]\,.
\end{align}
\end{subequations}
Finally, using the relations between $(\rho\rho), (\mu\rho)$ and $(\mu\nu)$ components, we can express the coefficients $\hat\phi_{\rho\rho}^{(2j)}, \hat\phi_{\mu\rho}^{(2j)}$ and $\hat\phi_{\mu\nu}^{(2j)}$ in terms of the single coefficient $\hat\phi_{\mu\nu}^{(0)}$ as 
\begin{subequations}
\begin{align}
\hat\phi_{\mu\nu}^{(2j)}&=C_j\left[(kL)^{2j}\hat\phi_{\mu\nu}^{(0)}-\frac{2j\,(kL)^{2(j-1)}L^{2}k^{\sigma}}{\Delta}\left(k_\mu \hat\phi_{\nu\sigma}^{(0)}+k_\nu \hat\phi_{\mu\sigma}^{(0)}-\left(\delta_{\mu\nu}+2(j-1)\frac{k_\mu k_\nu}{k^2}\right)\frac{k^\tau \hat\phi_{\sigma\tau}^{(0)}}{\Delta-1}\right)\right]\,,\\
    \hat\phi_{\mu\rho}^{(2j)}&=\frac{iLC_j\,k^{\sigma}}{2\Delta}\left[(kL)^{2j}\hat\phi_{\mu\sigma}^{(0)}-\frac{2j}{\Delta-1}\,k^{2(j-1)}L^{2j}k_\mu k^{\tau}\hat\phi_{\sigma\tau}^{(0)}\right]\,,\\
    \hat\phi_{\rho\rho}^{(2j)}&=-\,\frac{C_j}{4\Delta(\Delta-1)} k^{2j}L^{2(j+1)}\,k^\mu k^\nu\hat\phi_{\mu\nu}^{(0)}\,.
\end{align}
\end{subequations}
So far we have determined the expansion coefficients in terms of two undetermined quantities $\hat\phi^{(0)}_{\mu\nu}$ and $\hat\phi^{(2\Delta-d)}_{\mu\nu}$. Using the auxiliary condition \eqref{opi1}, we also get following constraints on these quantities 
\be
&&\delta^{\mu\nu}\hat\phi^{(0)}_{\mu\nu}=0\quad;\qquad\delta^{\mu\nu}\chi^{(2\Delta-d)}_{\mu\nu}\;=\;0\quad;\qquad\delta^{\mu\nu}\hat\phi^{(2j)}_{\mu\nu}+4\hat\phi^{\left(2j-2\right)}_{\rho\rho}=0\,.
\ee
where the 3rd identity is valid for all $j$ from $1$ to $\lfloor\Delta-\f{d}{2}\rfloor$. For the non integer $\Delta$, we also have a relation
\be
\delta^{\mu\nu}\hat\phi^{(2\Delta-d)}_{\mu\nu}\;=\;0\,.
\ee

\subsection{Matching with exact solution}

The full analytic solutions for $\phi_{\rho\rho},\,\phi_{\mu\rho}$ and $\phi_{\mu\nu}$ has one undetermined tensor $a_{\mu\nu}$, while the asymptotic solution is determined in terms of two constant tensors $\hat\phi_{\mu\nu}^{(0)}$ and $\hat\phi_{\mu\nu}^{\left(2\Delta-d\right)}$. Using the full analytic solution, we can determine the relation between $\hat\phi_{\mu\nu}^{(0)}$ and $\hat\phi_{\mu\nu}^{\left(2\Delta-d\right)}$. This will be useful later for computing the two point function of spin-2 CFT operators holographically.   

Using the relation between $c_1,\,b_\mu$ and $a_{\mu\nu}$ in \eqref{bmu}, the solution of $\phi_{\mu\nu}$ can be expressed in terms of the single tensor $a_{\mu\nu}$ as 
\be\label{phimnexp}
\phi_{\mu\nu}
&=&\rho^{\frac{d-4}{4}}\Bigg[a_{\mu\nu}K_\beta(k\sqrt{L\rho})+\frac{1}{k^2(d-\Delta)(\Delta-d+1)}\bigg(-a_{\sigma\uptau } k^\sigma k^\uptau  k_\mu k_\nu L \rho\, K_{\beta+2}(k\sqrt{L\rho})\non\\
&&k\sqrt{L\rho}\,K_{\beta+1}(k\sqrt{L\rho})\,k^\sigma\Big\{a_{\sigma\uptau } k^\uptau  \delta_{\mu\nu}+(\Delta-d+1)(a_{\sigma\mu} k_\nu+a_{\sigma\nu} k_\mu)\Big\}\bigg)\Bigg]\,.
\ee
Below, we shall be interested in the non integer conformal dimension $\Delta$. For this, the order of Bessel function- which is $\Delta-d/2$- will also be non integer. The expansion of the Bessel function $K_\nu(x)$ near zero for non integer order is given by
\be
\lim_{x\rightarrow 0}K_\nu(x) &=& \left[\Gamma(\nu)2^{\nu -1}\; x^{-\nu} -\f{\Gamma(\nu)2^{\nu-3}}{\nu -1}x^{2-\nu}+\f{\Gamma(\nu)2^{\nu-6}}{(\nu -1)(\nu-2)}x^{4-\nu} +\cdots\right]\non\\
&&+\left[\Gamma(-\nu)2^{-\nu -1}\; x^{\nu} +\f{\Gamma(-\nu)2^{-\nu-3}}{\nu +1}x^{2+\nu}+\f{\Gamma(-\nu)2^{-\nu-6}}{(\nu +1)(\nu+2)}x^{4+\nu} +\cdots\right]\,,
\ee
with the dots denoting further Frobenius terms in $x$- their form can be extrapolated by extending the pattern visible in the first few terms. Using this expression in \eqref{phimnexp} gives,
\be
\phi_{\mu\nu}&=& \left(\frac{\rho}{L}\right)^{\frac{d}{2}-\frac{\Delta}{2}-1}\left[\hat\phi^{(0)}_{\mu\nu}+\cdots+\left(\frac{\rho}{L}\right)^{\left(\Delta-\frac{d}{2}\right)}\,\hat\phi_{\mu\nu}^{\left(2\Delta-d\right)}+\cdots\right]\,,
\ee
where
\begin{subequations}
\label{hatphitildea}
\begin{align}
\hat\phi_{\mu\nu}^{\left(0\right)}
&=L^{\f{3d}{4}-\Delta-1}\,2^{\beta}k^{-\beta}\Bigg[2^{-1}\Gamma(\beta)\delta^\sigma_{(\mu} \delta^\uptau _{\nu)}+\frac{1}{k^2(d-\Delta)(\Delta-d+1)}
\non\\
&\Big(\Gamma(\beta+1)\left\{ k^\sigma k^\uptau  \delta_{\mu\nu}+(\Delta-d+1)k^\sigma(\delta^\uptau _\mu k_\nu+\delta^\uptau _\nu k_\mu)\right\}-\Gamma(\beta+2)2 k^{-2} k^\sigma k^\uptau  k_\mu k_\nu \Big)\Bigg] a_{\sigma\uptau }\,,\\
\hat\phi_{\mu\nu}^{\left(2\Delta-d\right)}&=L^{\Delta-\f{d}{4}-1}\,2^{-1-\beta}k^\beta\Gamma(-\beta) a_{\mu\nu}\,.
\end{align}
\end{subequations}
We want to express $\hat\phi^{\left(2\Delta-d\right)}_{\mu\nu}$ in terms of $\hat\phi^{\left(0\right)}_{\mu\nu}$. To do this, we first express $a_{\mu\nu}$ in terms of $\hat\phi^{\left(0\right)}_{\mu\nu}$. This can be done by decomposing $a_{\mu\nu }$ as 
\be
 a_{\mu\nu}&=&c_{\mu\nu }+\ell_\mu k_\nu +\ell_\nu  k_\mu+q k_\mu k_\nu +r \delta_{\mu\nu}\,.
\ee
Substituting this in \eqref{hatphitildea}, we find that the terms proportional to $\delta_{\mu\nu}, k_\mu,k_\nu$ and $k_\mu k_\nu$ can be gotten rid of by choosing 
\be
\ell_\mu=\frac{(d-2 \Delta )}{k^2\Delta} c_{\mu\nu}k^\nu\;\;;\quad 
q=\frac{(d-2 \Delta ) (d-2 \Delta +2)}{k^4 \Delta (\Delta -1) }c_{\mu\nu}k^\mu k^\nu\;\;;\quad 
r=-\frac{(d-2 \Delta)}{k^2\Delta(\Delta -1)}c_{\mu\nu}k^\mu k^\nu\,.
\ee
and we get the following relation
\be\label{cuv}
c_{\mu\nu}=\frac{2\beta}{L^{\f{3d}{4}-\Delta-1}2^{\beta}k^{-\beta}\Gamma(\beta+1)}\hat\phi_{\mu\nu}^{\left(0\right)}\,.
\ee
Using equations \eqref{hatphitildea} - \eqref{cuv}, we get following relation between $\hat\phi_{\mu\nu}^{\left(2\Delta-d\right)}$ and $\hat\phi^{\left(0\right)}_{\mu\nu}$
\be
\hat\phi_{\mu\nu}^{\left(2\Delta-d\right)}&=&A\,\mathcal{O}_{\mu\nu}{}^{\sigma\uptau }\,\hat\phi_{\sigma\uptau }^{\left(0\right)}\,,
\ee
where $\mathcal{O}_{\mu\nu}{}^{\sigma\uptau }$ is a tensor symmetric in indices $(\mu,\nu)$ as well as in $(\sigma,\uptau)$ and is given by
\be
    \mathcal{O}_{\mu\nu}{}^{\sigma\uptau }&=&\delta^\uptau _{(\mu}\delta^\sigma_{\nu)}+B\,k_\mu k_\nu k^\sigma k^\uptau +C\left(\delta^\uptau _\mu k^\sigma k_\nu+\delta^\uptau _\nu k^\sigma k_\mu \right)+D\,k^\uptau  k^\sigma\delta_{\mu\nu}\,,
\ee
where parentheses over indices denote symmetrization and where the constants $A,B,C,D$ are given by
\begin{align}
   A=\left(2^{-1}k\,L\right)^{2\beta}\frac{\Gamma(-\beta)}{\Gamma(\beta)},\quad B=\frac{(d-2\Delta)(d+2-2\Delta)}{k^4\,\Delta(\Delta-1)},\quad C=\frac{d-2\Delta}{ k^2\,\Delta},\quad D=-\frac{(d-2\Delta)}{k^2\,\Delta(\Delta-1)}\,.
\end{align}
Defining $\mathcal{O}^{\mu\nu\sigma\uptau}=\delta^{\mu\alpha}\delta^{\nu\beta}\mathcal{O}_{\alpha\beta}^{\;\;\;\;\;\sigma\uptau}$, we see that it is symmetric under the exchange $\mu\nu \leftrightarrow \sigma\uptau$ except for the term involving the coefficient $D$. To remedy this, we first note that the condition $\delta^{\sigma\uptau }\hat\phi_{\sigma\uptau }^{\left(0\right)}=0$ implies that we can add a term proportional to $\delta^{\sigma\uptau}$ to $ \mathcal{O}_{\mu\nu}{}^{\sigma\uptau }$ as
\be\label{phinphi0rel}
     \hat\phi_{\mu\nu}^{\left(2\Delta-d\right)}&=&A\Bigl(O_{\mu\nu}{}^{\sigma\uptau }+Q_{\mu\nu}\,\delta^{\sigma\uptau }\Bigl)\,\hat\phi_{\sigma\uptau }^{\left(0\right)}\;\;\equiv\;\; P_{\mu\nu}{}^{\sigma\uptau }\,\hat\phi_{\sigma\uptau }^{\left(0\right)}\,.
\ee
where $Q_{\mu\nu}$ is an arbitrary symmetric tensor. Since it is arbitrary, we can appropriately choose it to ensure that $\mathcal{O}^{\mu\nu\sigma\uptau}$ is symmetric under the exchange of $\mu\nu \leftrightarrow \sigma\uptau$. Choosing $Q_{\mu\nu}= Dk_\mu k_\nu +Q(k^2)\delta_{\mu\nu}$, we get 
\begin{align}
    P_{\mu\nu}{}^{\sigma\uptau }=A\left[\delta^{(\sigma}_{(\mu}\delta^{\uptau )}_{\nu)}+B\,k_\mu k_\nu k^\sigma k^\uptau +C\delta^{(\sigma}_{(\mu} k^{\uptau )} k_{\nu)}+D\, (k^\sigma k^\uptau \delta_{\mu\nu}+k_\mu k_\nu \delta^{\sigma\uptau })+Q(k^2)\,\delta_{\mu\nu}\delta^{\sigma\uptau }\right]\,.
\end{align}
This satisfies the desired relation $P^{\mu\nu\sigma\uptau }=P^{\sigma\uptau \mu\nu}$. Next, the condition $\delta^{\mu\nu}\hat\phi^{(2\Delta-d)}_{\mu\nu}=0$ implies 
\begin{align}
    \delta_{\mu\nu}\,P^{\mu\nu\sigma\uptau }=\frac{Q\,d(\Delta-1)\Delta-d+\Delta^2+\Delta}{\Delta(\Delta-1)}\,\delta^{\sigma\uptau }=0\,,
\end{align}
which fixes
\begin{align}
    Q=\frac{d-\Delta^2-\Delta}{d\,\Delta(\Delta-1)}\,.
\end{align}

\subsection{Regularized action and counter terms}
The classical action evaluated on-shell in AdS diverges and hence requires regularization. It is done by putting the boundary at $z=L\epsilon$ (where $\epsilon$ is an infinitesimal dimensionless quantity). Doing this, we obtain the following regularized action
\begin{eqnarray}
S_{reg}&=&\int_{\rho\ge L\epsilon} d^{d+1}x\;\sqrt{G}\biggl[\frac{1}{2}\nabla_M\phi^*_{NP}\nabla^M{\phi}^{NP} 
-\nabla_M{\phi^*}_{NP}\nabla^P\phi^{NM} +\frac{1}{2}\nabla_M{\phi^*}^{MN}\nabla_N {\phi}\nonumber\\
&& +\frac{1}{2}\nabla_M{\phi}^{MN}\nabla_N {\phi^*}-\frac{1}{2} \nabla_M{\phi^*}\nabla^M\phi+\frac{d}{L^2} (\phi^*_{MN}{\phi}^{MN} -\f{1}{2}{\phi^*}\phi)+\frac{m^2}{2} (\phi^*_{MN}{\phi}^{MN} -{\phi^*}\phi)\biggl]\non\\
&=&\frac{1}{2}\int_{\rho=L\epsilon} d^{d}x\;\sqrt{\gamma} \ n_M{\phi^*}_{NP}\nabla^M\phi^{NP}\non\\
&=&\frac{-1}{4}\int_{\rho=L\epsilon} d^{d}x\;\left(\frac{L}{\rho}\right)^{\frac{d}{2}+1}{\phi^*}_{NP}\nabla^\rho\phi^{NP}\,.\label{6.77}
\end{eqnarray}
In going to the second equality, we have used integration by parts and the bulk equations of motion \eqref{323er}. We have also used Ricci identities corresponding to spin-2 fields. For example, we have 
\begin{align}
    [\nabla_M,\nabla^{(P}]\,\phi^{N)M}-\nabla^{(P} \nabla_M\phi^{N)M}
=R^{(N}{}_{Q M}{}^{P)}\phi^{QM}
+R^{(P}{}_{Q}\,\phi^{N)Q}
\;=\;\frac{1}{L^{2}}\Big(G^{PN}\phi-(d+1)\phi^{PN}\Big)\,,
\end{align}
where the final equality is valid only in AdS. 

In going to the third equality in \eqref{6.77}, we have used the fact that the boundary hypersurface is defined by $\rho=$ constant. Hence, the unit normal space like vector and the induced boundary metric are given by 
\be
n_M =-\f{L}{2\rho} \delta_M^\rho\quad;\qquad
\gamma_{\mu\nu}(x) =L\f{\delta_{\mu\nu}}{\rho} \qquad,\qquad \sqrt{\gamma} =\left(\f{L}{\rho}\right)^{\f{d}{2}}\,.
\ee
Now, using the asymptotic solution, we can express the regularized action in the form 
\begin{align}
S_{reg}
=-\int d^dx \Bigg[&d_{(0)} \epsilon^{\f{d}{2}-\Delta}+d_{(2)} \epsilon^{\f{d}{2}-\Delta+1}+\dots+d_{(2\Delta-d)}\log(\epsilon)+\cdots\Bigg]\,,
\end{align}
where
\begin{subequations}
\begin{align}
d_{(2i)}&=\frac{1}{2L}(16a_i+8b_i+c_i)\quad;\qquad  0\le i<\Delta-\frac{d}{2}\,,\\
d_{(2\Delta-d)}&=\frac{1}{2L}\left((d-\Delta)\hat{\phi}_{\beta\nu}^{(0)}\hat{\chi}_{\mu\alpha}^{(2\Delta-d)}{}^*\delta^{\mu\nu}\delta^{\alpha\beta}+\Delta\hat{\chi}_{\beta\nu}^{(2\Delta-d)}\hat{\phi}_{\mu\alpha}^{(0)}{}^*\delta^{\mu\nu}\delta^{\alpha\beta}\right)\,,
\end{align}
\end{subequations}
with 
\begin{subequations}
\begin{align}
a_i&=\sum_{j,k\geq 0\atop j+k=i-2}^{\lfloor \Delta-\f{d}{2}\rfloor}(d-\Delta+2j+2)\hat{\phi}_{\rho\rho}^{(2j)}\hat{\phi}_{\rho\rho}^{(2k)}{}^*\,,\\
b_i&=\sum_{j,k\geq 0\atop j+k+1=i}^{\lfloor \Delta-\f{d}{2}\rfloor}(d-\Delta+2j+1)\hat{\phi}_{\rho\nu}^{(2j)}\hat{\phi}_{\rho\mu}^{(2k)}{}^*\delta^{\mu\nu}\,,\\
c_i&=\sum_{j,k\geq 0\atop j+k=i}^{\lfloor \Delta-\f{d}{2}\rfloor}(d-\Delta+2j)\hat{\phi}_{\beta\nu}^{(2j)}\hat{\phi}_{\mu\alpha}^{(2k)}{}^*\delta^{\mu\nu}\delta^{\alpha\beta}\,.
\end{align}
\end{subequations}
To express the regularized action in the boundary covariant form, we need to invert the series and express $\hat\phi^{(0)}_{MN}$ in terms of $\phi_{MN}$. Upto $O(\rho^j)$, the inverted relations are given by
\begin{subequations}
\begin{align}
\label{invphi}
\hat\phi^{(0)}_{\rho\rho}&=\left(\frac{\rho}{L}\right)^{\frac{\Delta-d+2}{2}}\sum_{r=0}^{j}b_r(L^2 k^2_{\gamma})^{r} \phi_{\rho\rho}\,,\\
\hat\phi^{(0)}_{\mu\rho}&=\left(\frac{\rho}{L}\right)^{\frac{\Delta-d+2}{2}}\sum_{r=0}^{j}b_r\left[(L^2 k^2_{\gamma})^{r} \phi_{\mu\rho}+4 r (L^2 k^2_{\gamma})^{r-1}\rho i k_\mu \phi_{\rho\rho}\right]\,,\\
\hat\phi^{(0)}_{\mu\nu}&=\left(\frac{\rho}{L}\right)^{(\frac{\Delta-d+2}{2})} \sum_{r=0}^{j} b_r\Bigg[(L^2 k^2_{\gamma})^{r} \phi_{\mu\nu}+4 r (L^2 k^2_{\gamma})^{r-1}\rho i (k_\mu \phi_{\nu\rho}+k_\nu \phi_{\mu\rho})\non\\
&\hspace*{.6in}-8r  (L^2 k^2_\gamma)^{r-2}(k_\gamma^2 \gamma_{\mu\nu}+2(r-1)k_\mu k_\nu)\rho^2\phi_{\rho\rho}\Bigg]\,,\label{serinvhatphimn}
\end{align}
\end{subequations}
where the coefficients $b_r$ are recursively defined as
\be
b_k&=& -\sum_{m+n=k\atop m\ge 1;n\ge 0} C_mb_n\quad;\qquad b_0=1\,.
\ee
We have also defined $k_\gamma^2= \f{\rho}{L}k^2$. This is a useful notation since in the position space, $k_\gamma^2$ becomes the boundary Laplacian $\Box_\gamma=\gamma^{\mu\nu}\p_\mu\p_\nu$.

In \eqref{serinvhatphimn}, $\hat\phi^{(0)}_{\mu\nu}$ is in terms of all the components $\phi_{\mu\nu}, \phi_{\mu\rho}$ and $\phi_{\rho\rho}$. To write down the counter term action, it is useful to express $\hat\phi_{\mu\nu}^{(0)}$ completely in terms of $\phi_{\mu\nu}$. It is given by
\begin{align}
% Useful series inversion
\hat\phi^{(0)}_{\mu\nu}
=&\left(\frac{\rho}{L}\right)^{\frac{\Delta-d}{2}+1}
\Biggl[
\phi_{\mu\nu}
-C_{1}\Biggl\{
\left(\frac{\rho}{L}\right)(kL)^{2}\phi_{\mu\nu}
-\left(\frac{\rho}{L}\right)\left(\frac{2L^{2}}{\Delta}\right) k^{\sigma}\bigl(k_{\mu}\phi_{\nu\sigma}+k_{\nu}\phi_{\mu\sigma}\bigr)\non\\
&+\left(\frac{\rho}{L}\right)^{2}\gamma_{\mu\nu}\frac{2L^{2}}{\Delta(\Delta-1)}\,k^{\sigma}k^{\uptau}\phi_{\sigma\uptau}
\Biggr\}
+\cdots
\Biggr]\,,
\end{align}
where dots denote higher order terms. Using the above expression, we can express the regularized action in a covariant form. Noting that the role of counterterms is to get rid of the divergent terms in the regularized action when we take the limit $\epsilon\to 0$, it follows that the counterterms are simply given by the negative of the divergent terms in
the regularized action. Noting this, the counter term action can be obtained to be (in position space)
\begin{align}
    S_{\rm ct} &= -S_{\rm reg}=\int d^{d}x\,\sqrt{\gamma}\;\left[
\frac{d-\Delta}{2L}\,\phi_{\beta\alpha}\phi^{*}_{\mu\nu}\,\gamma^{\mu\alpha}\gamma^{\nu\beta}+\cdots\right]\,,
\end{align}
The renormalised action is the sum of regularized and counterterm actions, i.e.,
\begin{align}
 S_{\rm ren} &= \lim_{\epsilon\to 0}\bigl(S_{\rm reg}+S_{\rm ct}\bigr)\,.
\end{align}

\subsection{One and two point functions}
The one point function of the boundary operator $O^{\mu\nu}$ which is dual to the bulk field $\phi^*_{\mu\nu}$ is given by
\be
    \left<O^{\mu\nu}(\textbf{k})\right>&=&\frac{\delta S_{ren}}{\delta{\hat\phi^*}{}^{(0)}_{\mu\nu}(-\textbf{k})}=\underset{\epsilon\to0}{lim}\frac{1}{\epsilon^{\frac{\Delta+2}{2}}}\frac{1}{\sqrt{\gamma}}\frac{\delta (S_{reg}+S_{ct})}{\delta{\phi^*}_{\mu\nu}(-\textbf{k})}\,.\label{5.98}
    \ee
Now, we have
\begin{subequations}
\begin{align}
\frac{\delta S_{reg}}{\delta {\phi^*}_{\mu\nu}}&=-\sqrt{\gamma}\gamma^{\mu\alpha}\gamma^{\nu\beta}\Big(\frac{1}{L}{\phi}_{\beta\alpha}+\frac{\rho}{L}\partial_\rho{\phi}_{\beta\alpha}\Big)\Big\rvert_{\rho=L\epsilon}\,,\\
    \frac{\delta S_{ct}}{\delta{\phi^*}_{\mu\nu}}&=\frac{1}{2L}\sqrt{\gamma}\gamma^{\mu\alpha}\gamma^{\nu\beta}\Bigl\{(d-\Delta)\phi_{\beta\alpha}+2C_1 L^2 k_{\gamma}^2\phi_{\beta\alpha}+\epsilon8C_1 L(ik_\beta\phi_{\alpha\rho}+ik_\alpha\phi_{\beta\rho})\Bigl\}\Big\rvert_{\rho=L\epsilon}\,.
    \end{align}
\end{subequations}
Using these in \eqref{5.98} and taking the boundary limit, we obtain the one point function
\be
   \left<O^{\mu\nu}(\textbf{k})\right>  
     =-\frac{1}{L}\left(\Delta-\frac{d}{2}\right)\delta^{\mu\alpha}\delta^{\nu\beta}\hat{\phi}^{\left(2\Delta-d\right)}_{\alpha\beta}(\textbf{k})\;=\;-\,\frac{1}{L}\left(\Delta-\frac{d}{2}\right)\delta^{\mu\alpha}\delta^{\nu\beta}\,P_{\alpha\beta}{}^{\sigma\uptau }(\textbf{k})\,\hat\phi_{\sigma\uptau }^{\left(0\right)}(\textbf{k})\,,
\ee
where we have used the relation \eqref{phinphi0rel}. 

\vspace*{.07in}The two-point function can now be derived by differentiating the one point function with respect to the source $\hat\phi_{\sigma\uptau }^{\left(0\right)}$
\begin{align}\label{2pointfn}
\left<{O^*}^{\sigma\uptau }(\textbf{p})O^{\mu\nu}(\textbf{k})\right>=&-\,(2\pi)^d \frac{\delta\left<O^{\mu\nu}(\textbf{k})\right>}{\delta\hat\phi^{(0)}_{\sigma\uptau }(-\textbf{p})}=\frac{1}{L}\left(\Delta-\frac{d}{2}\right)\,(2\pi)^d\delta^d(\textbf{p}+\textbf{k}) P^{\mu\nu\sigma\uptau }(\textbf{k})\,,
\end{align}
where we have used $\delta^{\mu\nu}k_\nu=k^\mu$.

Using the expression of $P^{\mu\nu\sigma\uptau}$ given earlier, we can express the two point function as 
\be
&&\hspace*{-.4in}\left<{O^*}^{\sigma\uptau }(\textbf{p})O^{\mu\nu}(\textbf{k})\right> \non\\
&=& \,(2\pi)^d\delta^d(\textbf{p}+\textbf{k})a_0 \left[\Pi_{\mu\sigma,\nu\uptau }+\frac{(d-\Delta)}{\Delta(\Delta-1)(d-1)}\frac{k_\mu k_\nu}{k^2}\left(E\,\delta_{\sigma\uptau }+F\,\frac{k_\sigma\,k_\uptau }{k^2}\right)\right]k^{2\Delta-d}\,,\label{5.109}
\ee
where $E,F$ and $\Pi_{\mu\nu,\sigma\uptau}$ are defined below equation \eqref{2.20iyt} and $a_0$ is an overall constant given by
\be
a_0 = \f{1}{L}\left(\Delta-\f{d}{2}\right)\left(\f{L}{2}\right)^{2\Delta-d}\f{\Gamma\left(\f{d}{2}-\Delta\right)}{\Gamma\left(\Delta-\f{d}{2}\right)}\,.\label{5.109wer}
\ee
The CFT two-point function of the non-conserved spin-2 operators in the momentum space has been studied in \cite{Marotta:2022jrp} and reviewed in section \ref{sec2review3er}. The above expression \eqref{5.109} matches precisely with the expected result given in \cite{Marotta:2022jrp}. Thus, our results for the CFT two point function of massive spin-2 operators computed using the holographic renormalization are in agreement with the expected boundary CFT results.

\section{3-point function from bulk theory} 
\label{s3a}

\subsection{Bulk theory}
In this section we consider a $d$ dimensional CFT 3-point function $\Bigl\langle O^{*\alpha\beta}J^\kappa O^{\gamma\delta} \Bigl\rangle $  
of a $U(1)$ conserved current $J^\mu$ and two spin-2 operators $O^{\mu\nu}$ and $O^{*\mu\nu}$ using the AdS/CFT prescription.
This 3-point function on the boundary can be evaluated from a $d+1$ dimensional bulk effective action in AdS whose cubic coupling terms are linear in the gauge field $A_M$ and quadratic in massive spin-2 fields, $\phi_{MN}$.  The most general such action in Euclidean signature describing the interaction between a $U(1)$ gauge field and a complex massive spin-2 field in $d+1$ dimension up to 3 derivative terms can be written as
\begin{eqnarray}
S&=&\int d^{d+1}x\;\sqrt{G}\biggl[-\,\frac{1}{16\pi G_N}(R-2\Lambda)+\f{1}{4}F_{MN}F^{MN}+\frac{1}{2}D_M\phi^*_{NP}D^M{\phi}^{NP} 
-D_M{\phi^*}_{NP}D^P\phi^{NM}\nonumber\\
&& -\frac{1}{2} D_M{\phi^*}D^M\phi+\frac{1}{2}D_M{\phi^*}^{MN}D_N {\phi} +\frac{1}{2}D_M{\phi}^{MN}D_N {\phi^*}+\frac{d}{L^2} (\phi^*_{MN}{\phi}^{MN} -\f{1}{2}{\phi^*}\phi)\non\\
&&+\frac{m^2}{2} (\phi^*_{MN}{\phi}^{MN} -{\phi^*}\phi)-i \alpha g {\phi^*}^{PM}F_{MN}\phi^N_{~P}+ig\beta_1 F^{MN}\left( \nabla_M\phi^*_{PQ}\nabla^P\phi_N^{\;\;Q}- \nabla_M\phi_{PQ}\nabla^P\phi_N^{*\;\;Q}   \right) \non\\
&&+ig\beta_2 F^{MN}\left( \nabla_M\phi^*_{PQ}\nabla_N\phi^{PQ}  \right)+ig\beta_3 F^{MN}\left( \nabla_P\phi^*_{MQ}\nabla^Q\phi_N^{\;\;P}- \nabla_P\phi_{MQ}\nabla^Q\phi_N^{*\;\;P}   \right)\biggl],
\label{action}
\end{eqnarray}
where the indices $M,N$ etc. run from $0$ to $d$ and $\Lambda$ is the cosmological constant. The $F_{MN} = \p_M A_N -\p_N A_M $ is the field strength of the gauge field $A_M $ and we have defined
\be
\phi &=& G^{MN}\phi_{MN}\quad;\qquad D_M=\nabla_M+igA_M\,,
\ee
  with $\nabla_M$ being the covariant derivative with respect to the Levi-Civita connection. The $\phi^*_{MN}$ is the complex conjugate of $\phi_{MN}$ and will be treated as an independent field. The cubic terms are parametrized by five independent parameters, namely $g, \alpha, \beta_1, \beta_2, \beta_3$,  matching the number of independent parameters that we find in the CFT analysis of corresponding 3 point function. One of these parameters is the gauge coupling constant $g$, and it multiples the terms introduced by minimal coupling. The $\alpha$ denotes the gyromagnetic coupling, whereas $\beta_i$ (for $i=1,2,3$) denote the higher derivative couplings. 
  % Upto three derivatives, the above action is the most general effective action one can write down involving complex massive spin two field and a gauge field in the AdS background. 
  One way to arrive at the higher derivative terms is to write down the most general cubic action involving the 3 derivatives and then use the lower order equations of motion to eliminate the redundant terms. This was explained in detail for the massive spin-1 field in \cite{Marotta:2024sce}.
   
\vspace*{.07in}Below, we shall only consider the minimal and gyromagnetic couplings. A systematic analysis of the higher derivative couplings for the massive spin-2 field requires causality analysis and will be postponed for a future work. Further, as mentioned in the introduction, we shall ignore the back reaction due to the presence of massive fields by working with the infinitesimal sources. An infinitesimal energy momentum tensor will also avoid the Velo-Zwanziger problem. The gauge field equation derived from \eqref{action} in the AdS background is given by
\begin{eqnarray}
\nabla_M\,F^{MN}=J^N\qquad\implies\qquad  \Bigl(\nabla_M\nabla^M +\frac{d}{L^2} \Bigl)A_N-\nabla_N\nabla_MA^M=J_N\,,
\end{eqnarray}
with the source current given by
\be
J^N &=&-\f{ig}{2}\phi^{*}_{MP}\nabla^N\phi^{MP}+\f{ig}{2}\nabla^N\phi^{*}_{MP}\phi^{MP}+ig\phi^{*}_{MP}\nabla^P\phi^{MN}-ig\nabla_M\phi^{*N}_{P}\phi^{PM}\non\\
&&+\f{ig}{2}\phi^*\nabla^N\phi -\f{ig}{2}\nabla^N\phi^*\;\phi-\f{ig}{2}\phi^{*MN}\nabla_M\phi+\f{ig}{2}\nabla_M\phi^{*MN}\phi+\f{ig}{2}\phi^{MN}\nabla_M\phi^{*}-\f{ig}{2}\nabla_M\phi^{MN}\phi^{*}\non\\
&&+ 2i \alpha g\nabla_M( {\phi^*}^{P[M}\phi^{N]}_{~P})-2ig\beta_1 \nabla_M\left( \nabla^{[M}\phi^*_{PQ}\nabla^{|P|}\phi^{N]Q}- \nabla^{[M}\phi_{PQ}\nabla^{|P|}\phi^{*N]Q}   \right) \non\\
&&-2ig\beta_2 \nabla_M\left( \nabla^{[M}\phi^*_{PQ}\nabla^{N]}\phi^{PQ}  \right)-2ig\beta_3 \nabla_M\left( \nabla_P\phi^{*[M}_{\;\;Q}\nabla^{|Q|}\phi^{N]P}- \nabla_P\phi^{[M}_{\;\;Q}\nabla^{|Q|}\phi^{*N]P}   \right)\,.\non\\
\ee
For computing the 3-point function, we shall only need the $O(g)$ terms in the above current. Since the trace of the spin-2 field vanishes at the zeroth order in the coupling $g$, we can set the trace $\phi$ to zero in the above expression. 

\vspace*{.07in}For computing the 3-point function, it is convenient to gauge fix $A_M(z,k)$. For similar analysis involving massive spin-1 fields, this was done in detail in \cite{Marotta:2024sce}. We shall follow the same approach here working in the axial gauge $A_0(z,k)=0$. The perturbative classical solution of the gauge field upto $O(g)$ is given, in momentum space, by \cite{Marotta:2024sce} 
\begin{eqnarray}
A_\mu(z,\,\textbf{k})=\mathbb{K}_{\mu}^{\;\;\nu}(z,\textbf{k}) A_{(0)\nu}(\textbf{k}) +\,\int dw \sqrt{G} \;{\cal G}_{\mu\nu}(z,\,w;\,\textbf{k})\,J^\nu(w,\,\textbf{k})\, , \label{ftr5a}
\end{eqnarray} 
where $A_{(0)\mu}(\textbf{k})$ denotes the boundary value of the gauge field and $\mathbb{K}_\mu^{\;\;\nu}(z,\textbf{k})$ and $\mathcal{G}_{\mu\nu}(z,w;k)$ denote the bulk-to-boundary and bulk-to-bulk propagators of the gauge field respectively. In the Poincar\'e coordinates, their expressions are given by \cite{Marotta:2024sce}
\begin{subequations}
\begin{align}
\mathbb{K}_{\mu\nu}(z,\textbf{k})&=c_0(k)z^{\f{d-2}{2}}K_{\f{d}{2}-1}(zk)\pi_{\mu\nu}\;+\; \f{k_\mu k_\nu}{k^2}\, , 
\\
 \mathcal{G}_{\mu\nu}(z,w;\textbf{k}) &= -L
    \begin{cases}
      (\hat{z}\hat{w})^{\f{d}{2}-1}I_{\f{d}{2}-1}(k z)K_{\f{d}{2}-1}(k w)\pi_{\mu\nu}+\f{\hat{z}^{d-2}}{d-2}\f{k_\mu k_\nu}{k^2},& \text{if } z< w\,,\\[.4cm]
     (\hat{z}\hat{w})^{\f{d}{2}-1}I_{\f{d}{2}-1}(k w)K_{\f{d}{2}-1}(k z)\pi_{\mu\nu}+\f{\hat{w}^{d-2}}{d-2}\f{k_\mu k_\nu}{k^2},              & \text{if } z > w\,,
         \end{cases}
      \end{align}
\end{subequations}
 where $\hat z=\f{z}{L}$ and $\pi_{\mu\nu}$ denotes the transverse projector
\be   
\pi_{\mu\nu}= \delta_{\mu\nu} -\frac{k_\mu\,k_\nu}{k^2}\quad;\quad \delta^{\mu\nu} k_\mu \pi_{\nu\sigma}=0\quad;\quad \pi_{\mu\nu} \,\delta^{\nu\uptau }\pi_{\uptau \sigma}=\pi_{\mu\sigma}\, .
\ee
with $k^2=\delta^{\mu\nu}\,k_\mu\,k_\nu$.

\vspace*{.07in}We shall use \eqref{ftr5a} to determine the 3-point function via holographic renormalization. Thus, we shall only need the free field classical solutions of the massive spin-2 fields, which can be expressed in terms of the Btb propagator as 
\begin{eqnarray}
\phi_{MN}(z,\,\textbf{k})={K}_{MN}{}^{\mu\nu}(z,\,\textbf{k})\, \hat\phi^{(0)}_{\mu\nu}(\textbf{k})\quad;\quad \phi^*_{MN}(z,\,\textbf{k})=\bar{{K}}_{MN}{}^{\mu\nu}(z,\,\textbf{k})\, \hat\phi^{*(0)}_{\mu\nu}(\textbf{k})\,.
\end{eqnarray}
The expressions of the Btb propagators ${K}_{MN}{}^{\mu\nu}(z;\textbf{k})$ and $\bar{{K}}_{MN}{}^{\mu\nu}(z;\textbf{k})$ in the Poincar\'e coordinates are given in appendix \ref{appenF}. The $\hat\phi^{(0)}_{\mu\nu}$ and $\hat\phi^{*(0)}_{\mu\nu}$ are related to the boundary values of the bulk fields $\phi_{\mu\nu}$ and $\phi^*_{\mu\nu}$, respectively. Note that we only need to specify the boundary components of the massive fields. The component involving the bulk direction, namely $\phi_{\mu z}$ and $\,\phi_{z z}$ get fixed in terms of the boundary components as discussed in section \ref{sec4}. The bulk fields $\phi_{MN}$ and $\phi^*_{MN}$ are dual to the non-conserved spin-2 boundary CFT operators with their mass $m$ related to the conformal dimension $\Delta$ of the boundary CFT operators by the relation $L^2\,m^2=\Delta(\Delta -d)$. On the other hand, the bulk field $A_M$ is dual to a boundary conserved current having conformal dimension $d-1$. 

\subsection{Three point function}\label{3pointsec}

In this subsection, we compute the 3-point function involving a gauge field and two massive spin-2 fields in the bulk and compare with the CFT 3-point function described in section \ref{sec2review3er}. The desired bulk 3-point function can be computed using the procedure of holographic renormalization following the approach described in \cite{Marotta:2024sce} in the momentum space for two massive spin-1 fields and one gauge field. Here, we need to follow the same steps, replacing the massive spin-1 fields with the massive spin-2 fields. The first step is to obtain the 1 point function of the bulk gauge field. This would be a function of the sources of all the fields in the theory. The functional derivative of this one point function with respect to the boundary sources of the massive spin-2 fields will give the desired 3-point function. The one point function of the bulk gauge field in the axial gauge $A_0=0$ is given in boundary momentum space by \cite{Marotta:2024sce}
\be
\langle \mathcal{J}^\mu(\textbf{p}_2)\rangle= \delta^{\uptau \lambda}\int_0^\infty d\sigma\sqrt{G}\;\mathbb{K}_{\lambda\kappa}(\sigma;\textbf{p}_2)J_{(0)}^\kappa(\sigma,\textbf{p}_2)\,.
\ee
Taking the functional derivative of the above expression with respect to the sources, we obtain 
\be
\langle O^*{}^{\alpha\beta}(\textbf{p}_1)\mathcal{J}^\kappa(\textbf{p}_2) O^{\gamma\delta}(\textbf{p}_3)\rangle 
= \frac{\delta^{\kappa\mu}(2\pi)^d (2\pi)^d\delta^2 }{\delta\hat\phi^{(0)}_{\alpha\beta}(-\textbf{p}_1)\delta\hat\phi{}^{*(0)}_{\gamma\delta}(-\textbf{p}_3) } \int_0^\infty d\sigma \sqrt{G}\, \mathbb{K}_{\mu\nu}(\sigma;\textbf{p}_2)J^\nu_{(0)}(\sigma,\textbf{p}_2)\,,\label{9191}
\ee
where $\hat\phi^{(0)}_{\mu\nu}$ and $\hat\phi^{*(0)}_{\mu\nu}$ are related to the boundary conditions of the bulk massive spin-2 fields $\phi_{MN}$ and $\phi^*_{MN}$, respectively. In AdS, these act as the sources of the boundary operators ${\cal O}^*_{\mu\nu}$ and ${\cal O}_{\mu\nu}$ respectively. The $J^\nu_{(0)}$ is the boundary component of the gauge field current $J^N$ evaluated on the free spin-2 field and is a function of the sources $\hat\phi^{(0)}_{\mu\nu}$ and $\hat\phi^{*(0)}_{\mu\nu}$. Its expression is given in equation \eqref{g349}. Here, we only consider the minimal and gyromagnetic terms, setting the higher derivative couplings $\beta_i$ to zero. For a detailed expansion of the current and contribution to the 3-point function from each interaction term, see Appendix \ref{G}. The contribution to the 3-point function from each term in the current can be organized in terms of the triple K integrals \eqref{B.51}. For stating the results, we shall again work with an index free notation multiplying both sides of \eqref{9191} with the auxiliary polarization tensors. Using the various triple-K identities (see, e.g., \cite{Bzowski:2013sza, Marotta:2022jrp} for the identities involving the triple K integrals), the transverse part of the desired 3-point function, after a tedious calculation, is found to be 
\be
 \llangle[\Bigl] \epsilon_1\cdot\mathcal{O}^*({\bf{p}_1}) \epsilon_2\cdot j({\bf p}_2) \epsilon_3\cdot\mathcal{O}({\bf p}_3)\rrangle[\Bigl]
=(\epsilon_2\cdot \pi_2\cdot p_1)A+ (\epsilon_2\cdot \pi_2\cdot \epsilon_1)B_1 +(\epsilon_2\cdot \pi_2\cdot \epsilon_3)B_2\,,
\label{adsrt8}
\ee
with the functions $A, B_1$ and $B_2$ having the same functional form as in \eqref{ansatzs} and \eqref{8103e} (see appendix \ref{sec5:s2} for detailed expressions in terms of the triple K integrals). However, the momentum independent coefficients are now given in terms of the bulk parameters as  
\begin{align}
&a_0^{(0,0)} = -gC;\quad\quad
a_1^{(0,1)} = gC\,\frac{(-1 + \alpha)}{ \Delta};\quad\quad a_1^{(1,0)} = -gC\,\frac{(-1 + \alpha)}{\Delta},\nonumber\\
&a_2^{(0,2)} = gC\,\frac{4 \alpha}{ (\Delta - 1)\Delta};\quad\quad a_2^{(2,0)} = gC\,\frac{4 \alpha}{ (\Delta - 1)\Delta};\quad\quad a_2^{(1,1)} = -gC\,\frac{4 \alpha}{ \Delta^2},\nonumber \\
&a_1^{(0,0)} = gC\, \frac{2 \left( (-2 + d)\alpha - 2 \Delta \right)(\Delta - 1)}{ \Delta^2};\quad\quad a_2^{(0,1)} = gC\,\frac{2 \left(d + 3 d \alpha + 2 \alpha (\Delta - 2) - 2 \Delta\right)}{ \Delta^2},\nonumber\\ 
&a_2^{(1,0)} = -gC\,\frac{2 \left(d + 3 d \alpha + 2 \alpha (\Delta - 2) - 2 \Delta\right)}{ \Delta^2},\nonumber\\
&a_2^{(0,0)} = gC\,\frac{2 (\Delta - 2)(-2 + d + 2 \Delta)(d + 2 d \alpha - 2(2 \alpha + \Delta))}{ (\Delta - 1) \Delta^2}, \nonumber\\
&a_1^{(1,1)} = 0;\quad\quad a_2^{(2,2)} = 0;\quad\quad a_2^{(1,2)} = 0;\quad\quad a_2^{(2,1)} = 0,\non
\end{align}
\begin{align}
&b_{1;1}^{(0,0)} = gC\,\frac{(-2 + \Delta) \left[ 3 d^2 (1 + \alpha) + 6 d (-2 + \alpha (-4 + \Delta) + \Delta) - 2 (-1 + \Delta)(\Delta + \alpha (12 + \Delta)) \right]}{3  \Delta^2},\nonumber\\
&b_{1;1}^{(2,1)} = 0;\quad\quad b_{1;0}^{(1,0)} = -gC\,\frac{(1 + \alpha)(-1 + \Delta)}{ \Delta};\quad\quad b_{1;1}^{(1,1)} = -gC\,\frac{(1 + \alpha)}{2  \Delta^2}, \nonumber \\[8pt]
&b_{1;1}^{(2,0)} = gC\,\frac{2(1 + \alpha)}{ \Delta};\quad\quad b_{1;0}^{(0,0)} = gC\,\frac{(1 + \alpha)(d - 2 \Delta)(-1 + \Delta)}{ \Delta},\nonumber\\ 
&b_{1;1}^{(0,1)} = gC\,\frac{(1 + \alpha)(-1 + d - \Delta)}{ \Delta};\quad\quad b_{1;1}^{(1,0)} = -gC\,\frac{2(1 + \alpha)(-1 + \Delta)(d + \Delta)}{ \Delta^2},\non
\end{align}
\begin{align}\label{b2formfac}
&b_{2;1}^{(0,0)} = gC\,\frac{(-2 + \Delta) \left[ 3 d^2 (1 + \alpha) + 6 d (-2 + \alpha (-4 + \Delta) + \Delta) - 2 (-1 + \Delta)(\Delta + \alpha (12 + \Delta)) \right]}{3  \Delta^2},\nonumber\\ 
&b_{2;1}^{(1,2)} = 0;\quad\quad b_{2;0}^{(0,1)} =gC\, \frac{(1 + \alpha)(-1 + \Delta)}{ \Delta};\quad\quad b_{2;1}^{(1,1)} = -gC\,\frac{(1 + \alpha)}{2 \Delta^2}, \nonumber \\[8pt]
&b_{2;1}^{(0,2)} =gC\, \frac{2(1 + \alpha)}{ \Delta};\quad\quad b_{2;0}^{(0,0)} =gC\, \frac{(1 + \alpha)(d - 2 \Delta)(-1 + \Delta)}{ \Delta}, \nonumber \\[8pt]
&b_{2;1}^{(1,0)} = -gC\,\frac{(1 + \alpha)(-1 + d - \Delta)}{ \Delta};\quad\quad b_{2;1}^{(0,1)} = gC\,\frac{2(1 + \alpha)(-1 + \Delta)(d + \Delta)}{ \Delta^2}, 
\end{align}
where, we have defined
\begin{eqnarray}
C= -\,\frac{2^{2-\frac{d}{2}}}{\Gamma\left(\frac{d}{2}-1\right)}\,\left[\frac{2^{\frac{d}{2}+1-\Delta}}{\Gamma\left(\Delta -\frac{d}{2}\right)}\right]^2\,L^{2\Delta-d-1}\,.
\end{eqnarray}
As discussed in appendix \ref{sec5:s2}, among the 30 form factors given above, only 5 are independent which we have chosen to be  $a_0^{(0,0)},\, a_1^{(1,1)}\,a_2^{(2,2)}, \,a_2^{(1,1)}$ and $b_{1;1}^{(1,1)}$. It is easy to verify that the above coefficients satisfy the relation between the different coefficients given in \eqref{6.109}-\eqref{secwardend}. This is a non trivial check of the relation between the bulk gyromagnetic coupling of massive spin-2 fields and the CFT parameters. 

\vspace*{.07in}In \eqref{adsrt8}, we have only determined the part of the 3-point function which gets contribution from the transverse part of the gauge field (first term in equation \eqref{3.21}). The longitudinal part is given by the second term in \eqref{3.21}. By equation \eqref{decnb4e}, the longitudinal part can be related to the two point function of the spin-2 operators. We shall now show that the longitudinal part of our 3-point function can also be determined from the bulk analysis and is consistent with \eqref{decnb4e}. 

\vspace*{.07in}Focussing on odd $d$, we start by noting that the 1-point function of the divergence of the boundary current is given in terms of the bulk current as \cite{Marotta:2024sce} 
\be
\llangle { p}_{2\mu}\mathcal{J}^\mu({\bf p}_2)\rrangle =  -\frac{2}{L}\,
\,\delta^{\mu\nu}\ \left(\frac{d}{2} -1\right) p_{2\mu} A_\nu^{(d -2)}\label{8.7yt}\,,
\ee
where $ A_\nu^{(d -2)}$ is the coefficient of $\rho^{\f{d-2}{2}}$ in the asymptotic expansion of the gauge field  
\be
 A_\mu(\rho,\textbf{x})\;=\; \sum_{j=0}^{\frac{d-3}{2}}   \left(\frac{\rho}{L}\right)^{j}\; A^{(2j)}_\mu(\textbf{x})\;\,+\;\,   \left(\frac{\rho}{L}\right)^{\frac{d-2}{2}}A_\mu^{(d-2)}(\textbf{x}) +\dots  
 \ee
 where the explicit expressions of $A^{(2j)}_\mu(x)$ can be found in \cite{Marotta:2024sce}.  

Now, for the action in \eqref{action}, the divergence of the gauge field can be computed, up to $O(g)$, via asymptotic analysis to be 
\begin{eqnarray}
\p^\mu A_\mu^{(d-2)} &=& \frac{2ig}{(d-2)} \left(\Delta -\frac{d}{2}\right)
\left(\phi_{\mu\nu}^{*(0)} \phi^{\mu\nu (2\Delta-d)}-
\phi_{\mu\nu}^{(0)} \phi^{*\mu\nu (2\Delta-d)}
\right)\, , 
\end{eqnarray}
In momentum space, this gives 
\be
&&\hspace*{-1cm}(d-2)\delta^{\mu\nu} { p}_\mu A_\nu^{(d-2)}({\bf p})\non\\%+2\delta^{\mu\nu} {\bf p}_\mu B_\nu^{(d-2)}({\bf p})\non\\
&=& 
g\left(\Delta-\f{d}{2}\right)\int \f{d^dk}{(2\pi)^d} \delta^{\mu\alpha}\delta^{\nu\beta}\biggl[ \hat\phi_{\mu\nu}^{*(0)}({\bf k}) \hat{\phi}_{\alpha\beta}^{(2\Delta-d)}({\bf p}-{\bf k})-\hat{\phi}_{\mu\nu}^{(0)}({\bf k}) \hat{\phi}_{\alpha\beta}^{*(2\Delta-d)}({\bf p}-{\bf k})    \biggl],\non\\
&=&g\left(\Delta-\f{d}{2}\right)\int \f{d^dk}{(2\pi)^d} \delta^{\mu\alpha}\delta^{\nu\beta}P_{\alpha\beta}{}^{\gamma\delta}({\bf p}-{\bf k})\biggl[ \hat\phi_{\mu\nu}^{*(0)}({\bf k}) \hat{\phi}_{\gamma\delta}^{(0)}({\bf p}-{\bf k})-\hat{\phi}_{\mu\nu}^{(0)}({\bf k}) \hat{\phi}_{\gamma\delta}^{*(0)}({\bf p}-{\bf k})    \biggl]\non\label{fgtyhoo}
\\
\ee
Using \eqref{8.7yt} and \eqref{fgtyhoo}, we now get 
\be
&&\Bigl\langle \mathcal{O}^{*\gamma\delta}({\bf p}_1){ p}_{2\mu}\mathcal{J}^\mu({\bf p}_2) \mathcal{O}^{\sigma\tau}({\bf p}_3)\Bigl\rangle \non\\[.2cm]
&=&-\f{1}{L}(d-2)\delta^{\mu\nu }{ p}_{2\mu}\f{\delta^2 {A}^{(d-2)}_\nu ({\bf p}_2)}{\delta\hat{\phi}^{(0)}_{\gamma\delta}(-{\bf p}_1)\hat{\phi}^{*(0)}_{\sigma\tau}(-{\bf p}_3)}(2\pi)^d(2\pi)^d\non\\[.2cm]
&=& -\f{1}{L} g\left(\Delta-\frac{d}{2}\right)\biggl[  P^{\sigma\tau\gamma\delta}(\textbf{p}_1)-P^{\sigma\tau\gamma\delta}(\textbf{p}_3) \biggl] (2\pi)^d\delta^d({\bf p}_1+{\bf p}_2+{\bf p}_3)  \non\\[.3cm]
&=&g \biggl[\llangle[\Bigl] \mathcal{O}^{*\sigma\tau}(-{\bf p}_3)\mathcal{O}^{\gamma\delta}({\bf p}_3)\rrangle[\Bigl] - \llangle[\Bigl] \mathcal{O}^{*\sigma\tau}({\bf p}_1)\mathcal{O}^{\gamma\delta}(-{\bf p}_1)\rrangle[\Bigl] \biggl] (2\pi)^d \delta^d({\bf p}_1+{\bf p}_2+{\bf p}_3)\label{11220}\,,
\ee
where we have used equation \eqref{2pointfn} for the 2-point function. 

\vspace*{.07in}The above result is consistent with the transverse Ward identity \eqref{decnb4e} as expected. Further, it is easy to check that the coefficients of 2 and 3-points given in \eqref{5.109wer} and \eqref{b2formfac} respectively, which are obtained using the holographic renormalization procedure, satisfy the expected CFT relation \eqref{214rtyd}. For even $d$, the same steps can be repeated and we obtain the same result \eqref{11220}.

 \section{Flat limit }
\label{sec7}
As in well known, in the flat limit, the AdS correlators morph into the flat space correlators \cite{9901076, Susskind:1998vk, Giddings:1999qu,9907129,0903.4437,1002.2641, Maldacena:2015iua, 2007.13745, Penedones:2010ue,Paulos:2016fap, Fitzpatrick:2011hu, 1201.6449,1201.6452,Hijano:2019qmi, Hijano:2020szl, Farrow:2018yni, 2204.06462,2106.04606, Fitzpatrick:2011ia,0904.3544, Goncalves:2014ffa, 0907.0151, Fitzpatrick:2010zm, Fitzpatrick:2011jn, 1912.10046, Caron-Huot:2021kjy, Chandorkar:2021viw, vanRees:2022itk, Duary:2022pyv }. Representations such as position, mellin or momentum provide different perspectives of the flat limit analysis. In particular, for analyzing energy poles, analytic structure of correlators and for dealing with fields with spins, the momentum representation is quite convenient and powerful \cite{Farrow:2018yni, 1912.10046,
Marotta:2024sce}. Moreover, for connecting with the CFT data, momentum space is the cleanest approach \cite{Marotta:2024sce}.

\vspace*{.07in}In this section, we shall consider the flat limit of the CFT 3-point function discussed in the previous section. Following \cite{Marotta:2024sce}, for taking the flat limit of AdS correlators in the momentum space, a convenient approach is to work in the bulk coordinate $\uptau$ defined by $z=L\,e^{\uptau/L}$ with the metric given by
\begin{align}
    ds^2_{AdS_{d+1}} = d\uptau ^2+ e^{-2\uptau /L}\delta_{\mu\nu} dx^\mu dx^\nu\,.
\end{align}
At the leading order in $L$, it reduces to the flat metric in Euclidean signature. By Wick rotating $\uptau =it$, we can obtain the corresponding Lorentzian signature metric.

\vspace*{.07in}In subsection \ref{Btb}, we shall consider the flat limit of the Btb propagator of the massive spin-2 field. This will dictate the polarization of the massive spin-2 field in the flat space. In subsection \ref{3ptflat}, we shall take the flat limit of the 3-point function. 

\subsection{Flat limit of massive spin-2 Btb propagator}
\label{Btb}

The Btb propagator of the massive spin-2 fields in the boundary momentum space and the bulk Poincar\'e coordinate $z$ is given in \eqref{f346}. In the $\uptau$ coordinate, the different components of the propagator can be obtained in the standard way as 
\be
    &K_{\uptau\uptau}{}^{\alpha\beta}(\uptau;\textbf{k})=\left(\frac{\d z}{\d\uptau}\right)^2K_{zz}{}^{\alpha\beta}(z(\uptau);\textbf{k}),\qquad
    K_{\uptau\nu}{}^{\alpha\beta}(\uptau ;\textbf{k})=\left(\frac{\d z}{\d\uptau }\right) K_{z\nu}{}^{\alpha\beta}(z(\uptau);\textbf{k}).
\ee
and $K_{\mu\nu}{}^{\alpha\beta}(\uptau ;\textbf{k})=K_{\mu\nu}{}^{\alpha\beta}(z(\uptau);\textbf{k})$. To take the flat limit, we need to know how various quantities behave as we take the limit $L\rightarrow\infty$. The relation $\Delta (\Delta-d)=m^2L^2$ implies that a finite mass in this limit can be obtained only if we simultaneously send the conformal dimension $\Delta$ to infinity along with $L$. More precisely, in the limit $L\rightarrow\infty$, we have
\begin{align}
    \Delta=m L+\frac{d}{2}+O\left(\frac{1}{L}\right)\,.
\end{align}
The Btb propagator also involves the Bessel function of second kind whose order depends upon the conformal dimension $\Delta$ and the argument depends upon the bulk coordinate $z$ (or equivalently $\uptau$ and $L$). Hence, both the order as well as arguments of the Bessel function must be taken to be large. This is known as the uniform expansion and gives \cite{Marotta:2024sce}
\begin{align}
  K_{\Delta-\frac{d}{2}+s}(kz)= \left(\frac{\pi}{2EL}\right)^{\tfrac{1}{2}} 
\left(\frac{k}{m+E}\right)^{-mL - s} e^{-EL-E\uptau }\left(1+O\left(\frac{1}{L}\right)\right)\,,
\end{align}
where $E=\sqrt{k^2+m^2}$ denotes the energy of the particle in the flat limit. 

\vspace*{.07in}Using the above relations, the different components of the bulk-to-boundary propagators can be easily worked out which in turn gives the flat limit of the classical solutions. To leading order, we have 
\begin{subequations}
\begin{align}
      &\phi_{\uptau \uptau }(\uptau,\textbf{k})=K_{\uptau \uptau }{}^{\alpha\beta}(\uptau,\textbf{k}) \,\hat\phi^{(0)}_{\alpha\beta}(\textbf{k})\;\;\to\;\; \frac{e^{-E(\uptau +L)}}{Z_{\phi}^{1/2}}\left(\frac{i k^{\alpha}}{m}\right)\left(\frac{i k^{\beta}}{m}\right)\,\hat\phi^{(0)}_{\alpha\beta}(\textbf{k})\,,\\[.3cm]
   &\phi_{\uptau \nu}(\uptau,\textbf{k})=K_{\uptau \nu}{}^{\alpha\beta}(\uptau,\textbf{k})\,\hat\phi^{(0)}_{\alpha\beta}(\textbf{k})\;\;\to\;\;\frac{e^{-E(\uptau +L)}}{Z_{\phi}^{1/2}}\left(\frac{i k^{\alpha}}{m}\right)\tilde\pi^{\beta}{}_\nu\,\hat\phi^{(0)}_{\alpha\beta}(\textbf{k})\,,\\[.3cm]
   &\phi_{\mu\nu}(\uptau,\textbf{k})=K_{\mu\nu}{}^{\alpha\beta}(\uptau,\textbf{k})\,\hat\phi^{(0)}_{\alpha\beta}(\textbf{k})\;\;\to\;\;\frac{e^{-E(\uptau +L)}}{Z_{\phi}^{1/2}}\tilde\pi^\alpha{}_{(\mu}\,\tilde\pi^{\beta}{}_{\nu)}\,\hat\phi^{(0)}_{\alpha\beta}(\textbf{k})\,.
\end{align}
\end{subequations}
where
\begin{align}
    &\tilde\pi_{\mu\nu}=\delta_{\mu\nu}+\frac{k_\mu k_\nu}{m(m+E)}\;\;\;,\qquad
    \frac{1}{Z_{\phi}^{1/2}}=\left(\frac{2^{-mL}\,e^{mL}\,(L(m+E))^{m L}}{(EL)^{1/2}\,(m L)^{mL-\frac{1}{2}}}\right)\,.
\end{align}
The above expressions give the classical Euclidean profile of the massive spin-2 field in the flat limit 
\begin{align}
     \phi_{ab}^{\text{flat}}(\uptau,\textbf{k})=\Phi_{ab}(\textbf{k}) \,e^{-E(\uptau +L)} \,,
\end{align}
where the indices $a,b$ run over the $(d+1)$ dimensional flat directions and we have defined 
\begin{align}
    &\Phi_{ab}=\Bigg(\frac{i k_{\alpha}}{m}\frac{i k_{\beta}}{m}\phi^{(0)}{}^{\alpha\beta}_R,\,\frac{i k_{\alpha}}{m}\tilde\pi_{\beta\nu}\phi^{(0)}{}^{\alpha\beta}_R,\,\tilde\pi_\alpha{}_\mu\,\tilde\pi_{\beta}{}_\nu\,\phi^{(0)}{}^{\alpha\beta}_R\Bigg)\;\;\;,\qquad
     \phi^{(0)}{}^{\mu\nu}_R =\frac{\hat\phi^{(0)}{}^{\mu\nu}}{Z_{\phi}^{1/2}}.
\end{align}
The factor $1/Z_\phi^{1/2} $ diverges as we take the limit $L\to\infty$. Therefore, in this limit we also need to take the AdS source $\hat\phi^{(0)}{}^{\mu\nu} \to 0$ to keep $\hat\phi^{(0)}{}^{\mu\nu}_R$ finite. This is possible since $\hat\phi^{(0)}{}^{\mu\nu}$ is a free parameter. The uplifted Euclidean momenta for the $(d+1)$ dimensional flat spin-2 massive field is 
\begin{align}
    q^a=(\pm i E,k^\mu)\;\;,\qquad q^2 =\delta^{ab}q_aq_b =-m^2\,.\label{11213}
\end{align}
The $\pm$ here corresponds to incoming vs outgoing fields.

\vspace*{.07in}We can express $\Phi_{ab}$ in a basis of suitable polarization tensors. For this, we first introduce a basis $\epsilon_{\mu\nu}^{(r)}$ for $\phi^{(0)}_{\mu\nu}{}_R$. The symmetry and tracelessness property of $\phi^{(0)}_{\mu\nu}{}_R$ requires
\begin{align}
\epsilon_{\mu\nu}^{(r)}=\epsilon_{\nu\mu}^{(r)}\;\;\;,\hspace{1cm} \delta^{\mu\nu}\epsilon_{\mu\nu}^{(r)}=0, \hspace{1cm}\mu,\nu\in\{1,2,\cdots,d\}.
\end{align}
The above conditions imply that the basis $\epsilon_{\mu\nu}^{(r)}$ satisfying these conditions has dimension $(\frac{d(d+1)}{2}-1)$ and hence $r \,\,\in\,\,\{1,2,\cdots,\frac{d(d+1)}{2}-1\}$. As will be clear shortly, we shall also choose $\epsilon_{\mu\nu}^{(r)}$ to be orthonormal, i.e.,
\begin{align}   \delta^{\mu\alpha}\delta^{\nu\beta}\epsilon_{\mu\nu}^{(r_1)}\epsilon_{\alpha\beta}^{(r_2)}=\delta^{r_1r_2},\hspace{1cm}r_1,r_2\in\left\{1,2,\cdots,\frac
   {d(d+1)}{2}-1\right\}.   
\end{align}
In terms of $\epsilon_{\mu\nu}^{(r)}$, we write
\begin{align}
    \phi^{(0)}_{\mu\nu}{}_R\;\; = \sum_{r=1}^{\frac{d(d+1)}{2}-1} \phi^{(r)}\,\epsilon_{\mu\nu}^{(r)}\,.
\end{align}
We can now express the flat space field $\Phi_{ab}$ as
\begin{align}
    \Phi_{ab}=\sum_{r=1}^{\frac{d(d+1)}{2}-1}\phi^{(r)}\,\varepsilon_{ab}^{(r)}\,,
\end{align}
where we have introduced the Euclidean $(d+1)$ dimensional flat polarisation tensors $\varepsilon^{(r)}_{ab}$ as 
\begin{align}
    \varepsilon_{ab}^{(r)}=\Bigg(\frac{i k^{\alpha}}{m}\frac{i k^{\beta}}{m}\epsilon_{\alpha\beta}^{(r)}\;,\,\frac{i k^{\alpha}\tilde\pi^{\beta}{}_\nu}{m}\epsilon_{\alpha\beta}^{(r)}\;,\,\tilde\pi^\alpha{}_\mu\,\tilde\pi^{\beta}{}_\nu\,\epsilon_{\alpha\beta}^{(r)}\Bigg)\,.
\end{align}
Using the definition of $ \varepsilon_{ab}^{(r)}$ and the properties of $\epsilon^{(r)}_{\mu\nu}$, we can show that $ \varepsilon_{ab}^{(r)}$ has the expected properties, namely it is symmetric and traceless and satisfies orthonormality, i.e., 
    \begin{align}\label{cond1vareps}
         &\varepsilon_{ab}^{(r)}= \varepsilon_{ba}^{(r)},\hspace{1cm}\delta^{ab} \varepsilon_{ab}^{(r)}=0,\hspace{1cm}\delta^{ac}\delta^{bd}\varepsilon_{ab}^{(r_1)}\varepsilon_{cd}^{(r_2)}=\delta^{r_1r_2}\,.
    \end{align}
Further, the Polarization tensors are also transverse and satisfy the completeness relation for massive spin-2 fields, i.e.,
\be
\varepsilon_{ab}^{(r)} \,q^b=0\,,\label{cond2vareps}
\ee
and
\be
         \sum_{r=1}^{\frac{d(d+1)}{2}-1} \varepsilon_{ab}^{(r)}\varepsilon_{cd}^{(r)}=\frac{\pi_{ac}\pi_{bd}+\pi_{ad}\pi_{bc}}{2}-\frac{1}{d}\pi_{ab}\pi_{cd}\quad,\qquad
    \pi_{ab}=\delta_{ab}+\frac{q_a q_b}{m^2}.\label{11.209}
     \ee
For a derivation of the completeness relation for massive spin-2 particles, see e.g., \cite{Koenigstein:2015asa}. 

\vspace*{.07in}It is also possible to construct the spin-2 polarization tensor in terms of the tensor product of spin-1 polarization vectors. Suppose $\varepsilon_a^{(\lambda)}$ denote the spin-1 polarization vectors (see \cite{Marotta:2024sce} for the construction of spin-1 polarization vectors in terms of the AdS quantities in flat limit). Define the object
\be
\epsilon^{(\lambda_1\lambda_2)}_{ab}=\frac{\varepsilon^{(\lambda_1)}_{a}\varepsilon^{(\lambda_2)}_{b}+\varepsilon^{(\lambda_2)}_{a}\varepsilon^{(\lambda_1)}_{b}}{2}-\frac{\pi_{ab}}{d}\varepsilon^{(\lambda_1)}_{c}\varepsilon^{(\lambda_2)c}\,,
\label{11.210}
\ee
where $\lambda_1, \lambda_2 \in (1,\cdots,d)$ and the spin-1 polarizations are given by \cite{Marotta:2024sce}
\begin{align}
    \varepsilon_a^{(\lambda)}=\left(\frac{ik^\nu}{m} \epsilon_\nu^{(\lambda)},\tilde\pi_{\mu}{}^\nu\epsilon_\nu{}^{(\lambda)}\right)\quad;\qquad \epsilon_\nu^{(\lambda)} = \delta^\lambda_\nu\,.\label{4.143fg}
\end{align}
Using the identities $\pi_{ab}q^a=0$ and $\delta^{ab}\pi_{ab}=d$, we can easily check that $\epsilon^{(\lambda_1\lambda_2)}_{ab}$ is symmetric, transverse and traceless.

Using the spin-1 completeness relation $\varepsilon_a^{(\lambda)}\varepsilon^{(\lambda')*a}=\delta^{\lambda\lambda'}$, one can show that the objects $\epsilon^{(\lambda_1\lambda_2)}_{ab}$ satisfy the relation 
\be
\epsilon^{(\lambda_1\lambda_2)}_{ab}\epsilon^{(\lambda_1'\lambda_2')*ab}= \frac{\delta^{\lambda_1\lambda_1'}\delta^{\lambda_2\lambda_2'}+\delta^{\lambda_1\lambda_2'}\delta^{\lambda_2\lambda_1'}}{2}-\frac{1}{d}\delta^{\lambda_1\lambda_2}\delta^{\lambda'_1\lambda'_2}\,.\label{12.212}
\ee

The spin-2 polarization can now be defined in terms of $\epsilon^{(\lambda_1\lambda_2)}_{ab}$ as
\be
\varepsilon^{(r)}_{ab}=\sum_{\lambda_1,\lambda_2=1}^dU^{(r)}_{\lambda_1\lambda_2}\epsilon^{(\lambda_1\lambda_2)}_{ab}\,,\label{11.213}
\ee
where, the matrix elements $U^{(r)}_{\lambda_1\lambda_2}$ are essentially the Clebsch Gordan (CG) coefficients in going from $\epsilon^{(\lambda_1\lambda_2)}_{ab}$ basis to $\varepsilon^{(r)}_{ab}$ basis. A canonical choice is $U^{(r)}_{\lambda_1\lambda_2}=\langle 1\lambda_1;1\lambda_2|2r\rangle$,
i.e.\ Clebsch--Gordan coefficients\footnote{ We use the standard notation $\langle j_1 m_1; j_2 m_2|JM\rangle$ to denote the Klebsch-Gordan coefficients for adding the angular momenta $j_1$ and $j_2$ with $z$ components $m_1$ and $m_2$ respectively to produce the angular momentum $J$ with $z$ component $M$.} for the standard angular momentum problem $1\otimes 1\to 2$ (in $D=4$ this is same as usual angular momentum
addition with $\lambda\in\{+1,0,-1\}$ and $r\in\{+2,+1,0,-1,-2\}$).
In general, we have following properties of $U^{(r)}_{\lambda_1\lambda_2}$
\begin{subequations}
\begin{align}
U^{(r)}_{\lambda_1\lambda_2} &= U^{(r)}_{\lambda_2\lambda_1}
\qquad\text{(symmetric)}, \label{eq:U-sym}\\
\delta^{\lambda\lambda'}U^{(r)}_{\lambda_1\lambda_2} &= 0
\qquad\text{(traceless)}, \label{eq:U-trace}\\
\sum_{\lambda,\lambda'=1}^{d}U^{(r)}_{\lambda_1\lambda_2}\,(U^{(r')}_{\lambda_1\lambda_2})^{*}
&=\delta^{rr'}
\qquad\text{(orthonormality on symmetric traceless subspace)}. \label{eq:U-ortho}
\end{align}
\end{subequations}
For Clebsch--Gordan coefficients, \eqref{eq:U-sym} follows from the exchange symmetry
$\langle j_1 m_1; j_2 m_2|JM\rangle=(-1)^{j_1+j_2-J}\langle j_2 m_2; j_1 m_1|JM\rangle$ with
$j_1=j_2=1$ and $J=2$, while \eqref{eq:U-trace} expresses orthogonality to the singlet
($J=0$) state and \eqref{eq:U-ortho} is the standard CG orthogonality relation.

\vspace*{.07in}Using the properties of $U^{(r)}_{\lambda_1\lambda_2}$, we can easily show that the polarization defined by \eqref{11.213} satisfy the orthonormality and completeness relations. For orthonormality, we note that 
\be
\varepsilon^{(r)}_{ab}\,\varepsilon^{(r')*ab}
&=&
\sum_{\lambda,\lambda'}\sum_{\kappa,\kappa'}
U^{(r)}_{\lambda\lambda'}\,(U^{(r)'}_{\kappa\kappa'})^{*}\;
\epsilon^{(\lambda\lambda')}_{ab}\epsilon^{(\kappa\kappa')*ab}\non\\
&=&\sum_{\lambda,\lambda'}
U^{(r)}_{\lambda\lambda'}\,(U^{(r)'}_{\lambda\lambda'})^{*}\non\\
&=& \delta^{rr'}\,,
\ee
where in going to the second equality, we have used equation \eqref{12.212}, \eqref{eq:U-sym} and \eqref{eq:U-trace}. In going to the last equality, we have used \eqref{eq:U-ortho}. 

The completeness relation also follows if we note that $\sum_s U^{s}_{\lambda\lambda'}\,(U^{s}_{\kappa\kappa'})^{*}$ acts as the projector onto the symmetric traceless subspace and hence it satisfies
\begin{equation}
\sum_r U^{(r)}_{\lambda\lambda'}\,(U^{(r)}_{\kappa\kappa'})^{*}
\equiv
P_{\lambda\lambda',\kappa\kappa'}
=
\frac12\Big(\delta^{\lambda\kappa}\delta^{\lambda'\kappa'}+\delta^{\lambda\kappa'}\delta^{\lambda'\kappa}\Big)
-\frac{1}{d}\,\delta^{\lambda\lambda'}\delta^{\kappa\kappa'}.
\end{equation}
Using the above identity together with the definition \eqref{11.210} and the completeness relation of spin-1 polarizations, it follows that the spin-2 completeness relation \eqref{11.209} is satisfied. 

\vspace*{.07in}The analysis done so far recovers the flat space massive spin-2 field in the Euclidean signature. However, we can also Wick rotate the results to the Lorentzian signature. The Lorentzian time $t$ is related to the Euclidean time $\uptau$ by $t=-i\uptau$, i.e., $x^0_L=-ix^0_E$ where the subscript $L$ and $E$ correspond to the Lorentzian and Euclidean respectively. The Wick rotated $(d+1)$ dimensional momenta and polarisation are given by\footnote{In general, under a Wick rotation to Minkowski signature, any $(n+k)$- rank Euclidean tensor $T^{0\cdots 0,\mu_1\cdots\mu_k}$ with $n$ contravariant $0$-components, picks up $n$ factors of $(-i)$, i.e., 
\begin{align}
    T_L{}^{\overbrace{0 \ldots 0}^{n \text{ times}} \,\mu_1 \ldots \mu_k}=(-i)^n\,T_E{}^{\overbrace{0 \ldots 0}^{n \text{ times}} \mu_1 \ldots \mu_k}\non.
\end{align}
}
\be
    q^0_L&=&-i q^0_E=E,\hspace{1cm} q^\mu_L=q^\mu_E;\non\\[.3cm]
    \varepsilon^{00}{}^{(r)}_L&=&(-i)^2\varepsilon^{00}{}^{(r)}_E,\hspace{1cm}\varepsilon^{0\nu}{}^{(r)}_L=(-i)\varepsilon^{0\nu}{}^{(r)}_E,\hspace{1cm}\varepsilon^{\mu\nu}{}^{(r)}_L=\varepsilon^{\mu\nu}{}^{(r)}_E\,.
\ee
Thus, the Lorentzian signature momenta and the polarization tensors in the AdS flat limit are given by 
\be
    q_L^a&=&(\pm E,k^\mu)\quad;\qquad
    \varepsilon_L^{ab}{}^{(r)}\;=\;\Bigg(\frac{ k^{\alpha}}{m}\frac{ k^{\beta}}{m}\epsilon_{\alpha\beta}^{(r)}\;\;,\,\frac{ k^{\alpha}\tilde\pi^{\beta}{}^\nu}{m}\tilde\pi^{\alpha}{}^\nu\epsilon_{\alpha\beta}^{(r)}\;\;,\,\tilde\pi^\alpha{}^\mu\,\tilde\pi^{\beta}{}^\nu\,\epsilon_{\alpha\beta}^{(r)}\Bigg)\,.
\ee
The Wick rotated Lorentzian $(d+1)$ dimensional flat polarisations also satisfy conditions \eqref{cond1vareps}, \eqref{cond2vareps}, \eqref{11.209} with $\delta^{ab}\to\eta^{ab}$.

\vspace*{.07in}The flat limit of the classical profile of the gauge field and its bulk-to-boundary propagators were given in \cite{Marotta:2024sce}. Here, we just note that in the flat limit $L\to\infty$ the transverse component of the massless spin-1 gauge field becomes
\begin{align}
    A_M^{\perp} \to A_a^{\text{flat}}{}^\perp=\mathbb{A}_a e^{-k(L+\uptau )}\,,
\end{align}
where
\be
    \mathbb{A}_a =\left(0,\pi_{\mu\nu}a^\nu{}^{(0)}_R\right)\quad;\qquad a^\nu{}^{(0)}_R=\frac{a^\nu{}^{(0)}}{Z_A^{1/2}}.
\ee
The factor $Z_A$ denotes a renormalization factor whose explicit expression can be found in \cite{Marotta:2024sce}. For $d>3$, as $L\to\infty$, $\frac{1}{Z_A^{1/2}}\to\infty$, therefore we need to also take $a^\nu{}^{(0)}\to 0$ to keep $a^\nu{}^{(0)}_R$ finite. In the flat limit, the Euclidean $(d+1)$ dimensional momentum for the massless spin-1 gauge field is identified with 
\be
q^a=(\pm ik,k^\mu)\quad;\qquad q^2=\delta_{ab}q^aq^b=0\label{11236}\,.
\ee 
The basis for the polarization vector in terms of the AdS quantities can be written as in \eqref{4.143fg}.

\subsection{Flat limit of CFT $3-$point function}
\label{3ptflat}

The Btb propagators dictate the flat space polarizations. In the bulk AdS calculation, the 3-point function in the momentum space is expressed as sum of terms involving the triple K integrals. In the large $L$ and large $\Delta$ limit, these triple K integrals take a simple form and give rise to the energy conserving delta functions \cite{Marotta:2024sce}  
\begin{align}
    J_{N\{k_i\}}\xrightarrow[]{\textbf{flat limit}} \left(\frac{\pi}{2}\right)^{3/2} L^{\frac{d-5}{2}+N}\,\frac{(m+E_1)^{mL+k_1}\,p_2^{\frac{d-3}{2}+k_2}\,(m+E_2)^{mL+k_3}}{(E_1\,E_3)^{1/2}}\,(2\pi i)\,\delta(E_1+E_2+E_3)\,.
\end{align}
We see that in the large $L$ limit, the energy conserving delta function emerges naturally. We can apply the above identity to the triple K integrals appearing in the 3-point function involving the spin-1 conserved current and two spin-2 non-conserved operators given in section \ref{3pointsec}. Alternatively, it is also possible to directly work with the 3-point function expressed as integrals over the modified Bessel function.  Both approaches give the same final result. Here, we shall follow the second approach. We give the details of the calculations in appendix \ref{flat3pointlimit} and state the main results here. In the large $L,\Delta$ limit, the AdS 3-point function takes the form 
\begin{equation}
\lim_{L \to \infty} \sqrt{Z_{\phi_1} Z_A Z_{\phi_3}}\, \epsilon^1_{\mu_1\mu_2}\epsilon^2_{\mu_3}\epsilon^3_{\mu_4\mu_5}A_3^{\mu_1 \mu_2; \mu_3,\mu_4\mu_5} 
=(2\pi)^{d+1}\delta^{d+1}(q_1+q_2+q_3)\, \varepsilon^1_{ab}\varepsilon^2_c\varepsilon^3_{de}\mathcal{M}_3^{ab;c;de}\,.
\label{6.99}
\end{equation}
In the above expression, $\varepsilon_{ab}$ and $\varepsilon_{a}$ are the flat space polarizations of massive spin-2 and massless spin-1 fields constructed out of the $d$ dimensional auxiliary polarizations $\epsilon_{\mu\nu}$ and $\epsilon_{\mu}$ as described in the previous subsection. The $A_3^{\mu_1 \mu_2; \mu_3;\mu_4\mu_5}$ is the momentum space CFT 3-point function and 
$\mathcal{M}_3^{ab;c;de}$ is the corresponding flat space scattering amplitude. The $q^a= (\pm E, p^\mu)$ denote the flat space momenta. The spatial momentum conserving delta function is present in the CFT 3-point function itself. The energy conserving delta function $\delta(E_1+E_2+E_3)$ arises in the flat limit after integration over the bulk coordinate (see appendix \ref{flat3pointlimit}). Putting the expression of $\mathcal{M}_3$ in \eqref{6.99}, we obtain
\begin{align}
&\hspace*{-.3in}\lim_{L \to \infty}\sqrt{Z_{A}Z_{\phi_1}Z_{\phi_3}}\epsilon^{(r)}_1{}_{\alpha\beta}\,\epsilon^{(\lambda)}_{\kappa}\,\epsilon^{(r)}_3{}_{\gamma\delta}\,\langle O^*{}^{\alpha\beta}(\textbf{p}_1)\,J^\kappa(\textbf{p}_2)\, O^{\gamma\delta}(\textbf{p}_3)\rangle\non\\
=&
-\,ig\,(2\pi) \delta(E_1+E_2+E_3)(2\pi)^d\delta^d(\textbf{p}_1+\textbf{p}_2+\textbf{p}_3)\non\\
&\left[\frac{1}{2} (\varepsilon_1{}_{ab} \, \varepsilon_3^{ab})(q_3-q_1)\cdot \varepsilon_2+\left\{(1+\alpha)\,\varepsilon_{2}^a\,\varepsilon_{1ab}\,\varepsilon_3^{bc}q_{2c}-(1\leftrightarrow 3)\right\}\right]\,.
\end{align}
 By comparing this $d$-dimensional CFT correlator in flat limit with the corresponding expected $(d+1)$ dimensional flat amplitude reviewed in appendix \ref{expflat3point}, we see that they match exactly provided we identify the flat space couplings $\hat g$ and the gyromagnetic ratio $\hat\alpha$ with their AdS counterparts $g$ and $\alpha$ respectively at tree level. This shows that the flat space limit of the CFT correlator involving non-conserved massive spin-2 operators correctly reproduces the interacting part of the corresponding tree level flat-space S-matrix.

\section{Discussion}
\label{s4}
In this paper, we have analysed $U(1)$ charged massive spin-2 fields in the context of holography. In particular, we have studied the holographic renormalization of massive spin-2 field and used it to compute the 2 point function in momentum space. The expression matches exactly with the expected CFT result. We have also computed the AdS 3-point correlator involving two complex massive spin-2 fields and a gauge field. Our analysis gives the relation between the gyromagnetic coupling in the bulk and the boundary OPE coefficients. We have also reproduced the expected 3-point function involving the gauge and massive spin-2 fields in the flat space by taking a suitable flat limit of AdS. 

\vspace*{.07in}The 3-point analysis we have done can't fix the value of the gyromagnetic ratio. One needs to analyse the 4-point function for that. In particular, a bootstrap type of analysis may help analyse the gyromagnetic coupling in the bulk. The relation between the bulk coupling and the boundary OPE coefficients, we have obtained in this work, will be useful for such an analysis. In this work, we have not considered the higher derivative couplings involving the massive spin-2 fields. At the level of the the 3-point function, it will be straightforward to include their contributions. They play an important role in the analysis of causality behaviour in theories of gravity coupled to massive higher spin fields. We have also worked with a specific choice 1 of the parameter $\xi$ which parametrises the space of possible kinetic terms of massive spin-2 fields in AdS. This gave the expected relation between mass $m$ of the bulk massive spin-2 field and the conformal dimension $\Delta $ of the dual boundary operator. However, since all values of $\xi$ are expected to give a consistent kinetic term in AdS, it would be interesting to explore the role played by other values of $\xi$ in the holographic setting. 

\vspace*{.07in}From the analysis done in \cite{Marotta:2024sce}, and in this work, we also see how to perform the holographic renormalization of any massive higher spin field in AdS. For the symmetric traceless tensors, the analysis will be very similar to the massive spin-2 case. The auxiliary conditions will relate the solutions of field components having all boundary indices with the components having one or more bulk indices. The calculation of two point function of such fields is expected to reproduce the result given in \cite{Marotta:2022jrp, Arkani-Hamed:2015bza} for arbitrary symmetric traceless massive higher spin fields in the momentum space. For the mixed symmetry fields, the analysis is expected to be more involved. In particular, the auxiliary conditions will be more complicated, and one may have to analyse the conditions on a case by case basis. The calculations of correlators involving the massive higher spin fields using holographic renormalization in an EFT approach is expected to proceed along similar lines. However, if we want to move beyond EFT, the analysis will become more involved due to the inclusion of all massive higher spin fields. It would be interesting to use the momentum space CFT techniques to explore this aspect.

In this work, we have considered the flat limit of only the bulk theory. The bulk limit $L\rightarrow \infty$ corresponds to a Carrollian limit in which we send the speed of light $c$ to zero.  on the boundary \cite{2406.19343}. The boundary CFT correlators reduce to Carrollian conformal correlators living on the null boundary in Minkowski spacetime in this limit. This has been analysed for the scalar fields in position \cite{2406.19343,Bagchi:2023fbj} as well as momentum spaces \cite{Marotta:2025qjh}. However, there have been very few studies of Carrollian theories for higher spin fields. The ultra relativistic Carrollian limit of CFTs provides a useful starting point to study the theories on the flat space null surfaces (for a discussion on limitations to this approach, see \cite{Marotta:2025qjh}). Since momentum space provides a powerful and systematic framework for taking the Carrollian limit, it would be useful to consider the Carrollian limit of the results in this paper.

\bigskip

{\bf Acknowledgement:} 
We thank Chandramouli Chowdhury, Raffaele Marotta and Kostas Skenderis for useful discussions. We also thank Raffaele Marotta for the comment on the draft. MV is also supported in part by the “Young Faculty Research Seed Grant'' (IITI/YFRSG-Dream Lab/2024-25/Phase-IV/02) and its PRIUS ( ITII/YFRSG-PRIUS/2023-24/Phase-I/03) component of IIT Indore. AS is supported by the Research Studentship in Mathematical Science, University of Southampton, and funding from the Government of Karnataka, India. VN is supported by a University of Edinburgh School of Mathematics Studentship.

\appendix
\section{Conventions and useful identities}
 \label{geo768}

In this appendix, we summarise our conventions and note some useful identities which have been used in this work. We closely follow the conventions of \cite{Marotta:2024sce} for the momentum space CFT. In particular, we shall denote the $d+1$ dimensional AdS indices by $M,\,N, P\dots$ which run from $0$ to $d$. The $d$ dimensional boundary indices will be denoted by the Greek letters $\mu,\,\nu,\alpha,\beta,\sigma,\cdots$. These run from 1 to $d$. The boundary position and momentum variables would be denoted by boldface letters (e.g., $\textbf{x}$ and $\textbf{p}$). The indices of $d+1$ dimensional flat space would be denoted by $a,b,c,\cdots$ which would from 1 to $d+1$. The symmetrization would include a factor of $\f{1}{n!}$. E.g., for 2nd rank tensor, we have
\begin{eqnarray}
P_{(\mu}\,Q_{\nu)}=\frac{1}{2} \Big(P_\mu\,Q_\nu+P_\nu\,Q_\mu\Big). 
\end{eqnarray}
The AdS metric in the Fefferman Graham and the Poincar\'e coordinates are given by

\begin{eqnarray}
\mbox{Poincar\'e}\quad:&& ds^2=\frac{L^2}{z^2} \big( dz^2+ \delta_{\mu\nu} \, dx^\mu \,dx^\nu\big)\quad;\qquad \sqrt{G} = \left(\f{L}{z}\right)^{d+1}\,,\label{stanpoin54}\\
\mbox{Fefferman Graham}\quad:&&ds^2=L^2\frac{d\rho^2}{4\rho^2}+L \frac{\delta_{\mu\nu}\, dx^\mu\,dx^\nu}{\rho} \quad;\qquad \sqrt{G} = \f{1}{2}\left(\f{L}{\rho}\right)^{\f{d+2}{2}}\,.\label{FG}
\end{eqnarray}
The two coordinates are related by $\rho=\frac{z^2}{L}$. For the AdS metric denoted $G_{MN}$, the various curvature quantities are given by
\begin{eqnarray}
R_{MNPQ}= \frac{G_{MQ}\,G_{NP}-G_{MP}G_{NQ}}{L^2}~~;~~R_{MN}=-\frac{d}{L^2}\,G_{MN}\quad;\quad R=-\frac{d(d+1)}{L^2}\,.
\end{eqnarray}
In our derivations, we need to work with covariant derivatives of the spin-2 fields in AdS. Below, we note the expressions of two covariant derivatives acting on the spin-2 fields in the Fefferman Graham coordinates. We start with following fully contracted covariant derivatives
\be
\nabla^M \nabla^N \phi_{M N}&=&\f{4}{L^4}(d-4)^2\rho^2\phi_{\rho\rho}+\f{(2-d)}{L^3}\rho\delta^{\mu\nu}  \phi_{\mu\nu}+\f{16(5-d)}{L^4}\rho^3\p_\rho \phi_{\rho\rho}+\f{2\rho^2}{L^3}\delta^{\mu\nu} \p_\rho \phi_{\mu\nu}\non\\[.25cm]
&&+\f{4(3-d)}{L^3}\rho^2\delta^{\mu\nu}\p_\mu\phi_{\nu\rho} +\f{16\rho^4}{L^4}\p_\rho \p_\rho \phi_{\rho\rho}+\f{8\rho^3}{L^3}\delta^{\mu\nu}\p_\mu\p_\rho \phi_{\rho \nu}+\f{\rho^2}{L^2}\delta^{\mu k}\delta^{\nu l}\p_\mu\p_\nu\phi_{kl}\,,\non\\[.3cm]
 \nabla^M\nabla_M\phi &=&\f{4}{L^2}\left(1-\f{d}{2}\right)\rho\p_\rho \phi +\f{4\rho^2}{L^2} \p_\rho\p_\rho \phi +\f{\rho}{L} \delta^{\mu\nu} \p_\mu\p_\nu\phi \,.
\ee

\paragraph{Components of $\nabla_M\nabla_N\phi$ }
\begin{align}
   & (\rho\rho)\,:\hspace{1cm}\nabla_\rho\nabla_\rho \phi
= \p_\rho^2 \phi +\frac{1}{\rho} \p_\rho \phi,\non\\
&(\mu\rho)\,:\hspace{1cm}\nabla_\mu  \nabla_\rho  \phi= \p_\mu  \p_\rho  \phi+\f{1}{2\rho}\p_\mu \phi,\non\\
&(\mu\nu)\,:\hspace{1cm}\nabla_\mu  \nabla_\nu  \phi=\p_\mu  \p_\nu  \phi-2\delta_{\mu\nu}\p_\rho \phi.
\end{align}

\paragraph{Components of $\nabla^P\nabla_M\phi_{PN}$ }
\be
(\rho\rho)\,:\hspace{1cm}\nabla^P\nabla_\rho  \phi_{P \rho } &=&\f{4\rho^2}{L^2}\p_\rho\p_\rho \phi_{\rho\rho}+\f{\rho}{L}\delta^{\mu\nu}\p_\mu\p_\rho \phi_{\nu\rho}+\f{2}{L^2}\Bigl(8-\f{5d}{2}\Bigl)\phi_{\rho\rho}+\f{2}{L}\delta^{\mu\nu}\p_\mu \phi_{\nu\rho}\non\\[.2cm]
&&+\f{3\delta^{\mu\nu}\phi_{\mu\nu}}{4\rho L}+\f{2(10-d)}{L^2}\rho\p_\rho \phi_{\rho\rho}+\f{1}{2L}\delta^{\mu\nu}\p_\rho \phi_{\mu\nu},\non\\
(\mu\rho)\,:\hspace{1cm} \nabla^P\nabla_\mu  \phi_{P \rho } 
&=&\f{4\rho^2}{L^2}\p_\rho\p_\mu \phi_{\rho \rho}+\f{\rho}{L}\delta^{kl}  \p_\mu\p_k \phi_{l\rho} +\f{2\rho}{L^2}\p_\rho \phi_{\mu\rho}+\f{1}{2L}\delta^{jk}\p_j \phi_{\mu k} \non\\
&&+\f{2(4-d)}{L^2}\rho\p_\mu \phi_{\rho \rho} +\f{2}{L^2}\Bigl(\f{1}{2}-d\Bigl) \phi_{\mu\rho} +\f{1}{2L}\delta^{jk}\p_\mu \phi_{jk},\non\\
(\rho\mu)\,:\hspace{1cm} \nabla^P\nabla_\rho  \phi_{P \mu } &=&\f{4\rho^2}{L^2}\p_\rho^2 \phi_{\rho \mu}+\f{\rho}{L}\delta^{jk}  \p_j\p_\rho \phi_{\mu k} +\f{3}{2L}\delta^{jk}\p_k\phi_{\mu j} +\f{2(7-d)}{L^2}\rho\p_\rho \phi_{\mu\rho}  \non\\
&&+\f{1}{L^2}({5}-4d)\phi_{\mu\rho}, \non\\
(\mu\nu)\,:\hspace{1cm} \nabla^P\nabla_\mu  \phi_{P \nu } 
&=&\f{4\rho^2}{L^2}\p_\rho\p_\mu \phi_{\rho \nu}+\f{\rho}{L}\delta^{kl}  \p_\mu\p_k \phi_{l\nu}-\f{8\rho^2}{L^3}\delta_{\mu\nu}\p_\rho \phi_{\rho\rho}-\f{2\rho}{L^3}\delta_{\mu\nu}\delta^{kl}\p_k \phi_{l\rho} \non\\
&&+\f{2(2-d)}{L^2}\rho\p_\mu \phi_{\rho \nu} -\f{(d+1)}{L^2} \phi_{\mu\nu} +\f{4(d-3)}{L^3} \rho \delta_{\mu\nu}\phi_{\rho\rho}. 
\ee

\paragraph{Components of $\nabla^P\nabla_P\phi_{MN}$ }

\be\label{Box}
(\rho\rho)\,:\hspace{1cm} \nabla^P\nabla_P  \phi_{\rho \rho } &=&\f{4\rho^2}{L^2}\p_\rho^2 \phi_{\rho\rho}+\f{\rho}{L}\delta^{\mu\nu}\p_\mu\p_\nu \phi_{\rho\rho}-\f{2(3d-8)}{L^2}\phi_{\rho\rho}+\f{2}{L}\delta^{\mu\nu}\p_\mu \phi_{\nu\rho}\non\\[.25cm]
&&+\f{\delta^{\mu\nu}\phi_{\mu\nu}}{2\rho L}-\f{2(d-10)}{L^2}\rho\p_\rho \phi_{\rho\rho},\non\\
(\mu\rho)\,:\hspace{1cm}  \nabla^P\nabla_P  \phi_{\mu\rho } &=&\f{4\rho^2}{L^2}\p_\rho^2 \phi_{\mu\rho}+\f{\rho}{L}\delta^{kl}\p_k\p_l \phi_{\mu\rho} -\f{2(2d-3)}{L^2}\phi_{\mu\rho}-\f{2(d-8)}{L^2}\rho\p_\rho \phi_{\mu\rho}\non\\
&&-\f{4\rho}{L^2}\p_\mu \phi_{\rho\rho}+\f{1}{L}\delta^{jk}\p_j \phi_{\mu k}, \non\\
(\mu\nu)\,:\hspace{1cm} \nabla^P\nabla_P  \phi_{\mu \nu } &=&\f{4\rho^2}{L^2}\p_\rho\p_\rho \phi_{\mu\nu}+\f{\rho}{L}\delta^{kl}\p_k\p_l \phi_{\mu\nu} -\f{4\rho}{L^2}\p_\mu \phi_{\nu\rho}-\f{4\rho}{L^2}\p_\nu \phi_{\mu\rho} \non\\
&&-\f{2(d-6)}{L^2}\rho\p_\rho \phi_{\mu\nu}-\f{2(d-1)}{L^2}\phi_{\mu\nu}  +\f{8\rho}{L^3}\delta_{\mu\nu} \phi_{\rho\rho} .
\ee

\section{Form factors of 3-point CFT correlator}
\label{sec5:s2}

In this appendix, we summarise the CFT$_d$ 3-point function of a conserved current with conformal dimension $d-1$ and two non-conserved spin-2 operators having arbitrary but same conformal dimension $\Delta$. The expressions will be given in terms of the triple-K integral
\begin{align}
    J_{N\{\beta_1,\beta_2,\beta_3\}}(k_1,k_2,k_3)
    &\equiv \int_0^\infty dz\, z^{\frac{d}{2}+N-1}
    \prod_{i=1}^3 k_i^{\Delta_i-\frac{d}{2}+\beta_i}\,
    K_{\Delta_i-\frac{d}{2}+\beta_i}(z k_i)\,.
\end{align}
The transverse part of the 3 point function of two spin-2 operators and a conserved vector current is given by (see equation \eqref{ansatzs})
\be
\mathcal{A}_{2,J,2}^\perp &=&(\epsilon_2\cdot \pi_2\cdot p_1)A+ (\epsilon_2\cdot \pi_2\cdot \epsilon_1)B_1 +(\epsilon_2\cdot \pi_2\cdot \epsilon_3)B_2\,,
\ee
where, the functions $A, B_1$ and $B_2$ are parametrized as
\be
A&=&z^2 A_0^{(0,0)}+\zeta _2 \xi _2 z A_1^{(0,0)}+ \frac{1}{2} \zeta _2^2 \xi _2^2 A_2^{(0,0)}+\zeta _1 \xi _2 z A_1^{(0,1)}+\zeta _2 \xi _1 z A_1^{(1,0)}+\zeta _1 \xi _1 z A_1^{(1,1)}\non\\
&&+\frac{1}{2} \zeta _1 \zeta _2 \xi _2^2 A_2^{(0,1)}+\frac{1}{4} \zeta _1^2 \xi _2^2 A_2^{(0,2)}+\frac{1}{2} \zeta _2^2 \xi _1 \xi _2 A_2^{(1,0)}+\frac{1}{2} \zeta _1 \zeta _2 \xi _1 \xi _2 A_2^{(1,1)}+\frac{1}{4} \zeta _1^2 \xi _1 \xi _2 A_2^{(1,2)}\non\\
&&+\frac{1}{4} \zeta _2^2 \xi _1^2 A_2^{(2,0)}+\frac{1}{4} \zeta _1 \zeta _2 \xi _1^2 A_2^{(2,1)}+\frac{1}{8} \zeta _1^2 \xi _1^2 A_2^{(2,2)}\,,\non\\[.3cm]
B_1&=&\xi _1 z {B}_{1;0}^{(1,0)}+\xi _2 z {B}_{1;0}^{(0,0)}+\zeta _2 \xi _2 \xi _1 {B}_{1;1}^{(1,0)}+\zeta _1 \xi _2 \xi _1 {B}_{1;1}^{(1,1)}+\zeta _2 \xi _2^2 {B}_{1;1}^{(0,0)}+\zeta _1 \xi _2^2 {B}_{1;1}^{(0,1)}\non\\
&&+\frac{1}{2} \zeta _2 \xi _1^2 {B}_{1;1}^{(2,0)}+\frac{1}{2} \zeta _1 \xi _1^2 {B}_{1;1}^{(2,1)}\,,\non\\[.3cm]
B_2&=&\zeta _1 z {B}_{2;0}^{(0,1)}+\zeta _2 z {B}_{2;0}^{(0,0)}
+\zeta _2 \zeta _1 \xi _2 {B}_{2;1}^{(0,1)}+\zeta _2 \zeta _1 \xi _1 {B}_{2;1}^{(1,1)}+\zeta _2^2 \xi _2 {B}_{2;1}^{(0,0)}+\zeta _2^2 \xi _1 {B}_{2;1}^{(1,0)}\non\\
&&+\frac{1}{2} \zeta _1^2 \xi _2 {B}_{2;1}^{(0,2)}+\frac{1}{2} \zeta _1^2 \xi _1 {B}_{2;1}^{(1,2)}\,,\non
\ee
with
\begin{equation}
z = \epsilon_{1}\!\cdot\!\epsilon_{3} , \qquad
\xi_{1} = \epsilon_{3}\!\cdot\! p_{2} , \qquad
\xi_{2} = \epsilon_{3}\!\cdot (p_{1}+p_{2}) , \qquad
\zeta_{1} = \epsilon_{1}\!\cdot\! p_{2} , \qquad
\zeta_{2} = \epsilon_{1}\!\cdot\! p_{1}\,.
\end{equation}
The 30 form factors appearing above are given by \cite{Marotta:2022jrp}
\be
A_2^{(2,2)} &=& a_2^{(2,2)}J_{5\{0,0,0\}}\non\\[.2cm]
A_2^{(2,1)}&=&  a_2^{(2,1)} J_{4\{-1,0,0\}}-a_2^{(2,2)} J_{5\{-1,1,0\}}\non\\[.2cm]
A_2^{(1,2)}&=&  a_2^{(1,2)} J_{4\{0,0,-1\}}+a_2^{(2,2)} J_{5\{0,1,-1\}}\non\\[.2cm]
A_2^{(2,0)}&=&  a_2^{(2,0)} J_{3\{-2,0,0\}}-a_2^{(2,1)} J_{4\{-2,1,0\}}+\f{1}{2}a_2^{(2,2)} J_{5\{-2,2,0\}}\non\\[.2cm]
A_2^{(0,2)}&=&  a_2^{(0,2)} J_{3\{0,0,-2\}}+a_2^{(1,2)} J_{4\{0,1,-2\}}+\f{1}{2}a_2^{(2,2)} J_{5\{0,2,-2\}}\non\\[.2cm]
 A_{2}^{(1,1)} &=& a_2^{(1,1)}J_{3\{-1,0,-1\}}-a_{2}^{(2,2)} J_{5\{-1,2,-1\}}+   \bigl(a_{2}^{(2,1)}-a_2^{(1,2)}\bigl)J_{4\{-1,1,-1\}}    \non\\[.2cm]
 A_{2}^{(1,0)} &=& \Bigl(a_2^{(2,0)}-a_2^{(1,1)}\Bigl)J_{3\{-2,1,-1\}}+  \f{1}{2}a_2^{(2,2)}J_{5\{-2,3,-1\}}  +a_2^{(1,0)}J_{2\{-2,0,-1\}}    \non\\
 &&+\Bigl(\f{1}{2}a_{2}^{(1,2)}-a_{2}^{(2,1)}\Bigl) J_{4\{-2,2,-1\}}\non\\[.2cm]
 A_{2}^{(0,1)} &=& \Bigl(a_2^{(1,1)}-a_2^{(0,2)}\Bigl)J_{3\{-1,1,-2\}}-  \f{1}{2}a_2^{(2,2)}J_{5\{-1,3,-2\}}  +a_2^{(0,1)}J_{2\{-1,0,-2\}}    \non\\
 &&+\Bigl(\f{1}{2}a_{2}^{(2,1)}-a_{2}^{(1,2)}\Bigl) J_{4\{-1,2,-2\}}\non\\[.2cm]
  A_{2}^{(0,0)} &=&a_2^{(0,0)}J_{1\{-2,0,-2\}} +\f{1}{4}a_2^{(2,2)}J_{5\{-2,4,-2\}} +\Bigl(\f{1}{2}a_{2}^{(0,2)}+\f{1}{2}a_{2}^{(2,0)}-a_{2}^{(1,1)}\Bigl) J_{3\{-2,2,-2\}}\non\\[.2cm]
 && +\f{1}{2}\bigl( a_2^{(1,2)}- a_2^{(2,1)}\bigl)J_{4\{-2,3,-2\}}   + \Bigl(a_2^{(1,0)}-a_2^{(0,1)}\Bigl)J_{2\{-2,1,-2\}} \non
 \ee
  \be
A_1^{(1,1)}&=&  a_1^{(1,1)} J_{3\{0,0,0\}}-\f{1}{2}a_2^{(2,2)} J_{4\{0,1,0\}}\non\\
 A_{1}^{(1,0)} &=&a_{1}^{(1,0)} J_{2\{-1,0,0\}}-\Bigl(a_{1}^{(1,1)}+\f{1}{2}a_2^{(2,1)}\Bigl) J_{3\{-1,1,0\}}+\f{1}{2} a_2^{(2,2)}J_{4\{-1,2,0\}}   \non\\[.2cm]
 A_{1}^{(0,1)} &=& a_{1}^{(0,1)} J_{2\{0,0,-1\}}+\Bigl( a_{1}^{(1,1)}-\f{1}{2}a_2^{(1,2)} \Bigl)J_{3\{0,1,-1\}}-\f{1}{2} a_2^{(2,2)}J_{4\{0,2,-1\}}   \non\\[.2cm]
  A_{1}^{(0,0)} &=& -\Bigl(a_{1}^{(1,1)}+\f{1}{2}a_2^{(2,1)}-\f{1}{2}a_2^{(1,2)} \Bigl)J_{3\{-1,2,-1\}}+  \bigl(a_{1}^{(1,0)}-a_1^{(0,1)}-\f{1}{2}a_2^{(1,1)}\bigl)J_{2\{-1,1,-1\}} \non\\
        &&+a_1^{(0,0)}J_{1\{-1,0,-1\}}+\f{1}{2}a_2^{(2,2)}J_{4\{-1,3,-1\}}   \non\\[.2cm]
A_0^{(0,0)}&=&a_0^{(0,0)} J_{1\{0,0,0\}}-a_1^{(1,1)} J_{2\{0,1,0\}}+\f{1}{4}a_2^{(2,2)} J_{3\{0,2,0\}}\non
\ee

\be
  B_{2;0}^{(0,1)} &=&b_{2;0}^{(0,1)}J_{1\{0,0,0\}} + \Bigl(\f{1}{2}a_2^{(2,1)}-b_{2;1}^{(1,1)}-b_{2;1}^{(1,2)}\Bigl)J_{2\{0,1,0\}}+\Bigl(\f{1}{2}a_2^{(2,1)}-b_{2;1}^{(1,1)}\Bigl)J_{2\{0,0,1\}}\non\\[.2cm]
  &&- \f{1}{2}a_{2}^{(2,2)}J_{3\{1,1,0\}}+ \Bigl(\f{1}{2}a_2^{(2,1)}-b_{2;1}^{(1,1)}+a_{1}^{(1,1)}\Bigl)J_{2\{1,0,0\}} \non\\[.3cm]
  B_{2;0}^{(0,0)} &=&b_{2;0}^{(0,0)}J_{0\{-1,0,0\}} -\Bigl(a_{1}^{(1,1)}+\f{1}{2}a_{2}^{(2,1)} \Bigl)J_{2\{0,1,0\}}+b_{2;0}^{(0,1)}J_{1\{-1,0,1\}}+  \bigl(b_{2;0}^{(0,1)}+a_1^{(1,0)}\bigl)J_{1\{0,0,0\}} \non\\
   &&+\f{1}{2}a_2^{(2,2)}J_{3\{0,2,0\}} +b_{2;1}^{(1,2)}J_{2\{-1,2,0\}}   \non\\[.2cm] 
  B_{1;0}^{(1,0)} &=&b_{1;0}^{(1,0)}J_{1\{0,0,0\}} + \Bigl(\f{1}{2}a_2^{(1,2)}+b_{1;1}^{(1,1)}-b_{1;1}^{(2,1)}\Bigl)J_{2\{0,1,0\}}+\Bigl(\f{1}{2}a_2^{(1,2)}+b_{1;1}^{(1,1)}\Bigl)J_{2\{1,0,0\}}\non\\[.2cm]
  &&+\f{1}{2}a_{2}^{(2,2)}J_{3\{0,1,1\}}+ \Bigl(\f{1}{2}a_2^{(1,2)}+b_{1;1}^{(1,1)}-a_{1}^{(1,1)}\Bigl)J_{2\{0,0,1\}} \non\\[.3cm]
  B_{1;0}^{(0,0)} &=&b_{1;0}^{(0,0)}J_{0\{0,0,-1\}} -\Bigl(a_{1}^{(1,1)}-\f{1}{2}a_{2}^{(1,2)} \Bigl)J_{2\{0,1,0\}}-  \bigl(b_{1;0}^{(1,0)}+a_1^{(0,1)}\bigl)J_{1\{0,0,0\}} \non\\
   &&-b_{1;0}^{(1,0)}J_{1\{1,0,-1\}}+\f{1}{2}a_2^{(2,2)}J_{3\{0,2,0\}} -b_{1;1}^{(2,1)}J_{2\{0,2,-1\}}   \non
   \ee

   \be
   B_{2;1}^{(1,2)} &=& b_{2;1}^{(1,2)}J_{3\{0,0,0\}}+ \f{1}{2}a_{2}^{(2,2)}J_{4\{1,0,0\}} \non\\
   B_{2;1}^{(1,1)} &=&b_{2;1}^{(1,1)}J_{3\{0,0,0\}} -\Bigl(b_{2;1}^{(1,2)}-b_{2;1}^{(1,1)}+\f{1}{2}a_2^{(2,1)} \Bigl)J_{3\{-1,1,0\}}+  \bigl(b_{2;1}^{(1,1)}-\f{1}{2}a_2^{(2,1)}\bigl)J_{3\{-1,0,1\}} \non\\
   &&-\f{1}{2}a_2^{(2,2)}J_{4\{0,1,0\}}   \non\\[.2cm] 
   B_{2;1}^{(1,0)} &=&b_{2;1}^{(1,0)}J_{1\{-2,0,0\}}+\f{1}{2}b_{2;1}^{(1,1)}J_{3\{0,0,0\}} +\f{1}{2}a_2^{(2,1)} J_{3\{-2,1,1\}}+  \f{1}{2}b_{2;1}^{(1,1)}J_{3\{-2,0,2\}} +b_{2;1}^{(1,1)}J_{3\{-1,0,1\}} \non\\[.2cm]
   &&+\Bigl(\f{1}{2}a_{2}^{(2,1)}-\f{1}{2}b_{2;1}^{(1,1)}+\f{1}{2}b_{2;1}^{(1,2)}\Bigl)J_{3\{-2,2,0\}} +\f{1}{4}a_{2}^{(2,2)}J_{4\{-1,2,0\}}+\f{1}{2}a_{2}^{(2,0)}J_{2\{-1,0,0\}} \non\\[.2cm] 
    B_{2;1}^{(0,1)} &=&b_{2;1}^{(0,1)}J_{1\{-1,0,-1\}}-\f{1}{2}a_{2}^{(2,2)}J_{4\{0,2,-1\}} +\f{1}{2}a_2^{(1,1)} J_{2\{0,0,-1\}}+  \Bigl(b_{2;1}^{(1,2)}+b_{2;1}^{(1,1)}-\f{1}{2}a_{2}^{(1,2)} \Bigl)J_{3\{0,1,-1\}} \non\\[.2cm]
   &&+\Bigl(b_{2;1}^{(1,1)}-\f{1}{2}a_{2}^{(2,1)}\Bigl)J_{3\{-1,2,-1\}} +\Bigl(b_{2;1}^{(1,2)}+b_{2;1}^{(1,1)}-\f{1}{2}a_{2}^{(2,1)}\Bigl)J_{3\{-1,1,0\}}-b_{2;1}^{(0,2)}J_{2\{-1,1,-1\}} \non\\[.2cm] 
    B_{2;1}^{(0,2)} &=&b_{2;1}^{(0,2)}J_{2\{0,0,-1\}} -b_{2;1}^{(1,2)}J_{3\{0,0,0\}} +  \f{1}{2}a_{2}^{(2,2)}J_{4\{1,1,-1\}} +\Bigl(\f{1}{2}a_{2}^{(1,2)}-b_{2;1}^{(1,2)}\Bigl)J_{3\{1,0,-1\}} \non
   \ee

  \be
  B_{2;1}^{(0,0)} &=&b_{2;1}^{(0,0)}J_{0\{-2,0,-1\}}-\f{1}{2}a_{2}^{(1,0)}J_{1\{-2,0,0\}} +  \Bigl(\f{1}{4}a_{2}^{(1,2)}-\f{1}{2}b_{2;1}^{(1,1)}-\f{1}{2}b_{2;1}^{(1,2)} \Bigl)J_{3\{-1,2,-1\}} \non\\[.2cm]
   &&+\Bigl(\f{1}{2}a_2^{(2,0)} -\f{1}{2}a_2^{(1,1)}\Bigl)J_{2\{-1,1,-1\}}+\Bigl(\f{1}{2}a_{2}^{(2,1)}-\f{1}{2}b_{2;1}^{(1,1)}-\f{1}{2}b_{2;1}^{(1,2)} \Bigl)J_{3\{-2,2,0\}}+\f{1}{4}a_{2}^{(2,2)}J_{4\{-1,3,-1\}}\non\\
   &&+\Bigl(b_{2;1}^{(1,0)}-\f{1}{2}a_{2}^{(1,0)}-b_{2;1}^{(0,1)}\Bigl)J_{1\{-2,1,-1\}} +\f{1}{2}b_{2;1}^{(0,2)}J_{2\{-2,2,-1\}}+\Bigl(\f{1}{2}a_{2}^{(2,1)}-\f{2}{3}b_{2;1}^{(1,1)}\Bigl)J_{3\{-2,3,-1\}}\non\\
   &&-\f{1}{2}b_{2;1}^{(1,1)}J_{3\{0,0,0\}} -\f{1}{2}b_{2;1}^{(1,1)}J_{3\{-1,0,1\}} -\f{1}{6}b_{2;1}^{(1,1)}J_{3\{1,0,-1\}}-\f{1}{6}b_{2;1}^{(1,1)}J_{3\{-2,0,2\}} \non 
  \ee

   \be
   B_{1;1}^{(2,1)} &=& b_{1;1}^{(2,1)}J_{3\{0,0,0\}}- \f{1}{2}a_{2}^{(2,2)}J_{4\{0,0,1\}} \non\\
   B_{1;1}^{(1,1)} &=&b_{1;1}^{(1,1)}J_{3\{0,0,0\}} +\Bigl(b_{1;1}^{(2,1)}+b_{1;1}^{(1,1)}+\f{1}{2}a_2^{(1,2)} \Bigl)J_{3\{0,1,-1\}}+  \bigl(b_{1;1}^{(1,1)}+\f{1}{2}a_2^{(1,2)}\bigl)J_{3\{1,0,-1\}} \non\\
   &&-\f{1}{2}a_2^{(2,2)}J_{4\{0,1,0\}}   \non\\[.2cm] 
   B_{1;1}^{(0,1)} &=&b_{1;1}^{(0,1)}J_{1\{0,0,-2\}}-\f{1}{2}b_{1;1}^{(1,1)}J_{3\{0,0,0\}} +\f{1}{2}a_2^{(1,2)} J_{3\{1,1,-2\}}-  \f{1}{2}b_{1;1}^{(1,1)}J_{3\{2,0,-2\}} -b_{1;1}^{(1,1)}J_{3\{1,0,-1\}} \non\\[.2cm]
   &&+\Bigl(\f{1}{2}a_{2}^{(1,2)}+\f{1}{2}b_{1;1}^{(1,1)}+\f{1}{2}b_{1;1}^{(2,1)}\Bigl)J_{3\{0,2,-2\}} -\f{1}{4}a_{2}^{(2,2)}J_{4\{0,2,-1\}}-\f{1}{2}a_{2}^{(0,2)}J_{2\{0,0,-1\}} \non\\[.2cm] 
    B_{1;1}^{(1,0)} &=&b_{1;1}^{(1,0)}J_{1\{-1,0,-1\}}+\f{1}{2}a_{2}^{(2,2)}J_{4\{-1,2,0\}} -\f{1}{2}a_2^{(1,1)} J_{2\{-1,0,0\}}+  \Bigl(b_{1;1}^{(2,1)}-b_{1;1}^{(1,1)}-\f{1}{2}a_{2}^{(2,1)} \Bigl)J_{3\{-1,1,0\}} \non\\[.2cm]
   &&-\Bigl(b_{1;1}^{(1,1)}+\f{1}{2}a_{2}^{(1,2)}\Bigl)J_{3\{-1,2,-1\}} +\Bigl(b_{1;1}^{(2,1)}-b_{1;1}^{(1,1)}-\f{1}{2}a_{2}^{(1,2)}\Bigl)J_{3\{0,1,-1\}}+b_{1;1}^{(2,0)}J_{2\{-1,1,-1\}} \non\\[.2cm] 
     B_{1;1}^{(2,0)} &=&b_{1;1}^{(2,0)}J_{2\{-1,0,0\}} +b_{1;1}^{(2,1)}J_{3\{0,0,0\}} +  \f{1}{2}a_{2}^{(2,2)}J_{4\{-1,1,1\}} +\Bigl(-\f{1}{2}a_{2}^{(2,1)}+b_{1;1}^{(2,1)}\Bigl)J_{3\{-1,0,1\}} \non
   \ee

 \be
  B_{1;1}^{(0,0)} &=&b_{1;1}^{(0,0)}J_{0\{-1,0,-2\}}+\f{1}{2}a_{2}^{(0,1)}J_{1\{0,0,-2\}} -  \Bigl(\f{1}{4}a_{2}^{(2,1)}+\f{1}{2}b_{1;1}^{(1,1)}-\f{1}{2}b_{1;1}^{(2,1)} \Bigl)J_{3\{-1,2,-1\}} \non\\[.2cm]
   &&+\Bigl(\f{1}{2}a_2^{(0,2)} -\f{1}{2}a_2^{(1,1)}\Bigl)J_{2\{-1,1,-1\}}+\Bigl(-\f{1}{2}a_{2}^{(1,2)}-\f{1}{2}b_{1;1}^{(1,1)}+\f{1}{2}b_{1;1}^{(2,1)} \Bigl)J_{3\{0,2,-2\}}+\f{1}{4}a_{2}^{(2,2)}J_{4\{-1,3,-1\}}\non\\
   &&+\Bigl(-b_{1;1}^{(0,1)}+\f{1}{2}a_{2}^{(0,1)}+b_{1;1}^{(1,0)}\Bigl)J_{1\{-1,1,-2\}} +\f{1}{2}b_{1;1}^{(2,0)}J_{2\{-1,2,-2\}}-\Bigl(\f{1}{2}a_{2}^{(1,2)}+\f{2}{3}b_{1;1}^{(1,1)}\Bigl)J_{3\{-1,3,-2\}}\non\\
   &&-\f{1}{2}b_{1;1}^{(1,1)}J_{3\{0,0,0\}} -\f{1}{2}b_{1;1}^{(1,1)}J_{3\{1,0,-1\}} -\f{1}{6}b_{1;1}^{(1,1)}J_{3\{-1,0,1\}}-\f{1}{6}b_{1;1}^{(1,1)}J_{3\{2,0,-2\}} \non 
  \ee
The above form factors are in terms of 30 momentum independent coefficients. However, for $\Delta_1=\Delta_3\equiv \Delta$, only 5 of them are independent while the rest can be expressed in terms of the independent ones. The expression of the 25 coefficients in terms of the 5 independent ones are given by \cite{Marotta:2022jrp}
\be
a_{1}^{(0,1)}&=&\frac{1}{8} (d+2) d^2 (\Delta +1) a_2^{(2,2)}+\left(-\frac{3 d^2}{4}-\frac{1}{2} d (\Delta +2)+\Delta +1\right) a_1^{(1,1)}+\left(\frac{d}{2 \Delta }+\frac{1}{\Delta }\right) a_0^{(0,0)}\non\\
&&+\frac{1}{8} (d-2) \Delta  a_2^{(1,1)}-d \Delta  b_{1;1}^{(1,1)}\non\\[.3cm]
a_1^{(0,0)}&=&\frac{(\Delta -1) \left(d^2-2 d+4 \Delta \right) a_0^{(0,0)}}{\Delta ^2}+\frac{d^2 \left(d^2-4\right) \left(\Delta ^2-1\right) a_2^{(2,2)}}{4 \Delta }+\frac{1}{4} (d-2)^2 (\Delta -1) a_2^{(1,1)}\non\\
&&-\frac{(d-2) (\Delta -1) \left(3 d^2+2 d (\Delta +1)-4 \Delta \right) a_1^{(1,1)}}{2 \Delta }-2 d (d-2) (\Delta -1) b_{1;1}^{(1,1)}
\label{6.109}
\ee

\be
a_{2}^{(1,2)}&=&\frac{d^2 (\Delta +1) a_2^{(2,2)}}{4 \Delta }-\frac{(3 d+2 \Delta -2) a_1^{(1,1)}}{2 \Delta }+\frac{a_0^{(0,0)}}{\Delta ^2}+\frac{1}{4} a_2^{(1,1)}-2 b_{1;1}^{(1,1)}\non\\[.4cm]
a_{2}^{(0,2)}&=&\frac{d \left(d^2+2 d-4\right) (\Delta +1) a_2^{(2,2)}}{4 (\Delta -1)}-\frac{\left(3 d^2+2 d (\Delta +1)-8 (\Delta +1)\right) a_1^{(1,1)}}{2 (\Delta -1)}+\frac{d a_0^{(0,0)}}{(\Delta -1) \Delta }\non\\
&&+\frac{(d-4) \Delta  a_2^{(1,1)}}{4 (\Delta -1)}-\frac{2 d \Delta  b_{1;1}^{(1,1)}}{\Delta -1}\non\\[.4cm]
a_{2}^{(0,1)}&=&\frac{\left(3 d^2+2 d (2 \Delta -5)+8 \Delta \right) }{2 \Delta ^2} a_0^{(0,0)}+\frac{(d+2) d^2 (\Delta +1) (3 d+4 \Delta -8) a_2^{(2,2)}}{8 \Delta }\non\\
&&+\left(-\frac{9 d^3}{4 \Delta }+d^2 \left(\frac{5}{\Delta }-\frac{9}{2}\right)+d \left(-2 \Delta +\frac{3}{\Delta }+5\right)+4 (\Delta -1)\right) a_1^{(1,1)}\non\\
&&+\frac{1}{8} \Bigl(3 d^2+4 d \Delta -18 d-8 \Delta +16\Bigl) a_2^{(1,1)}+d (-3 d-4 \Delta +6) b_{1;1}^{(1,1)}\non\\[.6cm]
a_2^{(0,0)}&=&\frac{\left(d^2-4\right) d^2 \left(\Delta ^2-\Delta -2\right) (d+4 \Delta -5) a_2^{(2,2)}}{4 (\Delta -1) \Delta }-\frac{2 (d-2) d (\Delta -2) (d+4 \Delta -4) b_{1;1}^{(1,1)}}{\Delta -1}\non\\
&&-\frac{(d-2) (\Delta -2) \left(14 d^2 (\Delta -1)+3 d^3+8 d \left(\Delta ^2-\Delta -1\right)-16 (\Delta -1) \Delta \right) a_1^{(1,1)}}{2 (\Delta -1) \Delta }\non\\
&&+\frac{(\Delta -2) \left(4 d^2 (\Delta -2)+d^3+d (12-8 \Delta )+8 (\Delta -1) \Delta \right) a_0^{(0,0)}}{(\Delta -1) \Delta ^2}\non\\
&&+\frac{(d-2) (\Delta -2) \left(d^2+4 d (\Delta -2)-8 \Delta +8\right) a_2^{(1,1)}}{4 (\Delta -1)}
\ee

\be
a_{1}^{(1,0)}=-a_1^{(0,1)}\quad;\quad a_{2}^{(2,1)}&=&-a_{2}^{(1,2)}\quad;\quad a_{2}^{(2,0)}=a_2^{(0,2)}\quad;\quad a_{2}^{(1,0)}=-a_2^{(0,1)}
\ee

\be
b_{1;0}^{(1,0)}&=& \left(1-\f{d}{2}-\f{1}{\Delta}\right)a_0^{(0,0)} + \f{1}{4}\Bigl(2d(3+\Delta^2)+3d^2\Delta-4(\Delta-1)^2\Bigl)a_1^{(1,1)} -\f{1}{8}\Delta \left(2+(d-2)\Delta\right)a_2^{(1,1)}\non\\
&&-\frac{1}{8} d^2 (\Delta +1) ((d+2) \Delta +2) {a}_2^{(2,2)}+d \Delta ^2 {b}_{1;1}^{(1,1)}\non\\[.6cm]
b_{1;0}^{(0,0)}&=&\frac{1}{8} d^2 \left(\Delta ^2-1\right) \left(d^2-2 d \Delta -4 (\Delta +2)\right) a_2^{(2,2)}-\frac{1}{8} (\Delta -1) \Delta  \left(-d^2+2 d (\Delta +2)-4 \Delta \right) a_2^{(1,1)}\non\\
&&+\frac{1}{4} (\Delta -1) \left(d^2 (4 \Delta +6)-3 d^3+4 d (\Delta ^2+2 \Delta +2)-8 (\Delta -1) \Delta \right) a_1^{(1,1)}\non\\
&&+\frac{(\Delta -1) \left(d^2-2 d (\Delta +2)+4 \Delta \right) a_0^{(0,0)}}{2 \Delta }+d \Delta  \left(-d \Delta +d+2 \Delta ^2-2\right) b_{1;1}^{(1,1)}
\ee

\be
b_{1;1}^{(2,1)}&=&\frac{1}{4} (3 d+2 \Delta +6) a_1^{(1,1)}-\frac{1}{8} d (d+4) (\Delta +1) a_2^{(2,2)}-\frac{a_0^{(0,0)}}{2 \Delta }-\frac{1}{8} \Delta  a_2^{(1,1)}+\Delta  b_{1;1}^{(1,1)}\non\\[.4cm]
b_{1;1}^{(2,0)}&=&\left(-\frac{3 d^2}{4}+d (3-2 \Delta )-\Delta ^2+1\right) a_1^{(1,1)}+\frac{1}{8} d (\Delta +1) \left(d^2+2 d (\Delta -1)+8 (\Delta -1)\right) a_2^{(2,2)}\non\\
&&+\left(\frac{d}{2 \Delta }-\frac{3}{\Delta }+1\right) a_0^{(0,0)}+\frac{1}{8} \Delta  (d+2 \Delta -6) a_2^{(1,1)}-\Delta  (d+2 \Delta -2) b_{1;1}^{(1,1)}\non\\[.4cm]
b_{1;1}^{(1,0)}&=&-\frac{(d+2) d^2 \left(\Delta ^2-1\right) (d+2 \Delta -4) a_2^{(2,2)}}{8 \Delta }+\left(\frac{d^2}{2 \Delta ^2}-\frac{d^2}{2 \Delta }-\frac{3 d}{\Delta ^2}+\frac{4 d}{\Delta }-d-\frac{2}{\Delta }+2\right) a_0^{(0,0)}\non\\
&&+\frac{(\Delta -1) \left(4 d^2 (2 \Delta -3)+3 d^3+4 d \left(\Delta ^2-3 \Delta -1\right)-8 (\Delta -1) \Delta \right) }{4 \Delta }a_1^{(1,1)}\non\\
&&-\frac{1}{8} (\Delta -1) \left(d^2+2 d (\Delta -3)-4 \Delta \right) a_2^{(1,1)}+d (\Delta -1) (d+2 \Delta -2) b_{1;1}^{(1,1)}\non\\[.4cm]
b_{1;1}^{(0,1)}&=&\frac{1}{16} (d-2) d^3 (\Delta +1) a_2^{(2,2)}+\left(\frac{d^2}{4 \Delta }-\frac{3 d}{2 \Delta }+\frac{2}{\Delta }\right) a_0^{(0,0)}+\frac{1}{16} \left(d^2-6 d+8\right) \Delta  a_2^{(1,1)}\non\\[.2cm]
&&-\frac{1}{8} (d-2) \left(3 d^2+2 d (\Delta -3)-8 \Delta +8\right) a_1^{(1,1)}+\frac{1}{2} \Delta  \left(-d^2+2 d+4 \Delta -4\right) b_{1;1}^{(1,1)}\non\\[.5cm]
b_{1;1}^{(0,0)}&=&\frac{(d+2) d^2 \left(\Delta ^2-\Delta -2\right) \left(d^2+4 d (\Delta -3)-16 \Delta +24\right) a_2^{(2,2)}}{16 \Delta }\non\\
&&+\frac{d (\Delta -2) \left(d^2+2 d (2 \Delta -7)-24 \Delta +32\right) a_0^{(0,0)}}{4 \Delta ^2}\non\\
&&-\f{1}{8\Delta}\biggl[2 d^3 \left(7 \Delta ^2-32 \Delta +36\right)+4 d^2 \left(2 \Delta ^3-23 \Delta ^2+53 \Delta -30\right)+3 d^4 (\Delta -2)\non\\
&&-16 d (\Delta -2)^2 (3 \Delta +1)+64 \Delta  \left(\Delta ^2-3 \Delta +2\right) \biggl]a_1^{(1,1)}\non\\
&&+\frac{1}{16} (d-4) (\Delta -2) \left(d^2+2 d (2 \Delta -5)-8 \Delta +8\right) a_2^{(1,1)}\non\\
&&+\frac{1}{6} (\Delta -2) \left(-6 d^2 (2 \Delta -5)-3 d^3+48 d (\Delta -1)+8 (\Delta -1) \Delta \right) b_{1;1}^{(1,1)}
%\label{6.12}
\ee
The corresponding coefficients for $B_2$ are given in terms of the coefficients of $B_1$ as
\be\label{secwardend}
b_{2;0}^{(0,1)}&=&-b_{1;0}^{(1,0)}\qquad;\quad b_{2;0}^{(0,0)}=b_{1;0}^{(0,0)}\non\\[.4cm]
b_{2;1}^{(1,2)}&=&-b_{1;1}^{(2,1)}\qquad;\qquad b_{2;1}^{(0,2)}=b_{1;1}^{(2,0)}\qquad;\qquad b_{2;1}^{(1,1)}=b_{1;1}^{(1,1)}\non\\[.4cm]
b_{2;1}^{(1,0)}&=&-b_{1;1}^{(0,1)}\qquad;\qquad b_{2;1}^{(0,1)}=-b_{1;1}^{(1,0)}\qquad;\qquad b_{2;1}^{(0,0)}=b_{1;1}^{(0,0)}
\ee

\section{Massive spin-2 Btb propagator in Poincar\'e Coordinates}
\label{appenF}
In section \ref{sec4}, we obtained the bulk-to-boundary propagator in the Fefferman Graham coordinates. Here, we shall give the expressions in the Poincar\'e coordinates which are useful in the evaluation of the 3-point function. Before giving the results in the Poincar\'e coordinates, we first note some useful relations involving the massive spin-2 Btb propagator. The fields $\phi_{MN}$ and $\phi^*_{MN}$ are treated as independent fields and their Fourier transforms along the boundary directions are defined in the same way, i.e., 
\begin{equation}
 \phi_{MN}(z,\textbf{k}) = \int d^d x\, e^{-i k \cdot x} \phi_{MN}(z,\textbf{x}) \quad;\qquad  \phi^*_{MN}(z,\textbf{k}) = \int d^d x\, e^{-i k \cdot x} \phi^*_{MN}(z,\textbf{x}) \,.
\end{equation}
For the Btb propagators we have
\begin{equation}
   K_{MN}^{\hspace{0.6cm}\mu\nu}(z,\textbf{k}) = \int d^d x\, e^{-i k \cdot (x-y)} K_{MN}^{\hspace{0.6cm}\mu\nu}(z,\textbf{x};\textbf{y})\,,
\end{equation}
with the Btb propagator defined by
\begin{equation}
  \phi_{MN}(z,\textbf{k}) = K_{MN}^{\hspace{0.6cm}\mu\nu}(z,\textbf{k})  \hat\phi^{(0)}_{\mu\nu}(\textbf{k})\,.
\end{equation}
The field $\phi^*_{MN}$ satisfies the same equation of motion as $\phi_{MN}$ and hence has the same bulk-to-boundary propagator, i.e., 
\begin{equation}
  \phi^*_{MN}(z,\textbf{k}) =\bar{K}_{MN}^{\hspace{0.6cm}\kappa\gamma}(z,\textbf{k})  \hat\phi^*{}^{(0)}_{\kappa\gamma}(\textbf{k})\quad;\qquad \bar{K}_{MN}^{\hspace{0.6cm}\kappa\gamma}(z,\textbf{k}) = K_{MN}^{\hspace{0.6cm}\kappa\gamma}(z,\textbf{k})\,.
\end{equation}
The Btb propagator also satisfies the following relations 
\be
  K_{\mu\nu}^{\hspace{0.35cm}\kappa\gamma}(z,-\textbf{k}) &=& K_{\mu\nu}^{\hspace{0.35cm}\kappa\gamma}(z,\textbf{k}) = K^*_{\mu\nu}{}^{\kappa\gamma}(z,\textbf{k})\,,  \non\\
   K_{zz}^{\hspace{0.35cm}\kappa\gamma}(z,-\textbf{k}) &=& K{}_{zz}^{\hspace{0.35cm}\kappa\gamma}(z,k)=K^*_{zz}{}^{\kappa\gamma}(z,\textbf{k}) \,,\non\\
  K_{z\mu}^{\hspace{0.35cm}\kappa\gamma}(z,-\textbf{k}) &=& -K{}_{z\mu}^{\hspace{0.35cm}\kappa\gamma}(z,\textbf{k}) = K^*_{z\mu}{}^{\kappa\gamma}(z,\textbf{k})\,.
\ee
The Btb propagator in the Poincar\'e $z$ coordinates is related to the Btb propagator in the Fefferman Graham $\rho$ coordinates as follows 
\begin{eqnarray}
   &K_{zz}^{\;\;\;\;\uptau \sigma}(z;\textbf{k})   &=\left(\frac{\partial\rho}{\partial z}\right)^2  K_{\rho\rho}^{\;\;\;\;\uptau \sigma}(\rho(z);\textbf{k}),
  \non \\
   &K_{z\mu}^{\;\;\;\;\uptau \sigma}(z;\textbf{k})&=\left(\frac{\partial\rho}{\partial z}\right)K_{\rho\mu}^{\;\;\;\;\uptau \sigma}(\rho(z);\textbf{k}),
  \non\\
   &K_{\mu\nu}^{\;\;\;\;\uptau \sigma}(z;\textbf{k})&=K_{\mu\nu}^{\;\;\;\;\uptau \sigma}(\rho(z);\textbf{k}).
\end{eqnarray}
By explicitly solving the equations in the Poincar\'e coordinates or using the above relations, we find that the Btb propagators for a massive spin-2 field in the Poincar\'e coordinates are given by
\be
K_{\mu\nu}^{\;\;\;\;\uptau \sigma}(z;\textbf{k})&=& \f{ L^{-\gamma}}{\alpha_1}\frac{z^{\f{d-4}{2}}}{L^{\f{d-4}{4}}} \biggl[  \delta_\mu^{\uptau }\delta_\nu^\sigma K_\beta(kz) - \f{4z \delta_{\mu\nu} k^\uptau  k^\sigma }{k (d+2\beta)(d+2\beta-2)}K_{\beta-1}(kz)+ \f{4z k^\uptau \delta_{(\mu}^{\;\;\sigma}  k_{\nu)} }{k (d+2\beta)}K_{\beta-1}(kz) \non\\
&&+ \f{4 k_\mu k_\nu k^\uptau  k^\sigma }{k^4 (d+2\beta)(d+2\beta-2)}\Bigl\{ (k^2\rho L+4\beta(\beta-1))K_{\beta}(kz)-2k(\beta-1)zK_{\beta+1}(kz)\Bigl\} \biggl]\,, \non\\[.3cm]
K_{z\mu}^{\;\;\;\;\uptau \sigma}(z;\textbf{k})&=&  \f{2\, L^{-\gamma}\,i  k^\uptau  }{\alpha_1(d+2\beta)}\left( \delta_\mu^{\;\sigma} K_\beta(kz) + \f{2z k^\sigma k_\mu}{k(2\beta+d-2)}K_{\beta-1}(kz) \right) \frac{z^{\f{d-2}{2}}}{L^{\f{d-4}{4}}} \,,  \non\\[.3cm]
K_{zz}^{\;\;\;\;\uptau \sigma}(z;\textbf{k})&=& - \f{4\, L^{-\gamma}\, k^\mu k^\nu }{\alpha_1(d+2\beta)(d+2\beta-2)} \frac{z^{\f{d}{2}}}{L^{\f{d-4}{4}}}  K_\beta(kz) \,,\label{f346}
\ee
where 
\begin{align}
\alpha_1= 2^{\beta -1} \Gamma (\beta ) k^{-\beta } L^{-\frac{\beta }{2}} \quad;\qquad\gamma=\frac{d}{4}-\frac{\beta}{2}-1\quad;\qquad\beta=\Delta-\frac{d}{2}.
\end{align}

\section{Some details of 3-point function computation}
\label{G}
The one point function of the gauge field computed using the holographic renormalization procedure is given by \cite{Marotta:2024sce}
\be
\langle \mathcal{J}^\mu(\textbf{x})\rangle= 
\,\delta^{\mu\uptau }\int d^{d} y\,dw \sqrt{G}\,{\mathbb K}_{\uptau }^{\;\;\nu}(w;\textbf{x},\textbf{y})J_\nu(\textbf{y},w)\,.
\ee
Fourier transforming the above expression, the 3-point function involving the conserved current and two non-conserved operators is given by taking the derivative with respect to the sources of the boundary operators as
\be
\langle O^*{}^{\alpha\beta}(\textbf{p}_1)\mathcal{J}^\kappa(\textbf{p}_2) O^{\gamma\delta}(\textbf{p}_3)\rangle &=& (2\pi)^d (2\pi)^d \frac{\delta^2\langle \mathcal{J}^\kappa(\textbf{p}_2)\rangle}{\delta\hat\phi^*{}^{(0)}_{\alpha\beta}(-\textbf{p}_1)\delta\hat\phi{}^{(0)}_{\gamma\delta}(-\textbf{p}_3) }\non\\
&=& \frac{\delta^{\kappa\mu}(2\pi)^d (2\pi)^d\delta^2 }{\delta\hat\phi^*{}^{(0)}_{\alpha\beta}(-\textbf{p}_1)\delta\hat\phi{}^{(0)}_{\gamma\delta}(-\textbf{p}_3) } \int_0^\infty d\sigma \sqrt{G}\, \mathbb{K}_{\mu\nu}(\sigma;\textbf{p}_2)J^\nu_{(0)}(\sigma,\textbf{p}_2)\,.\non
\ee
The $J^\nu_{(0)}$ is the boundary component of the gauge field current $J^N$ evaluated on the free spin-2 field and can be expressed in the form
 \be
J^{\nu}_{(0)}&=&ig \Biggl[\Bigg\{\frac{1}{2}\phi_{MP}\d^\nu \phi^*{}^{MP}-G^{00}\phi_{M0}\d_0\phi^*{}^{M\nu}-G^{\eta\sigma}\phi_{M\eta}\d_\sigma\phi^*{}^{M\nu}+\frac{2z}{L^2}\left(\phi^*{}^{0\nu}\phi_{00}+\phi^*{}^{\mu\nu}\phi_{\mu 0}\right)\non\\
&&-\frac{\alpha}{\sqrt{G}}\d_0\left(\sqrt{G}\, G^{00}\phi_{0P}\phi^*{}^{P\nu}\right)-\alpha G^{\mu\sigma}\d_\mu\left( \phi_{\sigma P}\phi^*{}^{P\nu}\right)\Bigg\}-\phi_{MN}\leftrightarrow\phi^*_{MN}\Biggl]\;\;+\;\;O(\p^3)\,,\non\\
\label{g349}
\ee
where we have used the auxiliary condition $\phi=0$, which is valid at the free field level. 

\vspace*{.07in}We shall now give the results of taking the functional derivatives for each of the terms in $J^{\nu}_{(0)}$ above. We consider each of the 6 terms inside the bracket in \eqref{g349} (together with the corresponding terms obtained by the interchange $\phi_{MN}\leftrightarrow\phi^*_{MN}$). 

\subsection*{First term: $J_1^\nu(x^0,x^\mu)=\frac{ig}{2}\left(\phi_{MK}\d^\nu \phi^*{}^{MK}-\phi^*{}_{MK}\d^\nu \phi^{MK}\right)$} 
The Fourier transform of the first term gives
\be
   J^\nu_{1}(x^0,\textbf{p}_2) &=& \frac{ig}{2}\int d^d x e^{-i p_2 \cdot x}\left(\phi_{MK}\d^\nu \phi^*{}^{MK}-\phi^*{}_{MK}\d^\nu \phi^{MK}\right)\non\\
   &=&\frac{ig}{2}G^{MN}G^{KS}G^{\nu\lambda}\int\frac{d^d k_1}{(2\pi)^d}\frac{d^d k_2}{(2\pi)^d}(2\pi)^d \delta^d(\textbf{k}_2+\textbf{k}_1-\textbf{p}_2)\Bigg[K_{MK}^{\hspace{0.4cm}\gamma\delta}(\textbf{k}_2)(ik_{1\lambda}) \bar K_{NS}^{\hspace{0.35cm}\alpha\beta}(\textbf{k}_1)\non\\
   &&\hspace{0.3cm}-\bar K_{MK}^{\hspace{0.45cm}\alpha\beta}(\textbf{k}_1)ik_{2\lambda} K_{NS}^{\hspace{0.35cm}\gamma\delta}(\textbf{k}_2)\Bigg]\hat\phi^{(0)}{}^*_{\alpha\beta}(\textbf{k}_1)\hat\phi^{(0)}_{\gamma\delta}(\textbf{k}_2)\,.
\ee
For each term appearing in the sum over indices, the result of the integration is given by 
\be
&& \hspace*{-.4in}\frac{(2\pi)^d (2\pi)^d\delta^2 }{\delta\hat\phi^*{}^{(0)}_{\alpha\beta}(-\textbf{p}_1)\delta\hat\phi{}^{(0)}_{\gamma\delta}(-\textbf{p}_3) }\left(\frac{ig}{2}\phi_{00}\d^\nu \phi^*{}^{00}-(\phi^*\leftrightarrow \phi)\right)\non\\
&=&-\,\frac{ig}{2}G^{00}G^{00}G^{\nu\lambda}(2\pi)^d \delta^d(\textbf{p}_1+\textbf{p}_2+\textbf{p}_3)
\Bigg[K_{00}^{\hspace{0.4cm}\gamma\delta}(\textbf{p}_3)ip_{1\lambda}  K_{00}^{\hspace{0.35cm}\alpha\beta}(\textbf{p}_1)- K_{00}^{\hspace{0.45cm}\alpha\beta}(\textbf{p}_1)ip_{3\lambda} K_{00}^{\hspace{0.35cm}\gamma\delta}(\textbf{p}_3)\Bigg]\,,\non
\ee
\be
&&\hspace*{-.4in} \frac{(2\pi)^d (2\pi)^d\delta^2 }{\delta\hat\phi^*{}^{(0)}_{\alpha\beta}(-\textbf{p}_1)\delta\hat\phi{}^{(0)}_{\gamma\delta}(-\textbf{p}_3) }\left(ig\phi_{\mu 0}\d^\nu \phi^*{}^{\mu 0}-(\phi^*\leftrightarrow \phi)\right)\non\\
&=& -ig\,G^{\mu\eta}G^{00}G^{\nu\lambda}(2\pi)^d \delta^d(\textbf{p}_1+\textbf{p}_2+\textbf{p}_3)\Bigg[K_{\mu 0}^{\hspace{0.4cm}\gamma\delta}(\textbf{p}_3)ip_{1\lambda}  K_{\eta 0}^{\hspace{0.35cm}\alpha\beta}(\textbf{p}_1)
   - K_{\mu 0}^{\hspace{0.45cm}\alpha\beta}(\textbf{p}_1)ip_{3\lambda} K_{\eta 0}^{\hspace{0.35cm}\gamma\delta}(\textbf{p}_3)\Bigg]\,,\non
\ee
\be
&&\hspace*{-.4in} \frac{(2\pi)^d (2\pi)^d\delta^2}{\delta\hat\phi^*{}^{(0)}_{\alpha\beta}(-\textbf{p}_1)\delta\hat\phi{}^{(0)}_{\gamma\delta}(-\textbf{p}_3) }\left(\frac{ig}{2}\phi_{\mu \kappa}\d^\nu \phi^*{}^{\mu \kappa}-(\phi^*\leftrightarrow \phi)\right)\non\\
&=& -\,\frac{ig}{2}G^{\mu\eta}G^{\kappa \sigma}G^{\nu\lambda}(2\pi)^d \delta^d(\textbf{p}_1+\textbf{p}_2+\textbf{p}_3)\Bigg[K_{\mu\kappa}^{\hspace{0.4cm}\gamma\delta}(\textbf{p}_3)ip_{1\lambda} K_{\eta \sigma}^{\hspace{0.35cm}\alpha\beta}(\textbf{p}_1)- K_{\mu\kappa}^{\hspace{0.45cm}\alpha\beta}(\textbf{p}_1)ip_{3\lambda} K_{\eta \sigma}^{\hspace{0.35cm}\gamma\delta}(\textbf{p}_3)\Bigg]\,,\non
\ee
\subsection*{Second term: $J_2^\nu(x^0,x^\mu)=-\,{ig}\left(G^{00}\phi_{M0}\d_0\phi^*{}^{M\nu}-G^{00}\phi^*_{M0}\d_0\phi^{M\nu}\right)$} 
The Fourier transform of the second term gives
\be
   J^\nu_{2}(x^0,\textbf{p}_2) 
   &=&-ig G^{00}\int d^d x e^{-i p_2 . x}\left(\phi_{M0}\d_0 (G^{MK}G^{\lambda\nu}\phi^*_{K\lambda})-\phi^*_{M0}\d_0 (G^{MK}G^{\lambda\nu}\phi_{K\lambda})\right)\non\\
   &=&-ig G^{00}\int\frac{d^d k_1}{(2\pi)^d}\frac{d^d k_2}{(2\pi)^d}(2\pi)^d \delta^d(\textbf{k}_2+\textbf{k}_1-\textbf{p}_2)\Bigg[K_{M0}^{\hspace{0.4cm}\gamma\delta}(\textbf{k}_2)\d_0 \left(G^{MK}G^{\lambda\nu}\bar K_{K\lambda}^{\hspace{0.35cm}\alpha\beta}(\textbf{k}_1)\right)\non\\
   &&\hspace{0.3cm}-\bar K_{M0}^{\hspace{0.45cm}\alpha\beta}(\textbf{k}_1)\d_0\left(G^{MK}G^{\lambda\nu} K_{K\lambda}^{\hspace{0.35cm}\gamma\delta}(\textbf{k}_2)\right)\Bigg]\hat\phi^{(0)}{}^*_{\alpha\beta}(\textbf{k}_1)\hat\phi^{(0)}_{\gamma\delta}(\textbf{k}_2)\,,
\ee
For different terms, the result of the integration is given by
\be
&& \hspace*{-.4in}\frac{(2\pi)^d (2\pi)^d\delta^2}{\delta\hat\phi^*{}^{(0)}_{\alpha\beta}(-\textbf{p}_1)\delta\hat\phi{}^{(0)}_{\gamma\delta}(-\textbf{p}_3) }\left(-ig G^{00}\phi_{00}\d_0 \phi^*{}^{0 \nu}-(\phi^*\leftrightarrow \phi)\right)\non\\
&=&ig G^{00}(2\pi)^d \delta^d(\textbf{p}_1+\textbf{p}_2+\textbf{p}_3)\non\\
&&\Bigg[K_{0 0}^{\hspace{0.4cm}\gamma\delta}(\textbf{p}_3)\d_0 \left(G^{00}G^{\lambda\nu} K_{0\lambda}^{\hspace{0.35cm}\alpha\beta}(\textbf{p}_1)\right)- K_{0 0}^{\hspace{0.45cm}\alpha\beta}(\textbf{p}_1)\d_0\left(G^{00}G^{\lambda\nu} K_{0\lambda}^{\hspace{0.35cm}\gamma\delta}(\textbf{p}_3)\right)\Bigg]\,,\non
\ee
\be
&& \hspace*{-.4in}\frac{(2\pi)^d (2\pi)^d\delta^2 }{\delta\hat\phi^*{}^{(0)}_{\alpha\beta}(-\textbf{p}_1)\delta\hat\phi{}^{(0)}_{\gamma\delta}(-\textbf{p}_3) }\left(-ig G^{00}\phi_{\mu 0}\d_0 \phi^*{}^{\mu \nu}-(\phi^*\leftrightarrow \phi)\right)\non\\
&=&ig G^{00}(2\pi)^d \delta^d(\textbf{p}_1+\textbf{p}_2+\textbf{p}_3)\non\\
&&\Bigg[K_{\mu 0}^{\hspace{0.4cm}\gamma\delta}(\textbf{p}_3)\d_0 \left(G^{\mu\kappa}G^{\lambda\nu} K_{\kappa\lambda}^{\hspace{0.35cm}\alpha\beta}(\textbf{p}_1)\right)- K_{\mu 0}^{\hspace{0.45cm}\alpha\beta}(\textbf{p}_1)\d_0\left(G^{\mu\kappa}G^{\lambda\nu} K_{\kappa\lambda}^{\hspace{0.35cm}\gamma\delta}(\textbf{p}_3)\right)\Bigg]\,,\non
\ee
\subsection*{Third term: $J_3^\nu(x^0,x^\mu)=-ig\left(G^{\eta\sigma}\phi_{M\eta}\d_\sigma\phi^*{}^{M\nu}-G^{\eta\sigma}\phi^*_{M\eta}\d_\sigma\phi^{M\nu}\right)$} 
The Fourier transform of the 3rd term gives
\be
   J^\nu_{3}(x^0,\textbf{p}_2) 
   &=&-ig G^{\eta\sigma} G^{\lambda\nu}G^{MK}\int d^d x e^{-i p_2 . x}\left(\phi_{M\eta}\d_\sigma\phi^*_{K\lambda}-\phi^*_{M\eta}\d_\sigma\phi_{K\lambda}\right)\non\\
   &=&-ig G^{\eta\sigma} G^{\lambda\nu}G^{MK}\int\frac{d^d k_1}{(2\pi)^d}\frac{d^d k_2}{(2\pi)^d}(2\pi)^d \delta^d(\textbf{k}_2+\textbf{k}_1-\textbf{p}_2)\Bigg[K_{M\eta}^{\hspace{0.4cm}\gamma\delta}(\textbf{k}_2)(ik_{1\sigma}) \bar K_{K\lambda}^{\hspace{0.35cm}\alpha\beta}(\textbf{k}_1)\non\\
   &&\hspace{0.3cm}-\bar K_{M\eta}^{\hspace{0.45cm}\alpha\beta}(\textbf{k}_1)ik_{2\sigma} K_{K\lambda}^{\hspace{0.35cm}\gamma\delta}(\textbf{k}_2)\Bigg]\hat\phi^{(0)}{}^*_{\alpha\beta}(\textbf{k}_1)\hat\phi^{(0)}_{\gamma\delta}(\textbf{k}_2)\,,
\ee
The different terms of the above expression are given by
\be
&& \hspace*{-.4in}\frac{(2\pi)^d (2\pi)^d\delta^2}{\delta\hat\phi^*{}^{(0)}_{\alpha\beta}(-\textbf{p}_1)\delta\hat\phi{}^{(0)}_{\gamma\delta}(-\textbf{p}_3) }\left(-ig G^{\eta\sigma} \phi_{0\eta}\d_{\sigma}\phi^*{}^{0\nu}-(\phi^* \leftrightarrow\phi)\right)\non\\
&=&ig G^{\eta\sigma} G^{\lambda\nu}G^{00}(2\pi)^d \delta^d(\textbf{p}_1+\textbf{p}_2+\textbf{p}_3)\Bigg[K_{0\eta}^{\hspace{0.4cm}\gamma\delta}(\textbf{p}_3)(ip_{1\sigma})  K_{0\lambda}^{\hspace{0.35cm}\alpha\beta}(\textbf{p}_1)- K_{0\eta}^{\hspace{0.45cm}\alpha\beta}(\textbf{p}_1)(ip_{3\sigma} )K_{0\lambda}^{\hspace{0.35cm}\gamma\delta}(\textbf{p}_3)\Bigg]\,,\non
\ee

\be
&& \hspace*{-.4in}\frac{(2\pi)^d (2\pi)^d\delta^2}{\delta\hat\phi^*{}^{(0)}_{\alpha\beta}(-\textbf{p}_1)\delta\hat\phi{}^{(0)}_{\gamma\delta}(-\textbf{p}_3) }\left(-ig G^{\eta\sigma} \phi_{\mu\eta}\d_{\sigma}\phi^*{}^{\mu\nu}-(\phi^* \leftrightarrow\phi)\right)\non\\
&=&ig G^{\eta\sigma} G^{\lambda\nu}G^{\mu \kappa}(2\pi)^d \delta^d(\textbf{p}_1+\textbf{p}_2+\textbf{p}_3)\Bigg[K_{\mu\eta}^{\hspace{0.4cm}\gamma\delta}(\textbf{p}_3)(ip_{1\sigma}) K_{\kappa\lambda}^{\hspace{0.35cm}\alpha\beta}(\textbf{p}_1)-K_{\mu\eta}^{\hspace{0.45cm}\alpha\beta}(\textbf{p}_1)(ip_{3\sigma} )K_{\kappa\lambda}^{\hspace{0.35cm}\gamma\delta}(\textbf{p}_3)\Bigg]\,,\non
\ee
\subsection*{Fourth term: $J_4^\nu(x^0,x^\mu)=ig\left(\phi^*{}^{MN}\phi_{MQ}-\phi^{MN}\phi^*_{MQ}\right)$} 
The Fourier transform of the 4th term gives
\be
   J^\nu_{4}(x^0,\textbf{p}_2) &=& ig\, G^{NS}G^{MK}\int d^d x e^{-i p_2 . x}\left(\phi^*_{KS}\phi_{MQ}-\phi_{KS}\phi^*_{MQ}\right)\non\\
   &=&ig\,G^{NS}G^{MK}\int\frac{d^d k_1}{(2\pi)^d}\frac{d^d k_2}{(2\pi)^d}(2\pi)^d \delta^d(\textbf{k}_2+\textbf{k}_1-\textbf{p}_2)\Bigg[\bar K_{KS}^{\hspace{0.4cm}\alpha\beta}(\textbf{k}_1) K_{MQ}^{\hspace{0.35cm}\gamma\delta}(\textbf{k}_2)\non\\
   &&\hspace{0.3cm}- K_{KS}^{\hspace{0.45cm}\gamma\delta}(\textbf{k}_2) \bar K_{MQ}^{\hspace{0.35cm}\alpha\beta}(\textbf{k}_1)\Bigg]\hat\phi^{(0)}{}^*_{\alpha\beta}(\textbf{k}_1)\hat\phi^{(0)}_{\gamma\delta}(\textbf{k}_2)\,,
\ee
The different terms of the above expression are given by
\begin{align}
& \frac{(2\pi)^d (2\pi)^d\delta^2 }{\delta\hat\phi^*{}^{(0)}_{\alpha\beta}(-\textbf{p}_1)\delta\hat\phi{}^{(0)}_{\gamma\delta}(-\textbf{p}_3) }\left(ig\, \phi^*{}^{0\nu}\phi_{00}-(\phi^*\leftrightarrow \phi)\right)\non\\
&=-ig\,G^{00}G^{\nu\lambda}(2\pi)^d \delta^d(\textbf{p}_1+\textbf{p}_2+\textbf{p}_3)\Bigg[ K_{0\lambda}^{\hspace{0.4cm}\alpha\beta}(\textbf{p}_1) K_{00}^{\hspace{0.35cm}\gamma\delta}(\textbf{p}_3)- K_{0\lambda}^{\hspace{0.45cm}\gamma\delta}(\textbf{p}_3)  K_{00}^{\hspace{0.35cm}\alpha\beta}(\textbf{p}_1)\Bigg]\,,\non
\end{align}

\begin{align}
& \frac{(2\pi)^d (2\pi)^d\delta^2 }{\delta\hat\phi^*{}^{(0)}_{\alpha\beta}(-\textbf{p}_1)\delta\hat\phi{}^{(0)}_{\gamma\delta}(-\textbf{p}_3) }\left(ig\, \phi^*{}^{\mu\nu}\phi_{\mu 0}-(\phi^*\leftrightarrow \phi)\right)\non\\
&=-ig\,G^{\mu\kappa}G^{\nu\lambda}(2\pi)^d \delta^d(\textbf{p}_1+\textbf{p}_2+\textbf{p}_3)\Bigg[ K_{\kappa\lambda}^{\hspace{0.4cm}\alpha\beta}(\textbf{p}_1) K_{\mu 0}^{\hspace{0.35cm}\gamma\delta}(\textbf{p}_3)- K_{\kappa\lambda}^{\hspace{0.45cm}\gamma\delta}(\textbf{p}_3)  K_{\mu 0}^{\hspace{0.35cm}\alpha\beta}(\textbf{p}_1)\Bigg]\,,\non
\end{align}

\subsection*{Fifth term: $J_5^\nu(x^0,x^\mu)=-\frac{ig\,\alpha }{\sqrt{G}}\d_0\Bigg(\sqrt{G}G^{00}\left( \phi_{0P}\phi^*{}^{P\nu}-\phi^*_{0P}\phi^{P\nu}\right)\Bigg)$}
The Fourier transform of the 5th term gives
\be
   J^\nu_{5}(x^0,\textbf{p}_2) 
   &=&- \frac{ig\,\alpha}{\sqrt{G}}\int d^d x e^{-i p_2 . x}\d_0\Bigg(\sqrt{G}G^{00}G^{\lambda\nu}G^{MP}\left( \phi_{0P}\phi^*_{M\lambda}-\phi^*_{0P}\phi_{M\lambda}\right)\Bigg)\non\\
   &=&-\frac{ig\,\alpha}{\sqrt{G}}\int\frac{d^d k_1}{(2\pi)^d}\frac{d^d k_2}{(2\pi)^d}(2\pi)^d \delta^d(\textbf{k}_2+\textbf{k}_1-\textbf{p}_2)\d_0\Bigg[\sqrt{G}G^{00}G^{\lambda\nu}G^{MP}\Bigg(K_{0P}^{\hspace{0.45cm}\gamma\delta}(\textbf{k}_2) \bar K_{M\lambda}^{\hspace{0.35cm}\alpha\beta}(\textbf{k}_1)\non\\
   &&\hspace{0.3cm}- \bar K_{0P}^{\hspace{0.4cm}\alpha\beta}(\textbf{k}_1) K_{M\lambda}^{\hspace{0.35cm}\gamma\delta}(\textbf{k}_2)\Bigg)\Bigg]\hat\phi^{(0)}{}^*_{\alpha\beta}(\textbf{k}_1)\hat\phi^{(0)}_{\gamma\delta}(\textbf{k}_2)\,,
\ee
The different terms are given by
\be
&&\hspace*{-.4in} \frac{(2\pi)^d (2\pi)^d\delta^2 }{\delta\hat\phi^*{}^{(0)}_{\alpha\beta}(-\textbf{p}_1)\delta\hat\phi{}^{(0)}_{\gamma\delta}(-\textbf{p}_3) }\left[-\frac{ig\,\alpha}{\sqrt{G}}\d_0\Bigg(\sqrt{G}G^{00}\left( \phi_{00}\phi^*{}^{0\nu}-\phi^*_{00}\phi^{0\nu}\right)\Bigg)\right]\non\\
&=&\frac{ig\,\alpha}{\sqrt{G}}(2\pi)^d \delta^d(\textbf{p}_1+\textbf{p}_2+\textbf{p}_3)\d_0\Bigg(\sqrt{G}G^{00}G^{00}G^{\lambda\nu}\Bigg[ K_{0\lambda}^{\hspace{0.4cm}\alpha\beta}(\textbf{p}_1) K_{00}^{\hspace{0.35cm}\gamma\delta}(\textbf{p}_3)- K_{0\lambda}^{\hspace{0.45cm}\gamma\delta}(\textbf{p}_3)  K_{00}^{\hspace{0.35cm}\alpha\beta}(\textbf{p}_1)\Bigg]\Bigg)\,,\non
\ee

\be
&&\hspace*{-.4in} \frac{(2\pi)^d (2\pi)^d\delta^2}{\delta\hat\phi^*{}^{(0)}_{\alpha\beta}(-\textbf{p}_1)\delta\hat\phi{}^{(0)}_{\gamma\delta}(-\textbf{p}_3) }\left[-\frac{ig\,\alpha}{\sqrt{G}}\d_0\Bigg(\sqrt{G}G^{00}\left( \phi_{\mu0}\phi^*{}^{\mu\nu}-\phi^*_{\mu 0}\phi^{\mu\nu}\right)\Bigg)\right]\non\\
&=&\frac{ig\,\alpha}{\sqrt{G}}(2\pi)^d \delta^d(\textbf{p}_1+\textbf{p}_2+\textbf{p}_3)\d_0\Bigg(\sqrt{G}G^{\mu\kappa}G^{00}G^{\lambda\nu}\Bigg[ K_{\kappa\lambda}^{\hspace{0.4cm}\alpha\beta}(\textbf{p}_1) K_{\mu 0}^{\hspace{0.35cm}\gamma\delta}(\textbf{p}_3)- K_{\kappa\lambda}^{\hspace{0.45cm}\gamma\delta}(\textbf{p}_3) K_{\mu 0}^{\hspace{0.35cm}\alpha\beta}(\textbf{p}_1)\Bigg]\Bigg)\,,\non
\ee

\subsection*{Sixth term: $J_6^\nu(x^0,x^\mu)=-ig\,\alpha \,G^{\kappa\sigma}\d_\kappa\left( \phi_{\sigma M}\phi^*{}^{M\nu}-\phi^*_{\sigma M}\phi^{M\nu}\right)$} 
Finally, the Fourier transform of the 6th term is
\be
    J^\nu_{6}(x^0,\textbf{p}_2)
    &=& -G^{MD}G^{\nu\lambda}\int d^dx e^{-i p_2 x}\left(ig\alpha G^{\kappa\sigma}\d_\kappa\left( \phi_{\sigma M}\phi^*{}_{D\lambda}-\phi^*_{\sigma M}\phi_{D\lambda}\right)\right)\nonumber\\
    &=&-G^{MD}G^{\nu\lambda}\int \frac{d^dk_1}{(2\pi)^d}\frac{d^dk_2}{(2\pi)^d} (2\pi)^d\delta^d(\textbf{k}_2+\textbf{k}_1-\textbf{p}_2)\bigg[ig\alpha G^{\kappa\sigma} i(k_2+k_1)_{\kappa}\nonumber\\
    &&\Big( K_{\sigma M}{}^{\gamma\delta}(\textbf{k}_2)\bar K_{D\lambda}{}^{\alpha\beta}(\textbf{k}_1)-\bar K_{\sigma M}{}^{\alpha\beta}(\textbf{k}_1)K_{D\lambda}{}^{\gamma\delta}(\textbf{k}_2)\Big)\bigg]\hat\phi^{(0)}{}^*_{\alpha\beta}(k_1)\hat\phi^{(0)}_{\gamma\delta}(\textbf{k}_2)\,,\non\\
\ee
with different terms given by
\be
&& \hspace*{-.4in}\frac{(2\pi)^d (2\pi)^d\delta^2}{\delta\hat\phi^*{}^{(0)}_{\alpha\beta}(-\textbf{p}_1)\delta\hat\phi{}^{(0)}_{\gamma\delta}(-\textbf{p}_3) }\left[-ig\alpha G^{\kappa\sigma}\d_\kappa\left( \phi_{\sigma 0}\phi^*{}^{0\nu}-\phi^*_{\sigma 0}\phi^{0\nu}\right)\right]\non\\
&=&ig\alpha\, G^{\kappa\sigma}G^{00}G^{\nu\lambda} (2\pi)^d\delta^d(\textbf{p}_1+\textbf{p}_2+\textbf{p}_3) i(p_1+p_3)_{\kappa}\Big( K_{\sigma 0}{}^{\gamma\delta}(\textbf{p}_3) K_{0\lambda}{}^{\alpha\beta}(\textbf{p}_1)- K_{\sigma 0}{}^{\alpha\beta}(p_1)K_{0\lambda}{}^{\gamma\delta}(\textbf{p}_3)\Big),\non
\ee

\be
&& \hspace*{-.4in}\frac{(2\pi)^d (2\pi)^d\delta^2}{\delta\hat\phi^*{}^{(0)}_{\alpha\beta}(-\textbf{p}_1)\delta\hat\phi{}^{(0)}_{\gamma\delta}(-\textbf{p}_3) }\left[-ig\alpha G^{\kappa\sigma}\d_\kappa\left( \phi_{\sigma \mu}\phi^*{}^{\mu\nu}-\phi^*_{\sigma \mu}\phi^{\mu\nu}\right)\right]\non\\
&=&ig\alpha\, G^{\kappa\sigma}G^{\mu\eta}G^{\nu\lambda} (2\pi)^d\delta^d(\textbf{p}_1+\textbf{p}_2+\textbf{p}_3) i(p_1+p_3)_{\kappa}\Big( K_{\sigma \mu}{}^{\gamma\delta}(\textbf{p}_3) K_{\eta\lambda}{}^{\alpha\beta}(\textbf{p}_1)- K_{\sigma \mu}{}^{\alpha\beta}(\textbf{p}_1)K_{\eta\lambda}{}^{\gamma\delta}(\textbf{p}_3)\Big)\,,\non
\ee

\section{Some details of flat limit computations}
\label{secf}

In this appendix, we give some details of the flat space limit. In taking the flat limit, we need to deal with the derivative of the Btb propagators. In subsection \ref{flbp}, we shall obtain an expression for the flat limit of the derivative of the Btb propagators. In subsection \ref{flat3pointlimit}, we shall determine the contributions of different terms in the expression of the 3-point function, to the flat limit. 

\subsection{Flat limit of derivatives of the Btb propagators}
\label{flbp}
To obtain the flat limit of the derivatives of the Btb propagators, we start with the following identity satisfied by the modified Bessel function of the second kind
\begin{equation}
\partial_z(z^m K_n(k z))=(m-n)z^{m-1}K_n(kz)-kz^mK_{n-1}(kz),\label{eq:BesselDer}
\end{equation}
Schematically, the general form of the bulk-to-boundary propagator in \eqref{f346} has the form
\begin{equation}
\label{eq:B2bSchematic}
K_{MN}{}^{AB}(z,\textbf{k})=\sum_n Q_{MN}{}^{AB}(n)\,z^{a(n)} \,K_{b(n)}(kz)\,,
\end{equation}
where $n$ is a summation index, $Q$s are independent of $z$, $a(n)$ is a $d$ dependent real number that does \textit{not} depend on any other parameter, while $b(n)$ depends on $\beta=\Delta-\f{d}{2}$. For the flat limit, the dependence on $z$ is important, and hence the exact form of the $Q$s or their tensor structure is not relevant. We also note that the large $L$ behaviours of $a(n)$ and $b(n)$ are given by 
\begin{align}\label{eq:abLargeL}
a(n)\;\;=\;\;\mathcal{O}(1)\quad;\qquad
b(n)\;\;=\;\;mL+\mathcal{O}(1).
\end{align}
Now, taking the derivative of \eqref{eq:B2bSchematic} and applying \eqref{eq:BesselDer} gives
\begin{align}
\frac{d}{dz}\left(K_{MN}{}^{AB}\right)=\sum_n Q_{MN}{}^{AB}(n)\left[\left(a(n)-b(n)\right)\,z^{a(n)-1} K_{b(n)}(kz)-k z^{a(n)} K_{b(n)-1}(kz)\right]\,.
\end{align}
We now recall that in the large $L$ limit, the modified Bessel function of the second kind behaves as \cite{Marotta:2024sce}
\begin{align}
\label{eq:KeyRule}
  K_{\Delta-\frac{d}{2}+c}(kz)= \left(\frac{\pi}{2EL}\right)^{\tfrac{1}{2}} 
\left(\frac{k}{m+E}\right)^{-mL - c} e^{-EL-E\uptau }\left(1+O\left(\frac{1}{L}\right)\right)\,.
\end{align}
The only $c$ dependence is in the exponent of the second term. Additionally, we also note that in the large $L$ limit, we have
\be
z=Le^{\frac{\uptau }{L}}= L+\uptau+ O\left(\f{1}{L}\right)\,.
\ee
Using the above identities together with \eqref{eq:abLargeL} gives, in the large L limit
\be
\frac{d}{dz}\left(K_{MN}{}^{AB}\right)&=&
\sum_nQ_{MN}{}^{AB}L^{a(n)}\left(-mK_{b(n)}(kz)-kK_{b(n)-1}(kz)\right)+O\left(\f{1}{L}\right)\non\\
&=&
-E\;\sum_nQ_{MN}{}^{AB}L^{a(n)}\left(\frac{\pi}{2EL}\right)^{\tfrac{1}{2}} 
\left(\frac{k}{m+E}\right)^{-b(n)} e^{-E(L+\uptau) } \,+O\left(\f{1}{L}\right)\,,\non
\\
\label{eq:DerB2b3}
\ee
where, in going to the second line, we have used equation \eqref{eq:KeyRule} and the identity
\be
m+\f{k^2}{m+E}\;\; =\;\; m+\f{E^2-m^2}{m+E} \;\;=\;\; E\,.
\ee
Next, following the same manipulations, we note that applying the flat limit directly to the bulk-to-boundary propagators gives 
\begin{align}
\label{eq:DerB2b4}
K_{MN}{}^{AB}=\sum_nQ_{MN}{}^{AB}L^{a(n)}\left(\frac{\pi}{2EL}\right)^{\tfrac{1}{2}} 
\left(\frac{k}{m+E}\right)^{-b(n)} e^{-E(L+\uptau) } \,+O\left(\f{1}{L}\right)\,.
\end{align}
Comparing \eqref{eq:DerB2b3} with \eqref{eq:DerB2b4}, we see that, at the leading order in $L$, the two expressions agree apart from a factor of $-E$. In other words, if $X_{MN}{}^{AB}$ is the flat limit of the bulk-to-boundary propagators and $Y_{MN}{}^{AB}$ is the flat limit of its derivative, then in coordinate space, we simply have
\begin{align}
Y_{MN}{}^{AB}\;\;=\;\;\p_\uptau X_{MN}{}^{AB}\;\;+\mathcal{O}\left(\frac{1}{L}\right)\;\;= \;\;\p_z X_{MN}{}^{AB}\;\;+\mathcal{O}\left(\frac{1}{L}\right)\,,
\end{align}
where, in writing the second equality, we have used 
\begin{align}
    \lim_{L\to\infty} \d_z  =  \lim_{L\to\infty} e^{-\uptau /L} \d_\uptau  = \d_\uptau\,, 
\end{align}
which shows that in the flat limit $L\rightarrow\infty$, taking the derivative with respect to $z$ and $\uptau$ coincide. 

\vspace*{.07in}The above result can be Wick rotated to Lorentzian signature by replacing $E\rightarrow -iE$ (or $\uptau =it$) which gives 
\begin{align}\label{derB2b}
Y^{(L)}_{MN}{}^{AB}=(-iE) X^{(L)}_{MN}{}^{AB}\;+\;\mathcal{O}\left(\frac{1}{L}\right).
\end{align}

\subsection{Flat limit of the 3-point function}\label{flat3pointlimit}

In this subsection, we consider each term of the current described in appendix \ref{G} and consider their contribution to the bulk 3-point function in the flat limit. We shall work in the bulk $\uptau$ coordinates. The 3-point function involves an integration over the bulk coordinate. In the limit $L\rightarrow\infty$, the integrand contains a factor $e^{-\uptau (E_1+E_{2}+E_3)}$ where $E_{1,3}=\sqrt{p_{1,3}^2+m^2}\equiv-iq^0_{1,3}$ and $E_2=p_2\equiv-iq^0_2$ (see equations \eqref{11213} and \eqref{11236}). This gives the desired energy conserving delta function as
\be
    \int_{-\infty}^\infty d\uptau \, e^{-\uptau (E_1+E_2+E_3)} e^{-L(E_1+E_2+E_3)}&=& \int_{-\infty}^\infty d\uptau \, e^{i\uptau (q^0_{1}+q^0_{2}+q^0_{3})} e^{i L(q^0_{1}+q^0_{2}+q^0_{3})}\non\\
    &=& 2\pi\delta(q^0_{1}+q^0_{2}+q^0_{3})\non\\
    &=& (2\pi i)\,\delta(E_1+E_2+E_3)
\ee
where we have set the exponent involving $L$ to 1 on the support of the delta function and the last equality corresponds to the Lorentzian signature. 

\vspace*{.07in}We now consider the $L\rightarrow\infty$ limit of each term in appendix \ref{G}. The contributions to the flat limit are as follows 
\subsection*{First term: $J_1^\nu(x^0,x^\mu)=\frac{ig}{2}\left(\phi_{MK}\d^\nu \phi^*{}^{MK}-\phi^*{}_{MK}\d^\nu \phi^{MK}\right)$} 
The 3 terms of $J_1^\nu$ contribute as follows
\begin{align}\label{f11}
&\lim_{L,\Delta\rightarrow\infty}\epsilon_1{}_{\alpha\beta}\,\epsilon_{\kappa}\,\epsilon_3{}_{\gamma\delta}\,\langle O^*{}^{\alpha\beta}(\textbf{p}_1)J_{1,1}^\kappa(\textbf{p}_2) O^{\gamma\delta}(\textbf{p}_3)\rangle\non\\
&=-g\,\epsilon_1{}_{\alpha\beta}\,\epsilon_{\kappa}\,\epsilon_3{}_{\gamma\delta}\,\int_{-\infty}^\infty d\uptau \Bigg(\frac{ k_{1}^{\alpha} \, k_{1}^{\beta} \, k_{3}^{\gamma} \, k_{3}^{\delta} \, \big( k_{3}^{\nu} - k_{1}^{\nu} \big) \, \pi^{\kappa}{}_{\nu} }
{ 2\, m^{4} \, \sqrt{Z_{A}Z_{\phi_1}Z_{\phi_3}}  }\,e^{-(E_1+E_2+E_3)(\uptau +L)}\Bigg)\,(2\pi)^d\delta^d(\textbf{p}_1+\textbf{p}_2+\textbf{p}_3)\non\\
&=-i(2\pi)^{d+1}\delta(E_{1}+E_{2}+E_{3})\delta^d(\textbf{p}_1+\textbf{p}_2+\textbf{p}_3)\frac{g}{2\, \sqrt{Z_{A}Z_{\phi_1}Z_{\phi_3}}} (\varepsilon_1{}_{00} \, \varepsilon_3^{00})\{(q_3-q_1)\cdot \varepsilon_2\}\,
\end{align}
\be
\label{f12}
&&\hspace*{-.3in}\lim_{L,\Delta\rightarrow\infty}\epsilon_1{}_{\alpha\beta}\,\epsilon_{\kappa}\,\epsilon_3{}_{\gamma\delta}\langle O^*{}^{\alpha\beta}(\textbf{p}_1)J_{1,2}^\kappa(\textbf{p}_2) O^{\gamma\delta}(\textbf{p}_3)\rangle\non\\
&=&-g\,(2\pi)^d\delta^d(\textbf{p}_1+\textbf{p}_2+\textbf{p}_3)\,\epsilon_1{}_{\alpha\beta}\,\epsilon_{\kappa}\,\epsilon_3{}_{\gamma\delta}\frac{ -1 }{\sqrt{Z_{A} Z_{\phi_1} Z_{\phi_3}}}\non\\
\,\,&&\int_{-\infty}^\infty d\uptau  \Bigg[\left(\frac{k_1^\alpha \tilde\pi_1^{\beta}{}_\mu+k_1^\beta\tilde\pi_1^{\alpha}{}_\mu}{2m}\right)\left(\frac{k_3^\gamma \tilde\pi_3^{\delta\mu}+k_3^\delta\tilde\pi_3^{\gamma\mu}}{2m}\right)\,(k_3^\nu-k_1^\nu)\pi^\kappa{}_\nu
\Bigg]e^{-(E_1+E_2+E_3)(\uptau +L)}\non\\
&=&-(2\pi)^{d+1}\,\delta(E_1+E_2+E_3)\delta^d(\textbf{p}_1+\textbf{p}_2+\textbf{p}_3)\frac{ig}{\, \sqrt{Z_{A}Z_{\phi_1}Z_{\phi_3}}} (\varepsilon_1{}_{0\mu} \, \varepsilon_3^{0\mu})(q_3-q_1)\cdot \varepsilon_2\,
\ee
\be
\label{f13}
&&\hspace*{-.3in}\lim_{L,\Delta\rightarrow\infty}\epsilon_1{}_{\alpha\beta}\,\epsilon_{\kappa}\,\epsilon_3{}_{\gamma\delta}\,\langle O^*{}^{\alpha\beta}(\textbf{p}_1)J_{1,3}^\kappa(\textbf{p}_2) O^{\gamma\delta}(\textbf{p}_3)\rangle\non\\
&=&-g\,(2\pi)^d\delta^d(\textbf{p}_1+\textbf{p}_2+\textbf{p}_3)\,\epsilon_1{}_{\alpha\beta}\,\epsilon_{\kappa}\,\epsilon_3{}_{\gamma\delta}\frac{ 1 }{ 8 \, \sqrt{Z_A Z_{\phi_1} Z_{\phi_3}} }\non\\
&&\int_{-\infty}^\infty d\uptau \Bigg( \,
\pi^{\kappa}{}_{\nu}
(k_3^{\nu}-k_1^{\nu}) \Big(
 (\tilde{\pi}_1^{\alpha m} \tilde{\pi}_1^{\beta a} 
  + \tilde{\pi}_1^{\alpha a} \tilde{\pi}_1^{\beta m})
 (\tilde{\pi}_3^{\gamma}{}_{m} \tilde{\pi}_3^{\delta}{}_{a} 
  + \tilde{\pi}_3^{\gamma}{}_{a} \tilde{\pi}_3^{\delta}{}_{m})
\Big)\,e^{-(E_1+E_2+E_3)(\uptau +L)}\Bigg)\non\\
&=&-(2\pi)^{d+1} \,\delta(E_1+E_2+E_3)\delta^d(\textbf{p}_1+\textbf{p}_2+\textbf{p}_3)\frac{ig}{2\, \sqrt{Z_{A}Z_{\phi_1}Z_{\phi_3}}} (\varepsilon_1{}_{\mu\nu} \, \varepsilon_3^{\mu\nu})\{(q_3-q_1)\cdot \varepsilon_2\}\,
\ee
\bigskip
Summing the contributions \eqref{f11}, \eqref{f12}, and \eqref{f13}, we obtain
\begin{align}
&\lim_{L,\Delta\rightarrow\infty}\epsilon_1{}_{\alpha\beta}\,\epsilon_{\kappa}\,\epsilon_3{}_{\gamma\delta}\,\langle O^*{}^{\alpha\beta}(\textbf{p}_1)J_{1}^\kappa(\textbf{p}_2) O^{\gamma\delta}(\textbf{p}_3)\rangle\;=\; \,-i\,\frac{g}{2} \Upsilon\;\varepsilon_1{}_{ab} \, \varepsilon_3^{ab}(q_3-q_1)\cdot \varepsilon_2\,
\end{align}
where, we have defined 
\be
\Upsilon \equiv (2\pi)^{d+1} \,\delta(E_1+E_2+E_3)\delta^d(\textbf{p}_1+\textbf{p}_2+\textbf{p}_3)\frac{1}{\, \sqrt{Z_{A}Z_{\phi_1}Z_{\phi_3}}}
\ee
\subsection*{Second term: $J_2^\nu(x^0,x^\mu)=-\,{ig}\left(G^{00}\phi_{M0}\d_0\phi^*{}^{M\nu}-G^{00}\phi^*_{M0}\d_0\phi^{M\nu}\right)$} 
For the $J_2$ contribution, we follow the same procedure. In this case, we need to use the flat limit of the derivative of the Btb propagators derived in \eqref{derB2b}. In the flat limit, the two terms of $J_2^\nu$ give the following contributions
\begin{align}
\lim_{L,\Delta\rightarrow\infty}\epsilon_1{}_{\alpha\beta}\,\epsilon_{\kappa}\,\epsilon_3{}_{\gamma\delta}\,\langle O^*{}^{\alpha\beta}(\textbf{p}_1)J_{2,1}^\kappa(\textbf{p}_2) O^{\gamma\delta}(\textbf{p}_3)\rangle
&{=}-ig\Upsilon\Bigl(\,\varepsilon_{2}^\mu\,\varepsilon^1_{\mu 0}\,\varepsilon_3^{00}\,q^1_0-(1\leftrightarrow 3)\Bigl)
\end{align}
\begin{align}
\lim_{L,\Delta\rightarrow\infty}\epsilon_1{}_{\alpha\beta}\,\epsilon_{\kappa}\,\epsilon_3{}_{\gamma\delta}\,\langle O^*{}^{\alpha\beta}(\textbf{p}_1)J_{2,2}^\kappa(\textbf{p}_2) O^{\gamma\delta}(\textbf{p}_3)\rangle
&=-ig\Upsilon\left(\,\varepsilon_{2}^\mu\,\varepsilon^1_{\mu \nu}\,\varepsilon_3^{\nu 0}\,q^1_0-(1\leftrightarrow 3)\right)
\end{align}
In writing the above expressions, we have used $q^1_M=(-E,k_\mu)=(-E,k^\mu)$ for the flat space momenta in the Lorentzian signature.

\subsection*{Third term: $J_3^\nu(x^0,x^\mu)=-ig\left(G^{\eta\sigma}\phi_{M\eta}\d_\sigma\phi^*{}^{M\nu}-G^{\eta\sigma}\phi^*_{M\eta}\d_\sigma\phi^{M\nu}\right)$} 
The two terms of $J_3^\nu$ give
\begin{align}
&\lim_{L,\Delta\rightarrow\infty}\epsilon_1{}_{\alpha\beta}\,\epsilon_{\kappa}\,\epsilon_3{}_{\gamma\delta}\,\langle O^*{}^{\alpha\beta}(\textbf{p}_1)J_{3,1}^\kappa(\textbf{p}_2) O^{\gamma\delta}(\textbf{p}_3)\rangle
=
-ig\Upsilon\left(\,\varepsilon_{2}^\mu\,\varepsilon^1_{\mu 0}\,\varepsilon_3^{0 \sigma}\,q^1_\sigma-(1\leftrightarrow 3)\right),\non\\[.3cm]
&\lim_{L,\Delta\rightarrow\infty}\epsilon_1{}_{\alpha\beta}\,\epsilon_{\kappa}\,\epsilon_3{}_{\gamma\delta}\,\langle O^*{}^{\alpha\beta}(\textbf{p}_1)J_{3,2}^\kappa(\textbf{p}_2) O^{\gamma\delta}(\textbf{p}_3)\rangle
=
-ig\Upsilon\left(\,\varepsilon_{2}^\mu\,\varepsilon^1_{\mu \nu}\,\varepsilon_3^{\nu \sigma}\,q^1_\sigma-(1\leftrightarrow 3)\right),
\end{align}

\subsection*{Fourth term: $J_4^\nu(x^0,x^\mu)=ig\left(\phi^*{}^{MN}\phi_{MQ}-\phi^{MN}\phi^*_{MQ}\right)$} 
The two terms of $J_4^\nu$ give
\begin{align}
\lim_{L,\Delta\rightarrow\infty}\epsilon_1{}_{\alpha\beta}\,\epsilon_{\kappa}\,\epsilon_3{}_{\gamma\delta}\,\langle O^*{}^{\alpha\beta}(\textbf{p}_1)J_{4,1}^\kappa(\textbf{p}_2) O^{\gamma\delta}(\textbf{p}_3)\rangle
&\,=\lim_{L\to\infty}\Upsilon\left[
\,\frac{g}{L}\,\varepsilon_{2}^\mu\,\varepsilon^1_{\mu 0}\varepsilon_3^{00}-(1\leftrightarrow 3)\right]=0,\non\\
\lim_{L,\Delta\rightarrow\infty}\varepsilon_1{}_{\alpha\beta}\,\varepsilon_{\kappa}\,\varepsilon_3{}_{\gamma\delta}\,\langle O^*{}^{\alpha\beta}(\textbf{p}_1)J_{4,2}^\kappa(\textbf{p}_2) O^{\gamma\delta}(\textbf{p}_3)\rangle
&\,=\,\lim_{L\to\infty}\Upsilon\left[
\frac{g}{L}\,\varepsilon_{2}^\mu\,\varepsilon^1_{\mu \sigma}\,\varepsilon_3^{\sigma 0}-(1\leftrightarrow 3)\right]=0.
\end{align}
This shows that the flat limit contribution from $J_4$ is subleading in $L$ compared to the contributions from $J_1, J_2$ and $J_3$ (as well as $J_5$ and $J_6$ given below) and thus vanishes in the flat limit. This is not surprising because the $J_4$ piece comes \textit{only} from the Christoffel connection terms in the action. Hence, in the flat limit, one would expect their contributions to vanish, which is precisely what happens.

\vspace*{.07in}Summing the contributions of $J_2, J_3$ and $J_4$, we get
\be
&&\hspace*{-.3in}\lim_{L,\Delta\rightarrow\infty}\epsilon_1{}_{\alpha\beta}\,\epsilon_{\kappa}\,\epsilon_3{}_{\gamma\delta}\,\langle O^*{}^{\alpha\beta}(\textbf{p}_1)\left(J_{2}^\kappa(\textbf{p}_2)+J_{3}^\kappa(\textbf{p}_2)+J_{4}^\kappa(\textbf{p}_2)\right) O^{\gamma\delta}(\textbf{p}_3)\rangle\non\\
&=&-ig\Upsilon\left(\,\varepsilon_{2}^a\,\varepsilon^1_{ab}\,\varepsilon_3^{bc}\,q^1_c-(1\leftrightarrow 3)\right)
\ee
where have used the axial gauge condition $\varepsilon_2^0=0$. 
\subsection*{Fifth term: $J_5^\nu(x^0,x^\mu)=-\frac{ig\,\alpha }{\sqrt{G}}\d_0\Bigg(\sqrt{G}G^{00}\left( \phi_{0P}\phi^*{}^{P\nu}-\phi^*_{0P}\phi^{P\nu}\right)\Bigg)$}
The two terms of $J_5^\nu$ give 
\begin{align}
\lim_{L,\Delta\rightarrow\infty}\epsilon_1{}_{\alpha\beta}\,\epsilon_{\kappa}\,\epsilon_3{}_{\gamma\delta}\,\langle O^*{}^{\alpha\beta}(\textbf{p}_1)J_{5,1}^\kappa(\textbf{p}_2) O^{\gamma\delta}(\textbf{p}_3)\rangle&\,=\,
-ig\Upsilon\left(
\,\alpha\,\varepsilon_2^{\mu}\,\varepsilon^1_{\mu 0}\,\varepsilon_{3}^{00}\,(q^1_0+q^3_0)-(1\leftrightarrow 3)\right),\non\\[.3cm]
\lim_{L,\Delta\rightarrow\infty}\epsilon_1{}_{\alpha\beta}\,\epsilon_{\kappa}\,\epsilon_3{}_{\gamma\delta}\,\langle O^*{}^{\alpha\beta}(\textbf{p}_1)J_{5,2}^\kappa(\textbf{p}_2) O^{\gamma\delta}(\textbf{p}_3)\rangle
&=
-ig\Upsilon\left(\,\alpha\,\varepsilon_2^{\mu}\,\varepsilon^1_{\mu \nu}\,\varepsilon_{3}^{\nu 0}\,(q^1_0+q^3_0)-(1\leftrightarrow 3)\right),
\end{align}
\subsection*{Sixth term: $J_6^\nu(x^0,x^\mu)=-ig\,\alpha \,G^{\kappa\sigma}\d_\kappa\left( \phi_{\sigma M}\phi^*{}^{M\nu}-\phi^*_{\sigma M}\phi^{M\nu}\right)$} 
The two terms of $J_6^\nu$ give
\begin{align}
\lim_{L,\Delta\rightarrow\infty}\epsilon_1{}_{\alpha\beta}\,\epsilon_{\kappa}\,\epsilon_3{}_{\gamma\delta}\,\langle O^*{}^{\alpha\beta}(\textbf{p}_1)J_{6,1}^\kappa(\textbf{p}_2) O^{\gamma\delta}(\textbf{p}_3)\rangle
&=
-ig\Upsilon\left(\,\alpha\,\varepsilon_2^{\mu}\,\varepsilon^1_{\mu 0}\,\varepsilon_{3}^{0\sigma}\,(q^1_\sigma+q^3_\sigma)-(1\leftrightarrow 3)\right),\non\\[.3cm]
\lim_{L,\Delta\rightarrow\infty}\epsilon_1{}_{\alpha\beta}\,\epsilon_{\kappa}\,\epsilon_3{}_{\gamma\delta}\,\langle O^*{}^{\alpha\beta}(\textbf{p}_1)J_{6,2}^\kappa(\textbf{p}_2) O^{\gamma\delta}(\textbf{p}_3)\rangle
&=-ig\Upsilon
\left(\,\alpha\,\varepsilon_2^{\mu}\,\varepsilon^1_{\mu \nu}\,\varepsilon_{3}^{\nu\sigma}\,(q^1_\sigma+q^3_\sigma)-(1\leftrightarrow 3)\right).
\end{align}
Summing the contributions from $J_5$ and $J_6$, we obtain
\be\label{f31}
\lim_{L,\Delta\rightarrow\infty}\epsilon_1{}_{\alpha\beta}\,\epsilon_{\kappa}\,\epsilon_3{}_{\gamma\delta}\,\langle O^*{}^{\alpha\beta}(\textbf{p}_1)\left(J_{5}(\textbf{p}_2)^\kappa+J_{6}^\kappa(\textbf{p}_2)\right) O^{\gamma\delta}(\textbf{p}_3)\rangle\,{=}\,-ig\Upsilon\left(\,\alpha\,\varepsilon_{2}^M\,\varepsilon^1_{ab}\,\varepsilon_3^{bc}\,q^1_c-(1\leftrightarrow 3)\right),\non
\ee
where we also used the orthogonality relation $\varepsilon_3^{NS}\,p^3_S=0$.

\vspace*{.07in}Summing all the contributions from $J_1$ through $J_6$, we finally obtain
\be
&&\hspace*{-.3in}\lim_{L,\Delta\rightarrow\infty}\epsilon_1{}_{\alpha\beta}\,\epsilon_{\kappa}\,\epsilon_3{}_{\gamma\delta}\,\langle O^*{}^{\alpha\beta}(\textbf{p}_1)\,J^\kappa(\textbf{p}_2)\, O^{\gamma\delta}(\textbf{p}_3)\rangle\non\\
&=&
-\,\frac{ig}{\sqrt{Z_{A}Z_{\phi_1}Z_{\phi_3}}}\, (2\pi)^{d+1}\delta^{d+1}(q_1+q_2+q_3)\non\\
&&\left[\frac{1}{2} (\varepsilon_1{}_{ab} \, \varepsilon_3^{ab})\{(q_3-q_1)\cdot \varepsilon_2\}+\left\{(1+\alpha)\,\varepsilon_{2}^a\,\varepsilon^1_{ab}\,\varepsilon_3^{bc}q^1_c-(1\leftrightarrow 3)\right\}\right].
\ee
In the above expression, the first term comes from $J_1$, the $\alpha$ independent term inside the curly bracket comes from $J_2$ and $J_3$, whereas $\alpha$ dependent term comes from $J_5$ and $J_6$.

\section{Expected 3-point amplitude in flat space}\label{expflat3point}
In this appendix, we review the 3-point function involving a $U(1)$ charged massive spin-2 field and the massless gauge field in the flat Minkowski spacetime. The 3-point Feynman amplitude is kinematically forbidden for real momenta. One may consider complex momenta or masses of incoming and outgoing massive fields to be different. For more details on this point, see appendix H of \cite{Marotta:2024sce}. Below, we summarise the expected 3-point function in $d+1$ dimensional flat spacetimes.  

\vspace*{.07in}The action describing the complex massive spin-2 field and a masssless field in flat spacetimes is given by\footnote{The 3rd term of the action in \eqref{A.8n2} is expressed slightly differently in \cite{Marotta:2019cip}. However, using the identity 
\be
D_a{\phi^*}_{bc}D^c\phi^{ba} &=&D_a \phi^{* a b} D^c\phi_{c b} +i \hat g{\phi^*}^{ca}F_{ab}\phi^b_{~c}\;,
\ee
the kinetic terms in \eqref{A.8n2} can be shown to agree with the expression used in \cite{Marotta:2019cip}. The gyromagnetic interaction terms also agree if we identify $1+\hat\alpha$ above with the gyromagnetic ratio given in \cite{Marotta:2019cip}. } (see, e.g., \cite{Marotta:2019cip} for the one and two derivative terms)
\begin{eqnarray}
S&=&\int d^{d+1}x\;\biggl[\f{1}{4}F_{ab}F^{ab}+\frac{1}{2}D_a\phi^*_{bc}D^a{\phi}^{bc} 
-D_a{\phi^*}_{bc}D^c\phi^{ba} -\frac{1}{2} D_a{\phi^*}D^a\phi\nonumber\\
&&+\frac{1}{2}D_a{\phi^*}^{ab}D_b {\phi} +\frac{1}{2}D_a{\phi}^{ab}D_b {\phi^*}+\frac{m^2}{2} (\phi^*_{ab}{\phi}^{ab} -{\phi^*}\phi)-i \,\hat \alpha \,\hat g\, {\phi^*}^{ca}F_{ab}\phi^b_{~c}\non\\
&&+i\beta_1\hat g\, F^{ab}\left(\p_a \phi^{*}_{cd}\p^c\phi_b{}^d-\p_a\phi_{cd}\p^c \phi^*_b{}^d\right)+i\beta_3\hat g\,F^{ab}\left(\p_c\phi^*_{ad}\p^d\phi_b{}^c-\p_c\phi_{ad}\p^d\phi^*_b{}^c\right)\nonumber\\
&&+i\beta_2\hat g\,F^{ab}\left(\p_a\phi^*_{cd}\p_b\phi^{cd}\right)\biggl]\,,
\label{A.8n2}
\end{eqnarray}
where $D_a=\d_a+i\hat gA_a$. 

\vspace*{.07in}The AdS action \eqref{action} reproduces exactly the above action in the flat limit $L\to\infty$. The 3-point interaction vertex involving the massless field $A^a$ having momentum and polarization $q^a_1$ and $\varepsilon_{1}^a$ respectively and the massive spin-2 fields $\phi_{ab}$ and $\phi^*_{ab}$ with momenta and polarizations $(q_2, \varepsilon_2^{ab} )$ and $(q_3, \varepsilon_3^{ab})$ is given by ( see also \cite{Marotta:2019cip} for the $\beta_i$ independent terms)
\be
&&\hspace*{-.2in}V_{e;ab;cd}(q_1,q_2,q_3) \non\\
&=& \frac{i\hat g}{2}\,  \Bigg[ \frac{1}{2} (\eta_{ca}\eta_{db} + \eta_{cb}\eta_{da} - 2\eta_{cd}\eta_{ba})(q_2 - q_3)_e 
+ \frac{1}{2}\eta_{ec}\eta_{ab}(q_2 - q_3)_d + \frac{1}{2}\eta_{e d}\eta_{ab}(q_2 - q_3)_c\nonumber\\
&&  
+ \frac{1}{2}\eta_{e a}\eta_{cd}(q_2 - q_3)_b 
+ \frac{1}{2}\eta_{e b}\eta_{cd}(q_2 - q_3)_a  - \frac{1}{2}\eta_{e c}(\eta_{db}q_{2a} + \eta_{a d}q_{2b})
- \frac{1}{2}\eta_{e d}(\eta_{cb}q_{2a} + \eta_{ca}q_{2b}) \nonumber\\
&& + \frac{1}{2}\eta_{e a}(q_{3c}\eta_{db} + \eta_{cb}q_{3d})
+ \frac{1}{2}\eta_{e b}(q_{3c}\eta_{da} + \eta_{ca}q_{3d}) + \frac{(1+\hat \alpha)}{2}\biggl\{\eta_{e a}(q_{1d} \eta_{c b} + q_{1c} \eta_{d b})\nonumber\\
&& 
+  \eta_{e b}(q_{1c} \eta_{d a} + q_{1d} \eta_{c a}) - \eta_{e d}(q_{1a} \eta_{c b} + q_{1b} \eta_{c a})
- \eta_{e c}(q_{1a} \eta_{db} + q_{1b} \eta_{a d})\biggl\}\nonumber\\
&&+2\beta_1\mathcal{F}^{mn}{}_{e}(q_1)\left\{q_{2(c}\eta_{d)(a}\eta_{b)}{}_{n}q_{3m}-q_{3(a}\eta_{b)(c}\eta_{d)n}q_{2m} \right\}+2\beta_2\mathcal{F}^{mn}{}_{e}(q_1)q_{3m}q_{2n}\eta_{c(a}\eta_{b)d}\nonumber\\
&&+2\beta_3\mathcal{F}^{mn}{}_{e}(q_1)\left\{\eta_{m(c}q_{2d)}\eta_{n(a}q_{3b)}-\eta_{m(a}q_{3b)}\eta_{n(c}q_{2d)}\right\}\biggl]\,,
\ee
where
\begin{align}
    \mathcal{F}^{mne}(q)=q^m\eta^{ne}-q^n \eta^{me}.
\end{align}
Contracting the above expression with the polarisation tensors and using their properties, we get the desired result
\begin{align}
   A_3 &=\varepsilon_1^e\,\varepsilon_2^{ab}\,\varepsilon_3^{cd}\,V_{e;ab;cd}(q_1,q_2,q_3) \non\\
&=i\hat g\left[\frac{1}{2} (\varepsilon_2{}_{ab} \, \varepsilon_3^{ab})\{(q_3-q_1)\cdot \varepsilon_1\}+\left\{(1+\hat \alpha)\,\varepsilon_{1}^a\,\varepsilon_{2ab}\,\varepsilon_3^{bc}q_{1c}-(1\leftrightarrow 3)\right\}\right]\nonumber\\
&\,\,\,\,\,+i\hat{g}\,(q_1^m\varepsilon_1^n-q_1^n\varepsilon_1^m)\biggl[\beta_1\left\{q_{3m} (q_{2}\cdot\varepsilon_3\cdot\varepsilon_{2})_{n}-q_{2m}(q_{3}\cdot\varepsilon_2\cdot\varepsilon_{3})_{n}\right\}+\beta_2\,q_{3m}q_{2n}\,(\varepsilon_{2}\cdot \varepsilon_{3})\nonumber\\
&\,\,\,\,\,+\beta_3\left\{(\varepsilon_3\cdot q_2)_{m}\,(\varepsilon_2\cdot q_3)_{n}\,-(\varepsilon_3\cdot q_2)_{n}\,(\varepsilon_2\cdot q_3)_{m}\right\}\biggl]\,.
\end{align}
The flat limit result obtained in the previous appendix matches exactly with the above expression for $\beta_i=0$.

\end{document}